\documentclass[3p,twocolumn]{elsarticle}

\usepackage{amssymb}
\usepackage{graphics}

\journal{Astroparticle Physics}

\begin{document}

\begin{frontmatter}


%%\title{This is Title\tnoteref{label1}}
%%\author{Name\corref{cor1}\fnref{label2}}
%%\ead{email address}

\title{
On the Relation between the True Directions of Neutrinos and
the Reconstructed Directions of Neutrinos in
L/E Analysis Performed by Super-Kamiokande Collaboration Part~2
\\
{\large --- Four Possible L/E Analyses for the Maximum Oscillations by
the Computer Numerical Experiment---}
}

\author[HU]{E.~Konishi\corref{cor1}}
\ead{konish@si.hirosaki-u.ac.jp}
\author[KU]{Y.~Minorikawa}
\author[MoU]{V.I.~Galkin}
\author[MeU]{M.~Ishiwata}
\author[SU1]{I.~Nakamura}
\author[HU]{N.~Takahashi}
\author[ky]{M.~Kato}
\author[SU2]{A.~Misaki}
\address[HU]{
Graduate School of Science and Technology, Hirosaki University, Hirosaki, 036-8561, Japan }    
\address[KU]{
Department of Science, School of Science and Engineering, Kinki University, Higashi-Osaka, 577-8502, Japan }
\address[MoU]{
Department of Physics, Moscow State University, Moscow, 119992, Russia}
\address[MeU]{
Department of Physics, Faculty of Science and Technology,
Meisei University, Tokyo, 191-8506, Japan}
\address[SU1]{
Comprehensive Analysis Center for Science, Saitama University, 
Saitama, 338-8570, Japan}
\address[ky]{
Kyowa Interface Science Co.,Ltd., Saitama, 351-0033, Japan }
\address[SU2]{
Inovative Research Organization, Saitama University, Saitama,
 338-8570, Japan}

 \cortext[cor1]{Corresponding author}
\begin{abstract}
In the previous paper (Part~1), we have verified that 
{\it the SK assumption on the direction} does not hold in 
the analysis of neutrino events occurred inside the SK detector,
which is the cornerstone for their analysis of zenith angle distributions of neutrino events.  
Based on the correlation between 
$L_{\nu}$ and $L_{\mu}$ 
(Figures~16 to 18 in Part~1) 
and the correlation between $E_{\nu}$ and $E_{\mu}$ 
(Figure~19 in Part~1), 
we have made four possible $L/E$ analyses, namely
$L_{\nu}/E_{\nu}$, $L_{\nu}/E_{\mu}$, $L_{\mu}/E_{\mu}$
 and $L_{\mu}/E_{\nu}$.
 Among four kinds of $L/E$ analyses, we have shown that only
 $L_{\nu}/E_{\nu}$ analysis can give the signature of maximum 
oscillations clearly, not only the first maximum oscillation 
but also the second and third maximum oscillation and etc.,
as they should be,
while the $L_{\mu}/E_{\mu}$ analysis which are really done 
 by Super-Kamiokande Collaboration cannot give any maximum 
oscillation at all.
It is thus concluded from those results that the experiments 
with the use of the cosmic-ray beam for neutrino oscillation,
 such as Super-Kamiokande type experiment, are unable to lead the 
maximum oscillation from their $L/E$ analysis, because the incident
neutrino cannot be observed due to its neutrality.
Therefore, we would suggest Super-Kamiokande Collaboration 
to re-analyze the zenith angle distribution of the neutrino events 
which occur inside the detector carefully,
 since $L_{\nu}$ and $L_{\mu}$ are alternative expressions of 
the cosine of the zenith angle for the incident neutrino 
and that for the emitted muon, respectively. 
 
\end{abstract}


\end{frontmatter}

% \linenumbers

\section{Introduction}
\footnote{In order to understand the text of our paper well, we strongly 
suggest 
readers to look at the same paper on the WEB where every figures are 
presented in colors, because figures with colors are strongly impressive 
compare with those with monochrome. In the figures with colors, we 
classify neutrino events by blue points and 
aniti-neutrino events by orange ones.}
The specification of the oscillation parameters for neutrino oscillation 
is entirely based on the survival probability for a given flavor (Eq.(1)) 
in which two physical quantities to be measured, namely, the directions 
of incident neutrinos and their energies are included. 
However, these two quantities cannot be measured directly due to their 
neutrality and Super-Kamiokande Collaboration are forced to introduce
{\it the SK assumption on the direction}.

 As shown in Figures~11 to 13 and/or Figures~16 to 18 of the preceding 
paper\cite{Konishi2},
 we have shown that {\it the SK assumption on the direction}
% \footnote{
%The definition of {\it the SK assumption on the direction} is explicitly 
%given in the preceding paper\cite{Konishi2}, citing from the original 
%papers by Super-Kamiokande Collaboration.}
requiring that the directions of the incident neutrinos are the 
same as those of the emitted muons does not hold 
in the case of the neutrino events with the highest quality,
namely the single ring muon events due to quasi-elastic scattering(QEL) 
among {\it Fully Contained Events}.

 Also, in Figure~19 of the same paper, we have shown that the 
energies of the incident neutrinos cannot be determined uniquely
from those of the emitted muons.

Compared  Figures~11 to 18 with Figures~19 in the preceding 
paper\cite{Konishi2}, it is easily understood 
that the approximated $E_{\nu}$ by Eq.(9) in \cite{Konishi2}
does not bring a fatal error into the survival probability for a 
given flavor (Eq.(1) in the present paper)
in spite of the unsuitable theoretical treatment,
while {\it the SK assumption on the direction}
introduces the fatal error into the $L/E$ analysis.

In the present paper, we examine how does the variable $L/E$
in the survival probability for a given flavor influences over the 
results around $L/E$ analyses. In our computer numerical experiments,
it is possible to analyze four different kinds of $L/E$ distribution, namely, $L_{\nu}/E_{\nu}$, $L_{\nu}/E_{\mu}$,
$L_{\mu}/E_{\mu}$ and $L_{\mu}/E_{\nu}$ distributions, where
$L_{\nu}$, $L_{\mu}$, $E_{\nu}$ and $E_{\mu}$
 denote the flight length of neutrino,
 the corresponding flight length of emitted muon,
 the incident neutrino energy and the emitted muon energy in QEL, 
respectively.

%%%%%%%%%%%%%%%%%%%%%%%%%%%%%%%%%%%%%%%%%%%%%%%%%%%%%
\section{$L/E$ Distributions in Our Computer Numerical Experiment}

\begin{figure}
\begin{center}
\vspace{-1.2cm}
\rotatebox{90}{%
\resizebox{0.47\textwidth}{!}{%
  \includegraphics{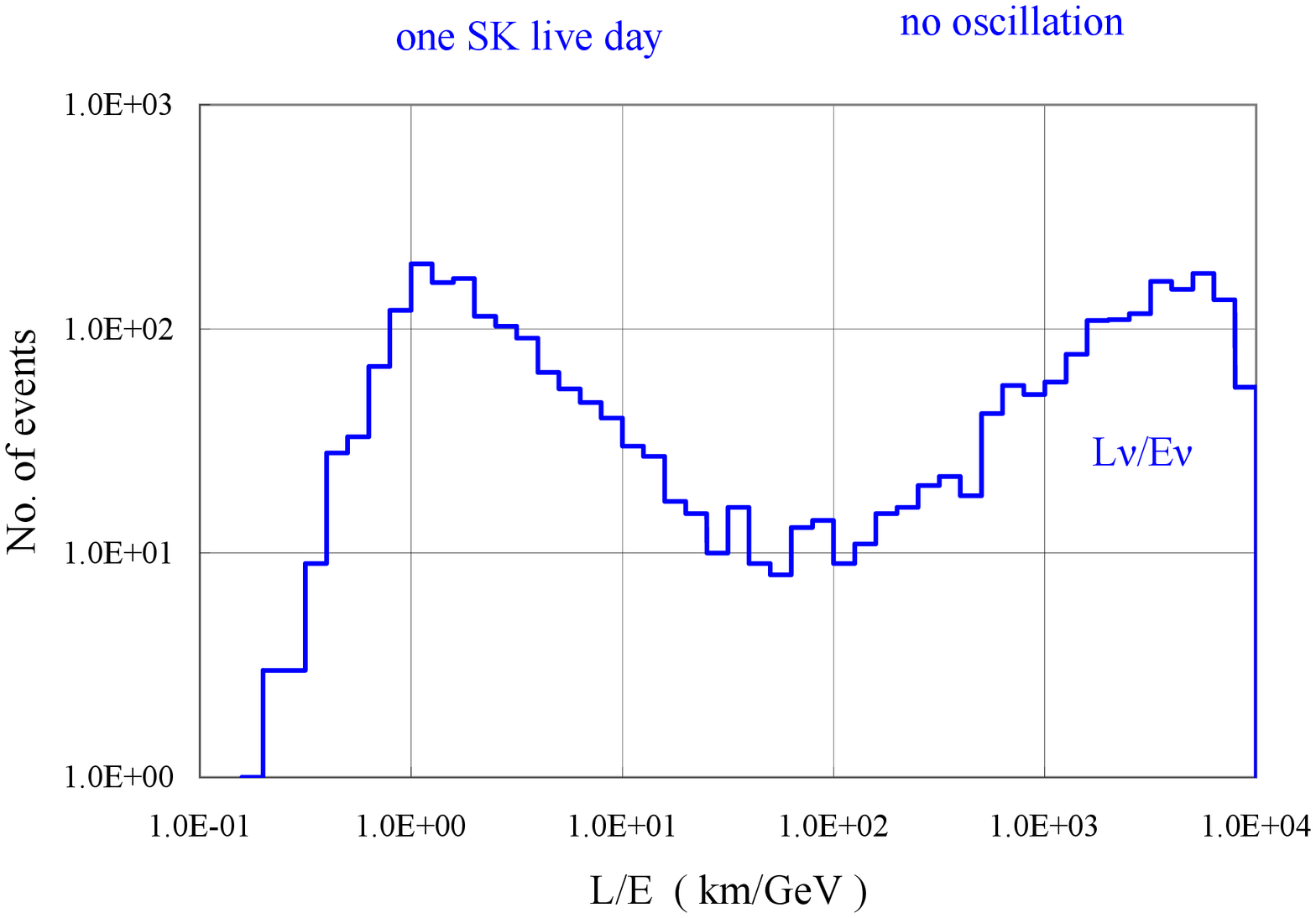}
}}
\vspace{-1.5cm}
\caption{$L_{\nu}/E_{\nu}$ distribution without oscillation
for 1489.2 live days (one SK live day).}
\label{figJ015}       % Give a unique label
\vspace{-0.5cm}
\hspace*{0.2cm}
\rotatebox{90}{%
\resizebox{0.44\textwidth}{!}{%
  \includegraphics{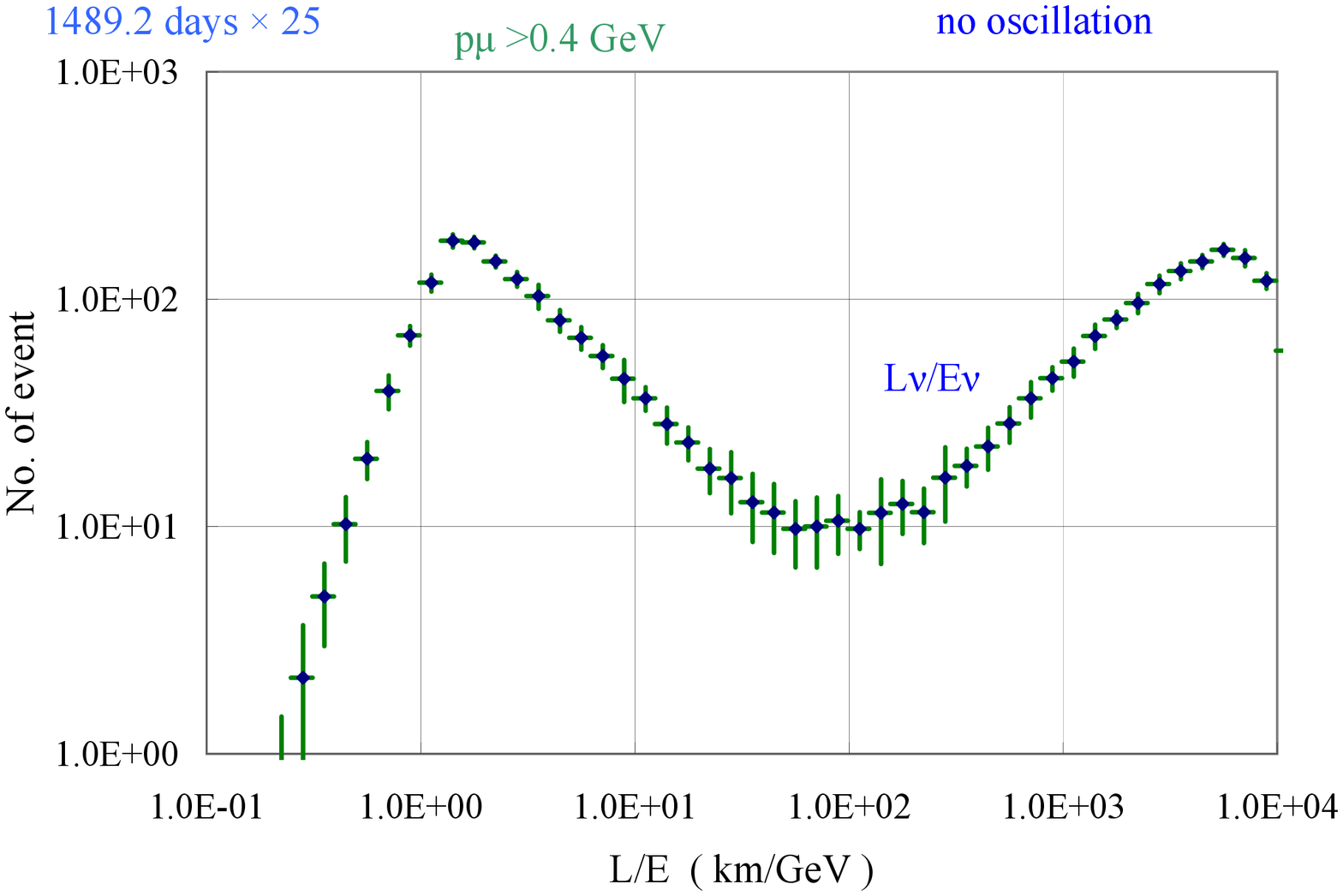}
}}
\vspace{-1.5cm}
\caption{$L_{\nu}/E_{\nu}$ distribution without oscillation
for 37230 live days (25 SK live days).}
\label{figJ016}       % Give a unique label
\vspace{-0.2cm}
\hspace*{-0.2cm}
\rotatebox{90}{%
\resizebox{0.4\textwidth}{!}{%
  \includegraphics{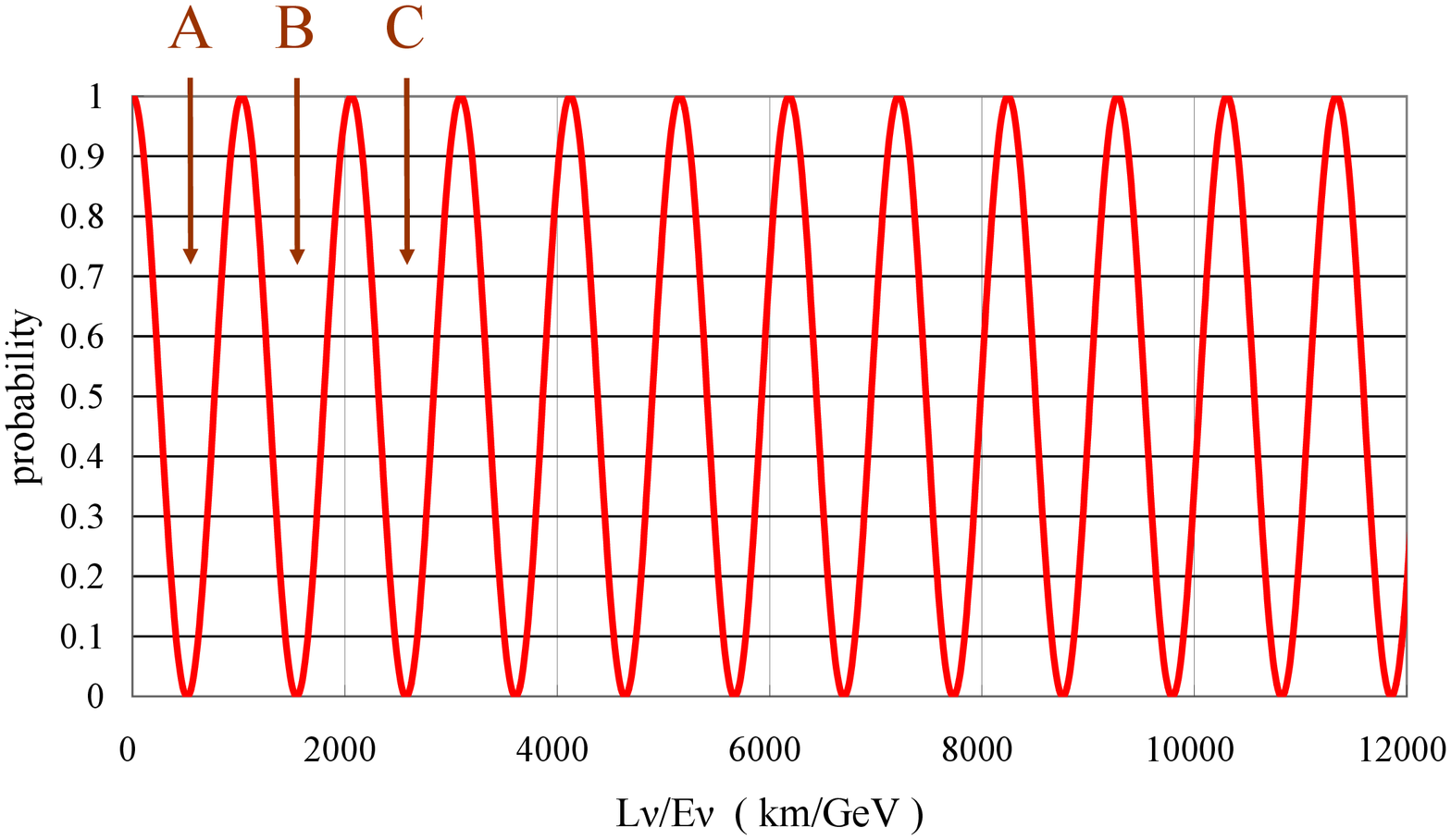}
}}
\vspace{-1.5cm}
\caption{Survival probability of
$P(\nu_{\mu} \rightarrow \nu_{\mu})$
as a function of $L_{\nu}/E_{\nu}$
under the neutrino oscillation parameters
obtained by Super-Kamiokande Collaboration.
A, B and C represent the first, the second and the third
maximum oscillation, respectively
.}
\label{figJ017}
\end{center}
\end{figure}

Here, we explain the procedure of our computer numerical experiment 
roughly (See details in the Appendices in the preceding paper
\cite{Konishi2}).  At first we construct a hypothetical SK detector in the 
computer. We randomly sample a neutrino event with both certain energy 
and zenith angle from neutrino spectra 
at the opposite side of the Earth, imaging the injection of the neutrino
 concerned into the detector. 
We pursue the neutrino concerned up to the inside 
detector where the neutrino interaction is 
expected\footnote{
Exactly speaking, instead of the description in the text, we adopt to 
sample the neutrino events from the neutrino interaction spectra inside 
the detector, which is mathematically equivalent to the procedure described in the text.
}. 
In the interaction (QEL) occurred inside the detector, 
we "measure" the muon energy from the random sampling of $Q^2$ for the
neutrino concerned and we pursue the muon concerned up to 
the end of the detector, taking into account of all 
physical processes due to the muon and judging whether the muon 
concerned belongs to {\it Fully Contained Event} or  
{\it Partially Contained Events}. 
There, we adopt {\it Fully Contained Events} only.
As the result of a series of these procedure, 
we are able to know a series of a pair of the neutrino with the known 
primary energy, $E_{\nu}$, 
 and the zenith angle, $cos\theta_{\nu}$ or $L_{\nu}$,
 and the produced muon with the energy $E_{\mu}$ 
and the zenith angle, $cos\theta_{\mu}$ or $L_{\mu}$.
In our computer numerical experiments, every physical process is 
treated stochastically and physical results are thus obtained, 
taking account of the stochastic characters inherent in their 
processes exactly.  Namely in this sense, there is one-to-one 
correspondence between "measured" neutrinos and "measured" their 
daughters muons, 
while one cannot generally specify the parent neutrinos from the 
measured muons in the real experiment.
%
% There is exactly one-to-one correspondence for the neutrinos and their 
%daughter muons among {\it Fully Contained Events}.

Our computer numerical experiments are carried out in the unit of
1489.2 days.  
The live days of 1489.2  is the total live days for the analysis of the 
neutrino events generated inside the detector used by Super-Kamiokande 
Collaboration\cite{Ashie2}.
Hereafter, we call 1489.2 live days as one SK live day.
We repeat one SK live day experiment as much as 25 times,
namely, the total live days for our computer numerical experiments 
is 37230 live days (25 SK live days).

\subsection{$L_{\nu}/E_{\nu}$ distribution}
\subsubsection{For null oscillation}

 In Figure~\ref{figJ015}, we show $L_{\nu}/E_{\nu}$ 
distribution without oscillation for one experiment (1489.2 live days)
 among twenty five computer numerical experiments (25 SK live days).
In those numerical experiments, there are statistical uncertainties only
which are due to both the stochastic character in the physical
 processes concerned and the geometry of the detector.
Therefore we add the standard deviation as for the statistical 
uncertainty around their average in the forthcoming graphs, if neccessary. 
 In Figure~\ref{figJ016}, we show the 
statistical uncertainty, the standard deviations around their average 
values through twenty five experiments.    
Similarly for
other possible combinations of $L$ and $E$ ( 
$L_{\nu}/E_{\mu}$, $L_{\mu}/E_{\mu}$ and $L_{\mu}/E_{\nu}$) for 
37230 live days (25 SK live days) we did so.
%, we added the statistical uncertainties, standard 
%deviations, around their averages.
%%%%%%%%%

\begin{figure}
\begin{center}
\vspace{-1.5cm}
\rotatebox{90}{%
\resizebox{0.45\textwidth}{!}{%
  \includegraphics{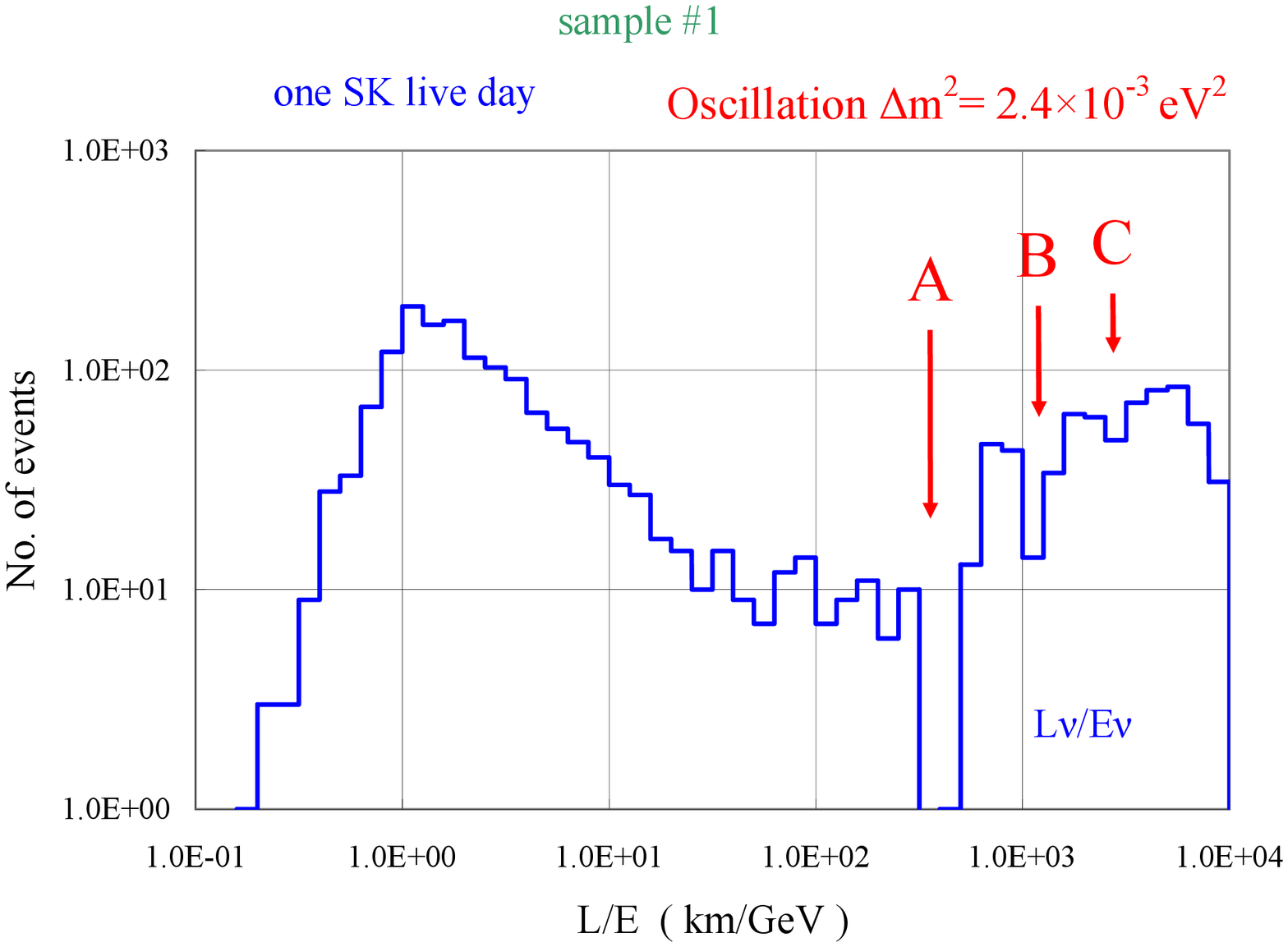}
}}
\vspace{-1.5cm}
\caption{$L_{\nu}/E_{\nu}$ distribution with oscillation
for 1489.2 live days (one SK live day), sample No.1.}
\label{figJ018}
\vspace{-0.5cm}
\rotatebox{90}{%
\resizebox{0.45\textwidth}{!}{%
  \includegraphics{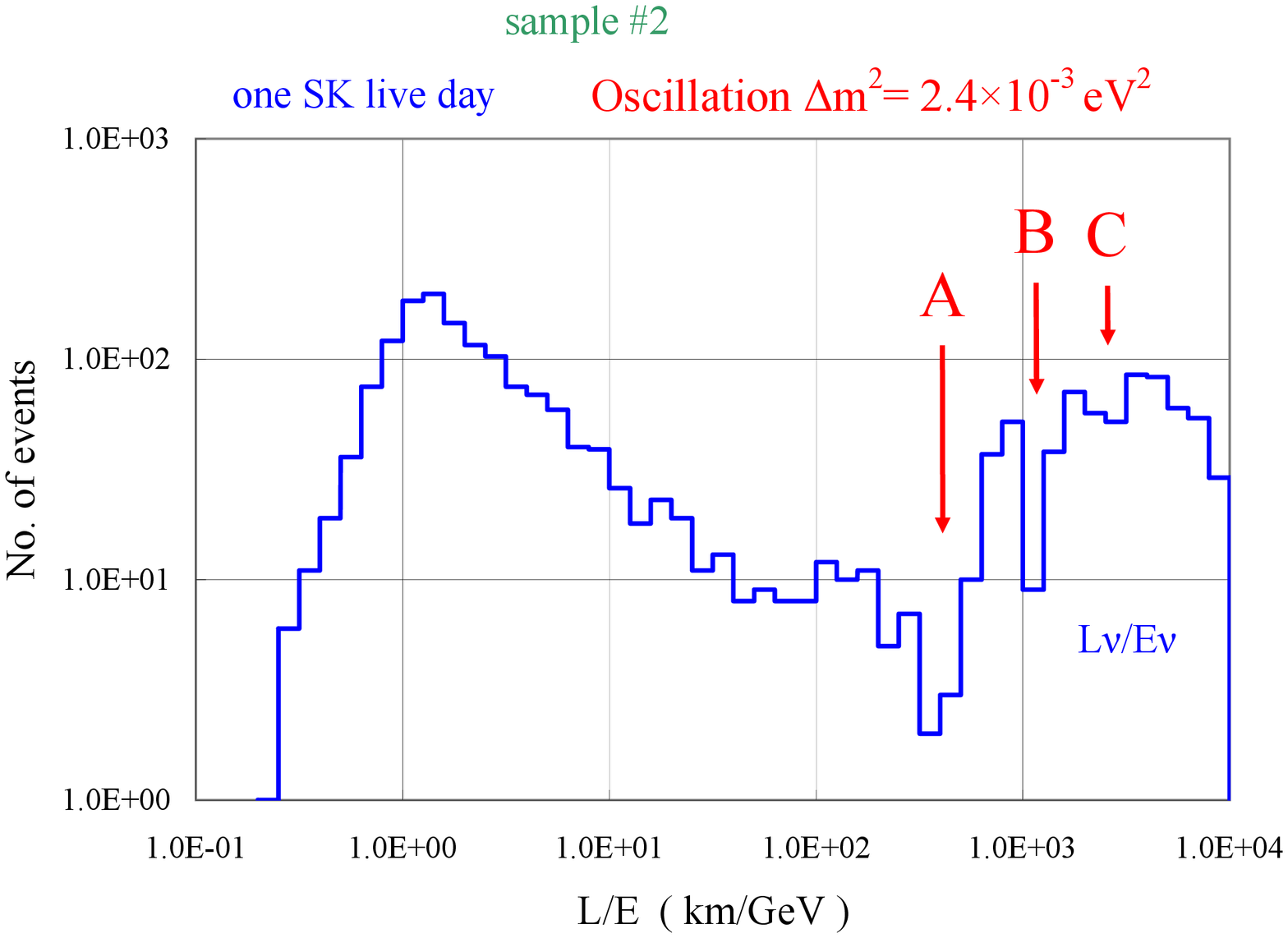}
}}
\vspace{-1.5cm}
\caption{$L_{\nu}/E_{\nu}$ distribution with oscillation
for 1489.2 live days (one SK live day), sample No.2.}
\label{figJ019}      
\vspace{-0.5cm}
\rotatebox{90}{%
\resizebox{0.45\textwidth}{!}{%
  \includegraphics{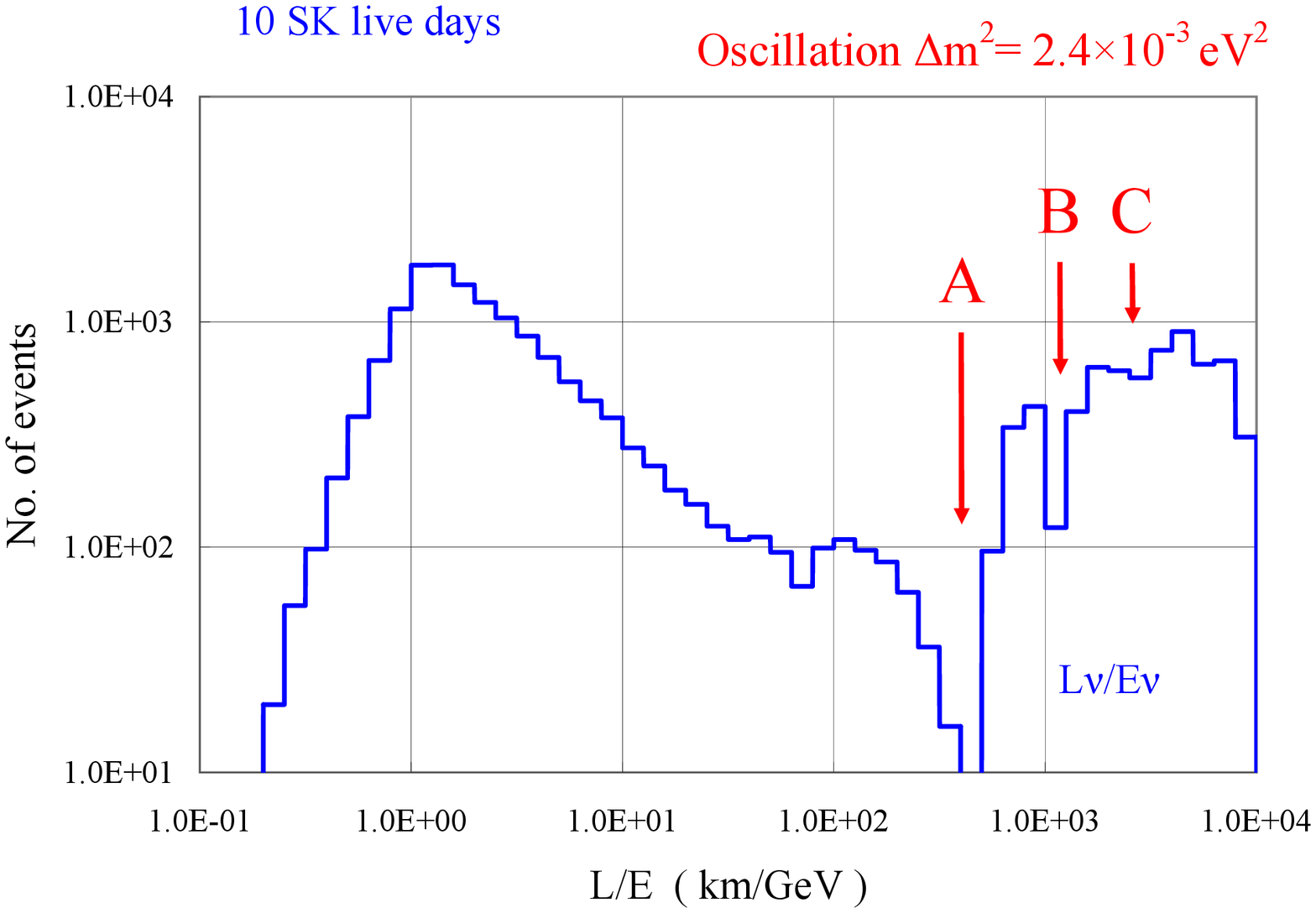}
}}
\vspace{-1.5cm}
\caption{$L_{\nu}/E_{\nu}$ distribution with oscillation
for 14892 live days (10 SK live days).}
\label{figJ020}       % Give a unique label
\end{center}
\end{figure}

In Figures~\ref{figJ015} and \ref{figJ016},
both distributions show the sinusoidal-like
 character for $L_{\nu}/E_{\nu}$ distribution, namely, 
the appearance of the top and the bottom, even for null
 oscillation. 
In this case, it should be noticed that their distributions are 
expressed in a logarithmic scale.
The uneven histograms in Figure~\ref{figJ015}, comparing 
 with those in Figure~\ref{figJ016}, show that the statistics of 
 Figure~\ref{figJ015} is not enough compared with that of 
Figure~\ref{figJ016}.
Roughly speaking, smaller parts of $L_{\nu}/E_{\nu}$ 
correspond to the contribution from downward neutrinos, 
larger parts of $L_{\nu}/E_{\nu}$ correspond to those from upward 
neutrinos and $L_{\nu}/E_{\nu}$ near the minimum 
correspond to the horizontal neutrinos, although the real situation is
more complicated, because the backscattering effect in QEL
as well as the azimuthal angle effect in QEL
could not be neglected as shown in 
the preceding paper\cite{Konishi2}.
From Figure~\ref{figJ016}, we understand that the bottom around
 70~km/GeV
corresponds to the contribution mainly from the horizontal-like direction
and has no relation with neutrino oscillation in any sense.

\subsubsection{For oscillation (SK oscillation parameters)}

% If we adopt SK neutrino oscillation parameters, then we could expect 
%the in the following equation:
The survival probability of a given flavor, such as $\nu_{\mu}$,
 is given by
$$
P(\nu_{\mu} \rightarrow \nu_{\mu}) 
= 1- sin^2 2\theta \cdot sin^2
(1.27\Delta m^2 L_{\nu} / E_{\nu} ).         \,\,\,(1)  
$$                                                    

%$$
% sin^2(1.27\Delta m^2 L_{\nu} / E_{\nu}) = 1.0 .  \,\,\,\,\,\,(5)  
%$$                                                    
\noindent Then, for maximum oscillations under SK neutrino oscillation 
parameters\cite{Ashie2},  we have
%for maximum oscillation.
% The Eq.(5) holds in the following case:
$$
 1.27\Delta m^2 L_{\nu} / E_{\nu} = (2n+1)\times\frac{\pi}{2},  \,\,\,\,\,\,(2)  
$$                                                    
where $sin^2 2\theta = 1.0$ and
$\Delta m^2 = 2.4\times 10^{-3}\rm{eV^2}$.
From Eq.(1), we have the following values of
$L_{\nu} / E_{\nu}$ for maximum oscillations.
\begin{eqnarray}
     L_{\nu}/E_{\nu} 
  &= 515 {\rm km/GeV}\,\,\,\,\, for\,\, n=0   \,\,\, (3-1)\nonumber\\
 &= 1540 {\rm km/GeV}\,\,\,   for\,\, n=1   \,\,\, (3-2)\nonumber\\ 
 &= 2575 {\rm km/GeV}\,\,\,   for\,\, n=2   \,\,\, (3-3) \nonumber\\
&{\rm and}\,{\rm so}\,{\rm on.}\nonumber
%\nonumber
\end{eqnarray}
In Figure~\ref{figJ017}, we give the survival probability 
$P(\nu_{\mu} \rightarrow \nu_{\mu})$ as a function of 
$L_{\nu} / E_{\nu}$ under the neutrino oscillation parameters obtained 
by Super-Kamiokande Collaboration.
 In cosmic ray experiments, the energy spectrum of the incident neutrinos 
is convoluted into the survival probability.

\begin{figure}
\begin{center}
\vspace{-2.0cm}
\hspace*{-0.5cm}
\rotatebox{90}{%
\resizebox{0.4\textwidth}{!}{%
  \includegraphics{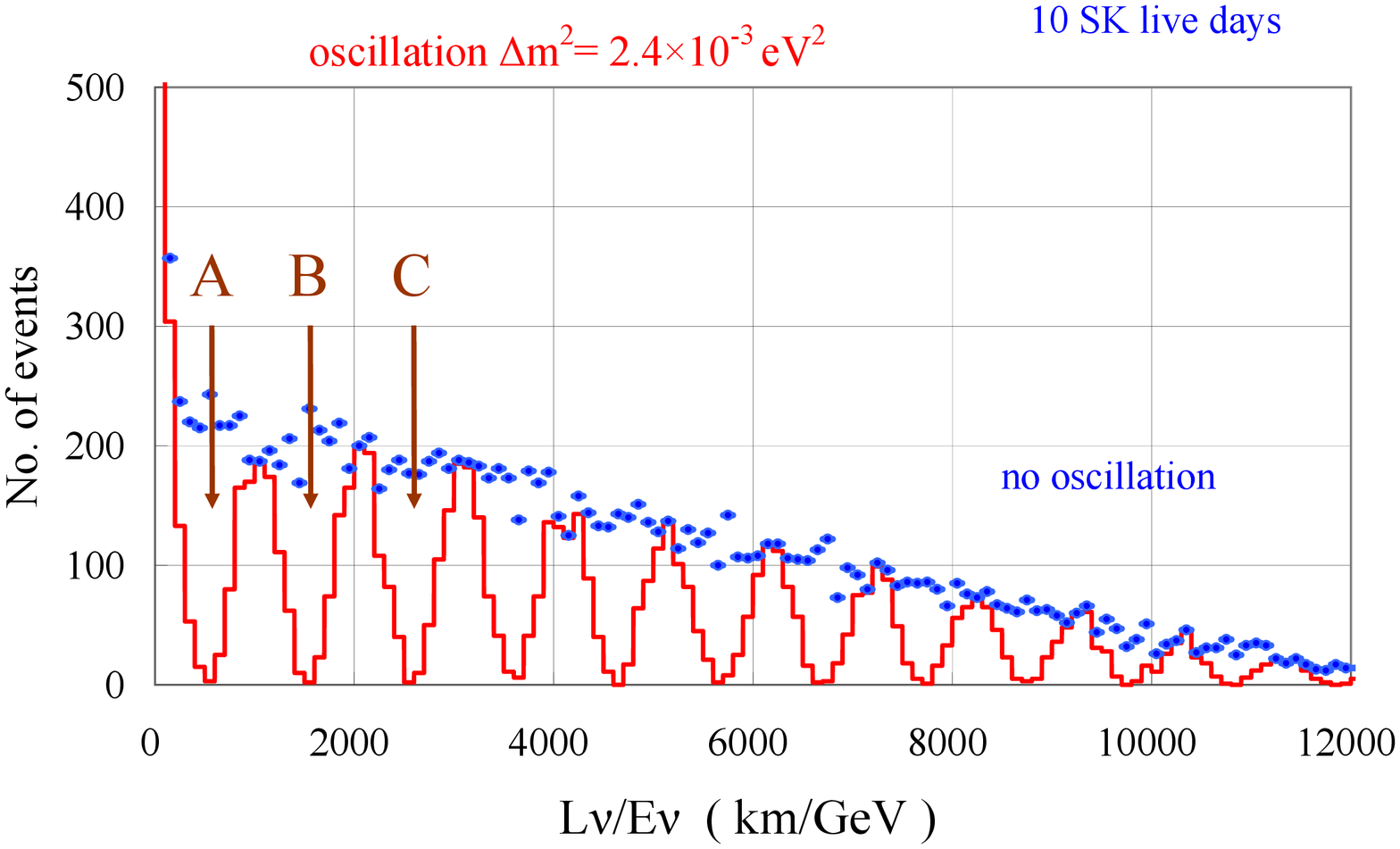}
}}
\vspace{-1.5cm}
\caption{$L_{\nu}/E_{\nu}$ distribution
with and without oscillation for 14892 live days (10 SK live days).}
\label{figJ024}       % Give a unique label

\vspace{-1.0cm}
\rotatebox{90}{%
\resizebox{0.45\textwidth}{!}{%
  \includegraphics{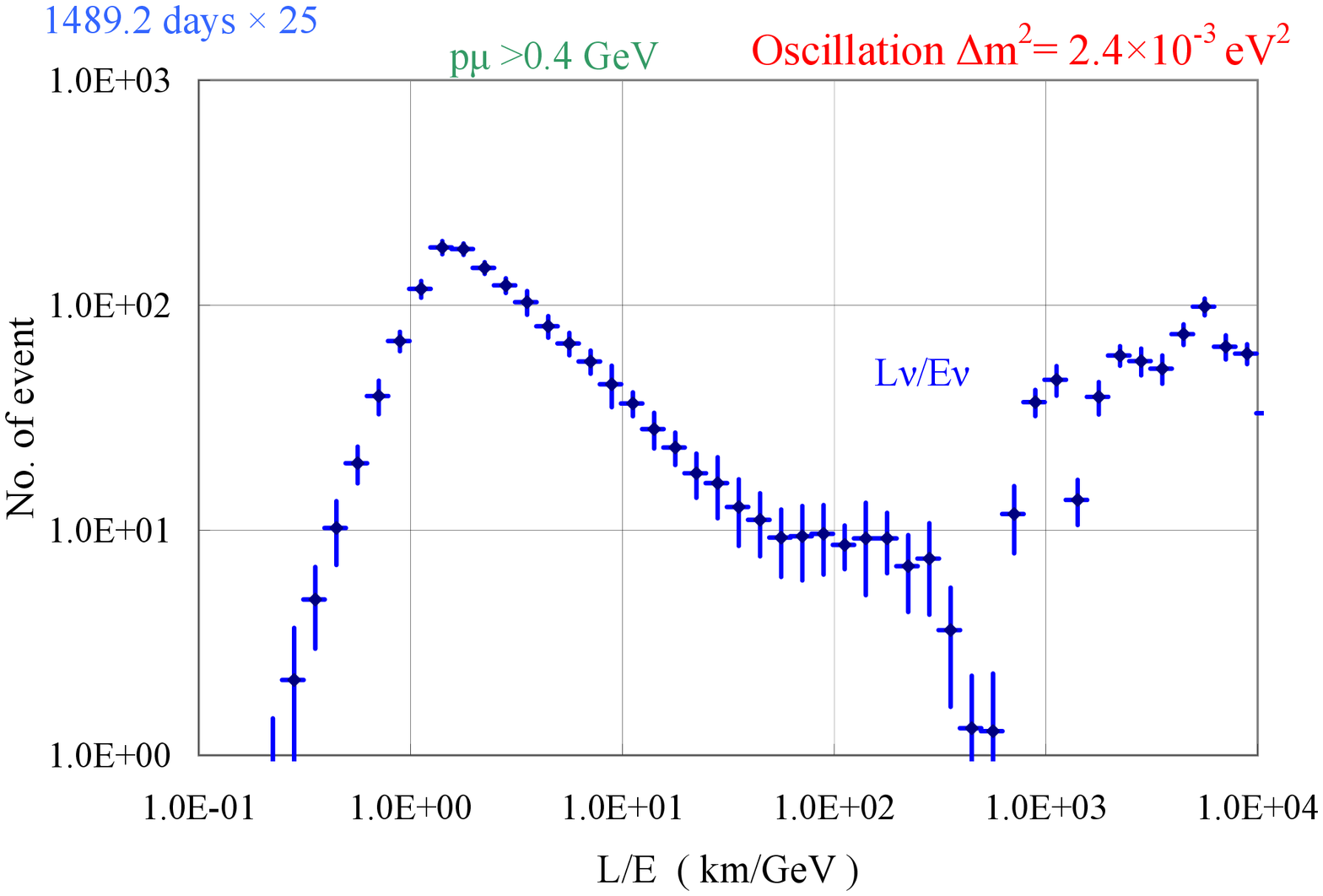}
}}
\vspace{-1.5cm}
\caption{$L_{\nu}/E_{\nu}$ distribution with standard deviations
with oscillation for 37230 live days (25 SK live days).}
\label{figJ021}
\vspace{-0.5cm}
\end{center}
\end{figure}

In Figure~\ref{figJ018}, we give one example of our $L_{\nu} / E_{\nu}$ 
distribution for one SK live day (1489.2 live days)\cite{Ashie2}
among twenty five sets of 
the computer numerical experiments in the unit of one SK live day. 
In Figure~\ref{figJ019}, we give another example for one SK 
live day. Arrows A, B and C represent locations for
the first, the second and the third maximum oscillation which are 
given in Eqs. (3-1), (3-2) and (3-3), respectively. By the definition 
of our computer numerical experiments, there are 
no experimental error bars 
in $L_{\nu} / E_{\nu}$ distributions in Figures~\ref{figJ018} and 
\ref{figJ019}.

 In Figure~\ref{figJ020}, we show the $L_{\nu} / E_{\nu}$
 distribution for 14892 live 
days (10 SK live days). Compared 
Figure~\ref{figJ020} with Figures~\ref{figJ018} and \ref{figJ019},
  it is clear that  
$L_{\nu} / E_{\nu}$ distribution in Figure~\ref{figJ020} becomes 
smoother due to larger statistics. 

In Figure~\ref{figJ024}, we give $L_{\nu} / E_{\nu}$ distribution for 
14892 live days (10 SK live days) in a linear scale 
together with the corresponding one without oscillation.
The $L/E$ distribution with oscillation in 
Figure~\ref{figJ024} is a representation in linear scale 
and it is the same as that in Figure~\ref{figJ020} in logarithmic scale.

However, it is clear in Figure~\ref{figJ024} that 
magnitudes of freqency indicated by arrows A, B and C are  
almost same\footnote{
 Superkamiokande collaboration mention that the 
second or more higher maximum oscillations are not observed due to their 
experimental condition\cite{Ashie1}.
  However, according to our results, they could have 
observed the second well and 
even higher maximum oscillations,
 if they could have really occured like the first maximum oscillation.
 See arrows A, B and C in Fgure~\ref{figJ024}.
}.
It should be noticed from Fgure~\ref{figJ024} that
$L_{\nu} / E_{\nu}$ distribution without oscillation represents 
the envelop of the corresponding distribution with oscillation.
 Namely, $L_{\nu} / E_{\nu}$ distribution with oscillation is 
equivalent to the $L_{\nu} / E_{\nu}$ distribution without oscillation 
multiplied by the survival probability for a given flavor (Eq.(6)).  

\begin{figure}
\begin{center}
\vspace{-3cm}
\hspace*{-2cm}
\rotatebox{90}{%
\resizebox{0.5\textwidth}{!}{%
  \includegraphics{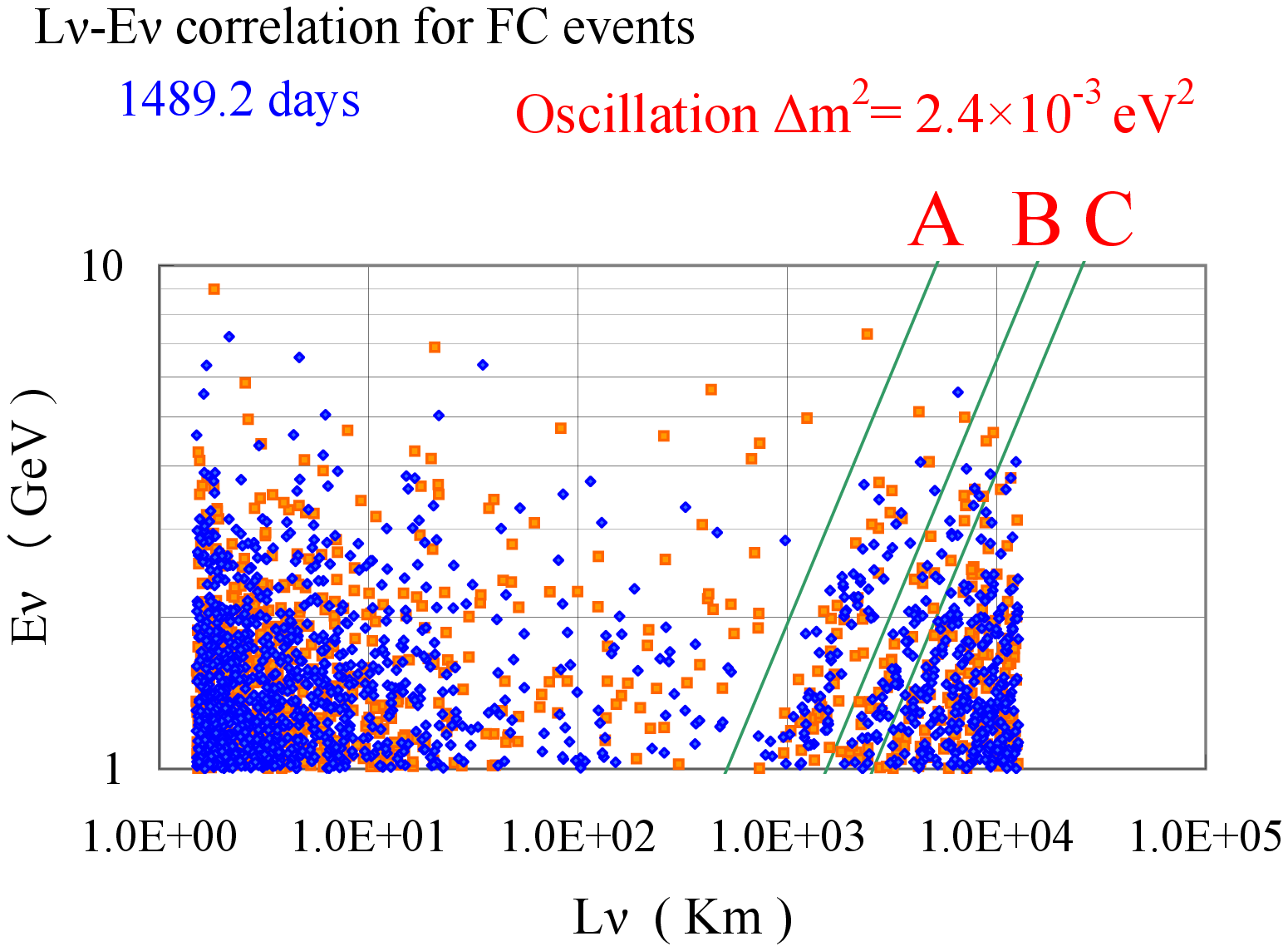}
}}
\vspace{-3.2cm}
\caption{Correlation diagram between $L_{\nu}$ and $E_{\nu}$
with oscillation for 1489.2 live days (one SK live day).}
\label{figJ022}     
\vspace{-1.0cm}
\hspace*{-2cm}
\rotatebox{90}{%
\resizebox{0.5\textwidth}{!}{%
  \includegraphics{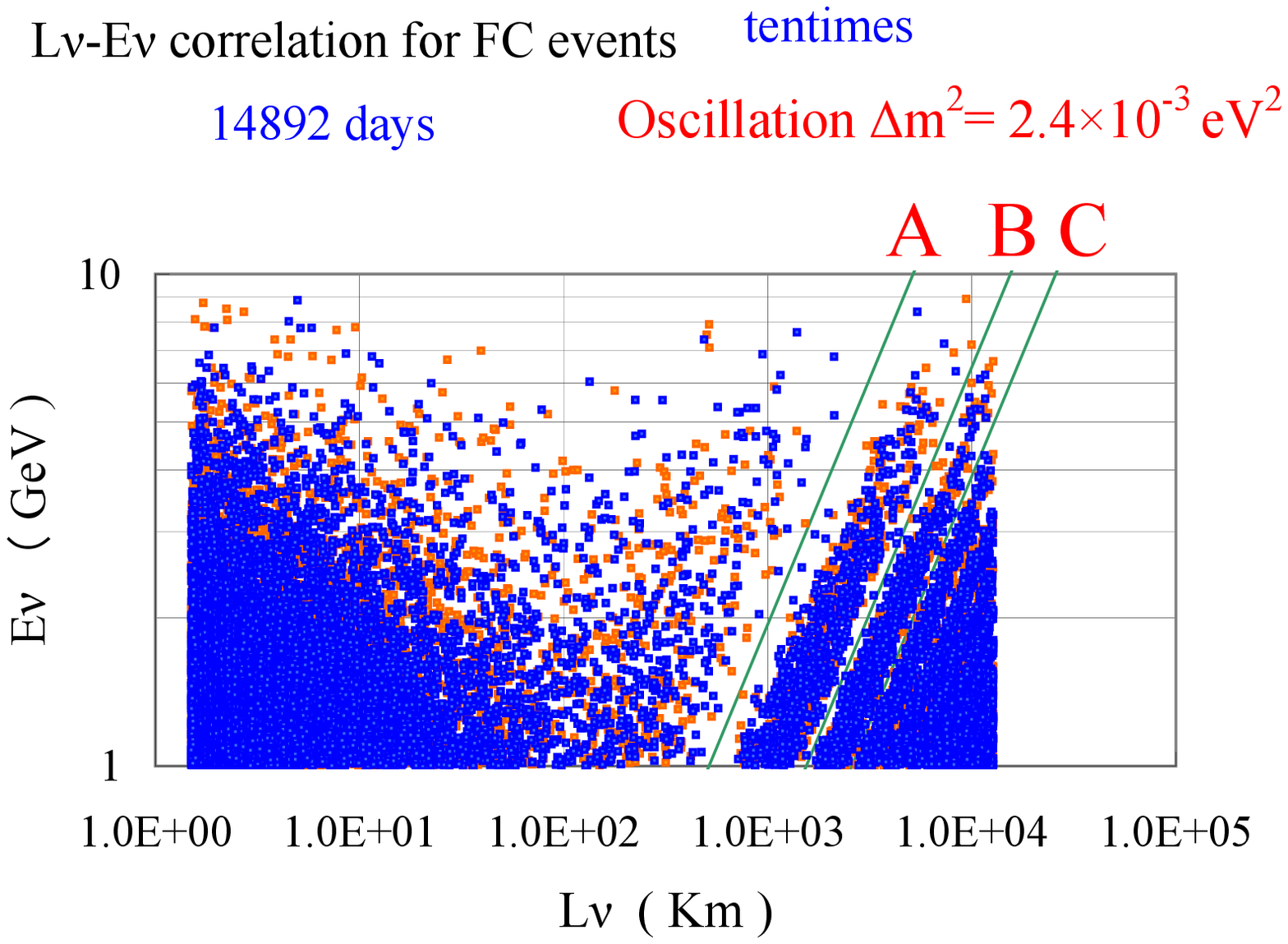}
}}
\vspace{-3.2cm}
\caption{Correlation diagram between $L_{\nu}$ and $E_{\nu}$
with oscillation for 14892 live days (10 SK live days).}
\label{figJ023}       % Give a unique label
%\end{center}
%\end{figure}

%\begin{figure}
%\begin{center}
\vspace{-1.0cm}
\rotatebox{90}{%
\resizebox{0.4\textwidth}{!}{%
  \includegraphics{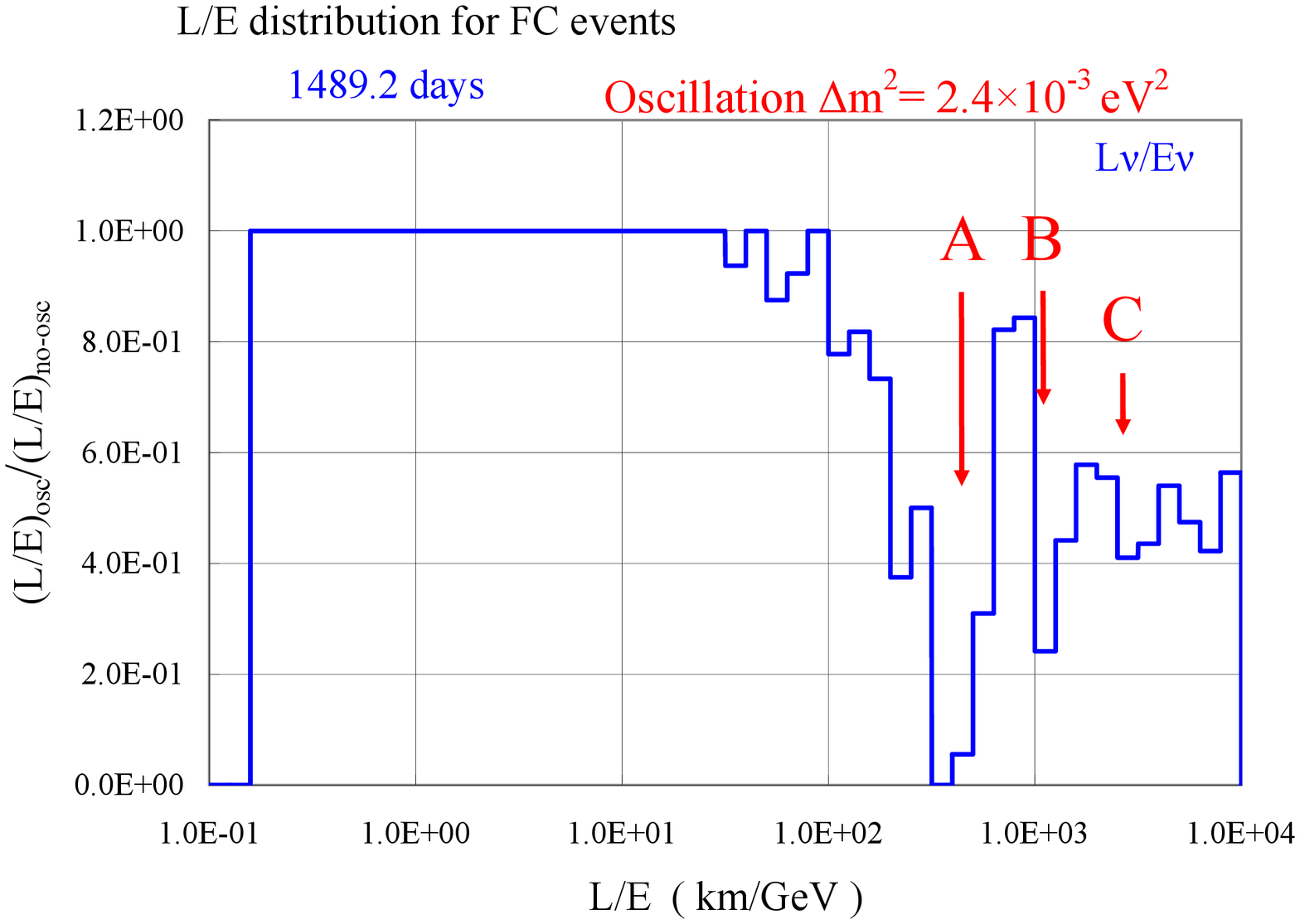}
}}
\vspace{-1.7cm}
\caption{Ratios of 
$(L_{\nu}/E_{\nu})_{osc}/(L_{\nu}/E_{\nu})_{null}$
 for 1489.2 live days (one SK live day).}
\label{figJ025}       % Give a unique label

\vspace{-1.0cm}
\rotatebox{90}{%
\resizebox{0.4\textwidth}{!}{%
  \includegraphics{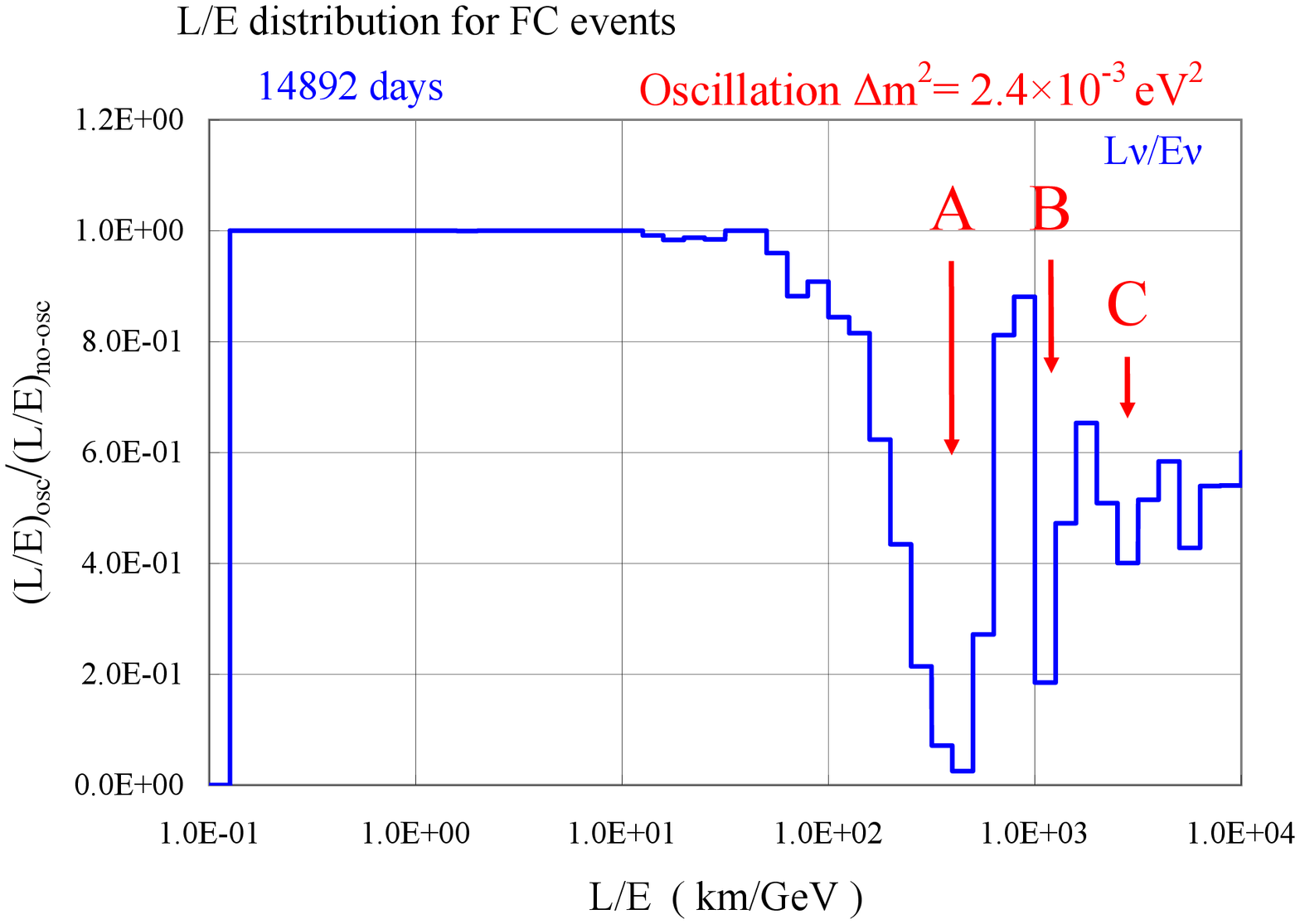}
}}
\vspace{-1.7cm}
\caption{Ratios of 
$(L_{\nu}/E_{\nu})_{osc}/(L_{\nu}/E_{\nu})_{null}$
 for 14892 live days (10 SK live days).}
\label{figJ026}       % Give a unique label
\vspace{-1.5cm}
\end{center}
\end{figure}

 We have repeated the 
computer numerical experiment for one SK live day as 
much as twenty five times independently,
in both cases with oscillation and without 
oscillation.
In Figure~\ref{figJ021}, 
we can add the statistical uncertainty (standard deviation 
in this case) around their average, because every one SK live 
day experiment among twenty five sets of the experiments 
fluctuates one by one due to their stochastic character 
in their physical processes and geometrical conditions 
of the detectors concerned. 

 In order to make the image of the maximum 
oscillations in $L_{\nu} / E_{\nu}$
 distributions clearer, we show the 
correlations between $L_{\nu}$ and $E_{\nu}$
 in Figures~\ref{figJ022} and \ref{figJ023}, 
which correspond to Figures~\ref{figJ018} and \ref{figJ020},
 respectively.
 In Figure~\ref{figJ022} for one SK live day,
we can observe vacant regions for  the events concerned 
assigned by A, B and C. 
In Figure~\ref{figJ023} for ten SK live days, the existence of 
the vacant regions for the events concerned becomes 
 clearer due to larger statistics. 

We give ratios of $(L_{\nu}/E_{\nu})$ 
 distribution with oscillation to that without oscillation 
for 1489.2 live days (one SK live day) in 
Figure~\ref{figJ025} and for 14892 live days (10 SK live 
days) in Figure~\ref{figJ026}, respectively.

The situation shown in Figures~\ref{figJ018}
 to \ref{figJ023} shows definitely that our computer 
numerical experiments are carried out exactly from the view 
point of the stochastic treatment to the matter,
if neutrino oscillation parameters obtained by Super-Kamiokande
are correct. 

\subsection{$L_{\nu}/E_{\mu}$ distribution}

\begin{figure}
\begin{center}
\vspace{-2cm}
\rotatebox{90}{%
\resizebox{0.4\textwidth}{!}{%
  \includegraphics{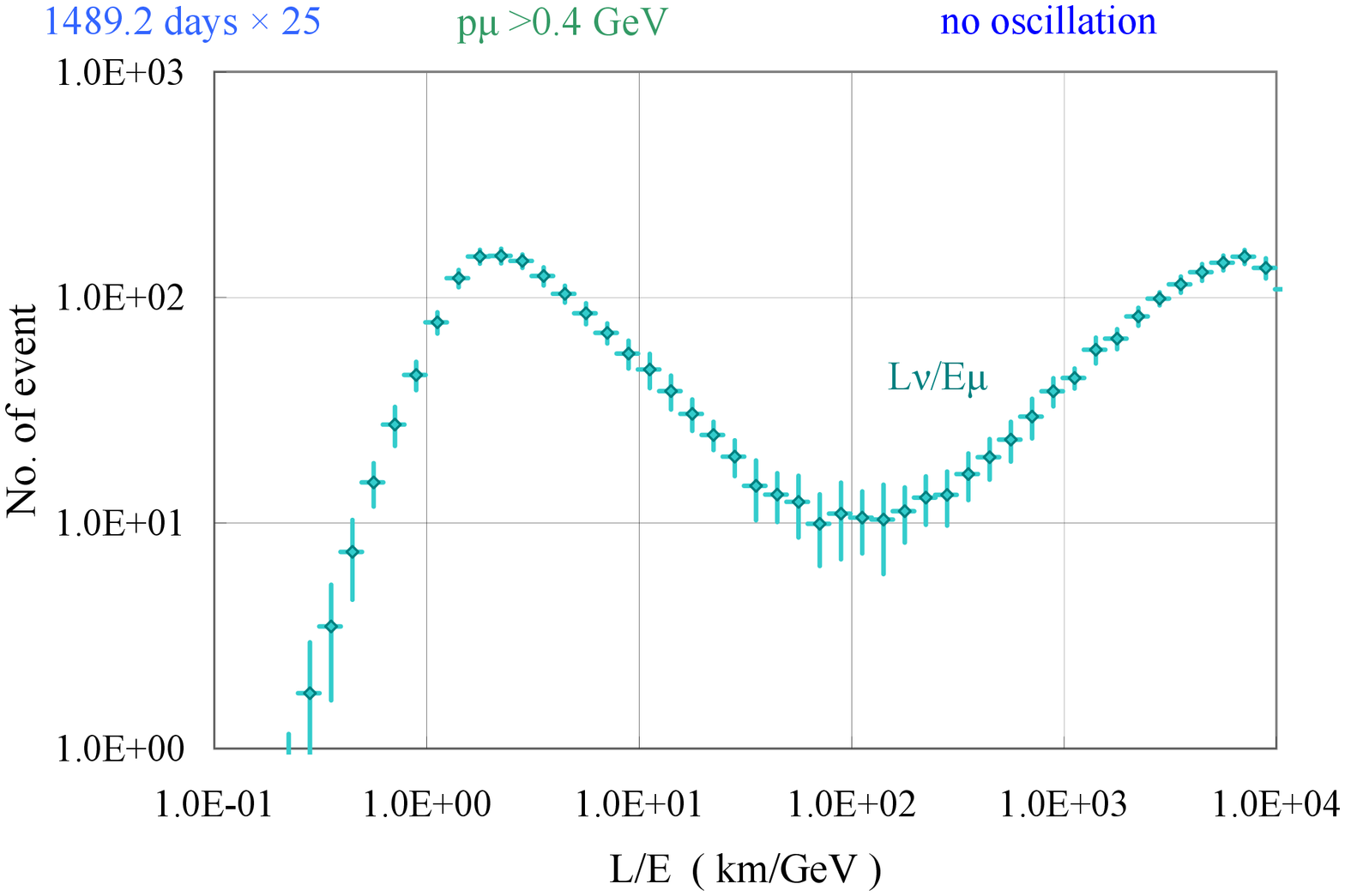}
%LnuEmufcNO_ten.eps}
}}
\vspace{-1.5cm}
\caption{$L_{\nu}/E_{\mu}$ distribution without
oscillation for 37230 days (25 SK live days).}
\label{figJ041}
\vspace{-1.0cm}
\rotatebox{90}{%
\resizebox{0.4\textwidth}{!}{%
  \includegraphics{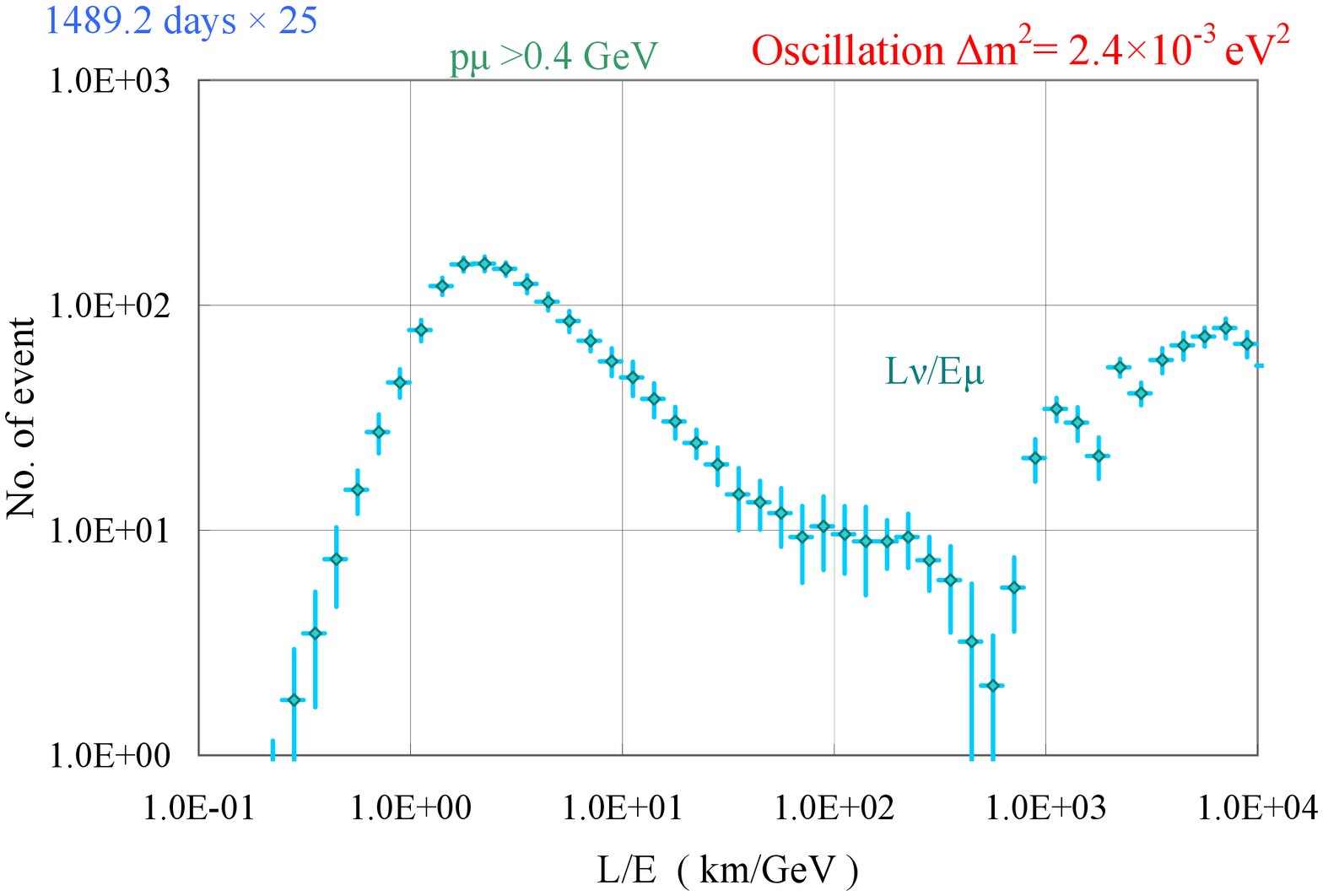}
}}
\vspace{-1.5cm}
\caption{$L_{\nu}/E_{\mu}$ distribution with 
oscillation for 37230 days (25 SK live days).}
\label{figJ042}
\vspace{-1cm}
\hspace*{-0.5cm}
\rotatebox{90}{%
\resizebox{0.4\textwidth}{!}{%
  \includegraphics{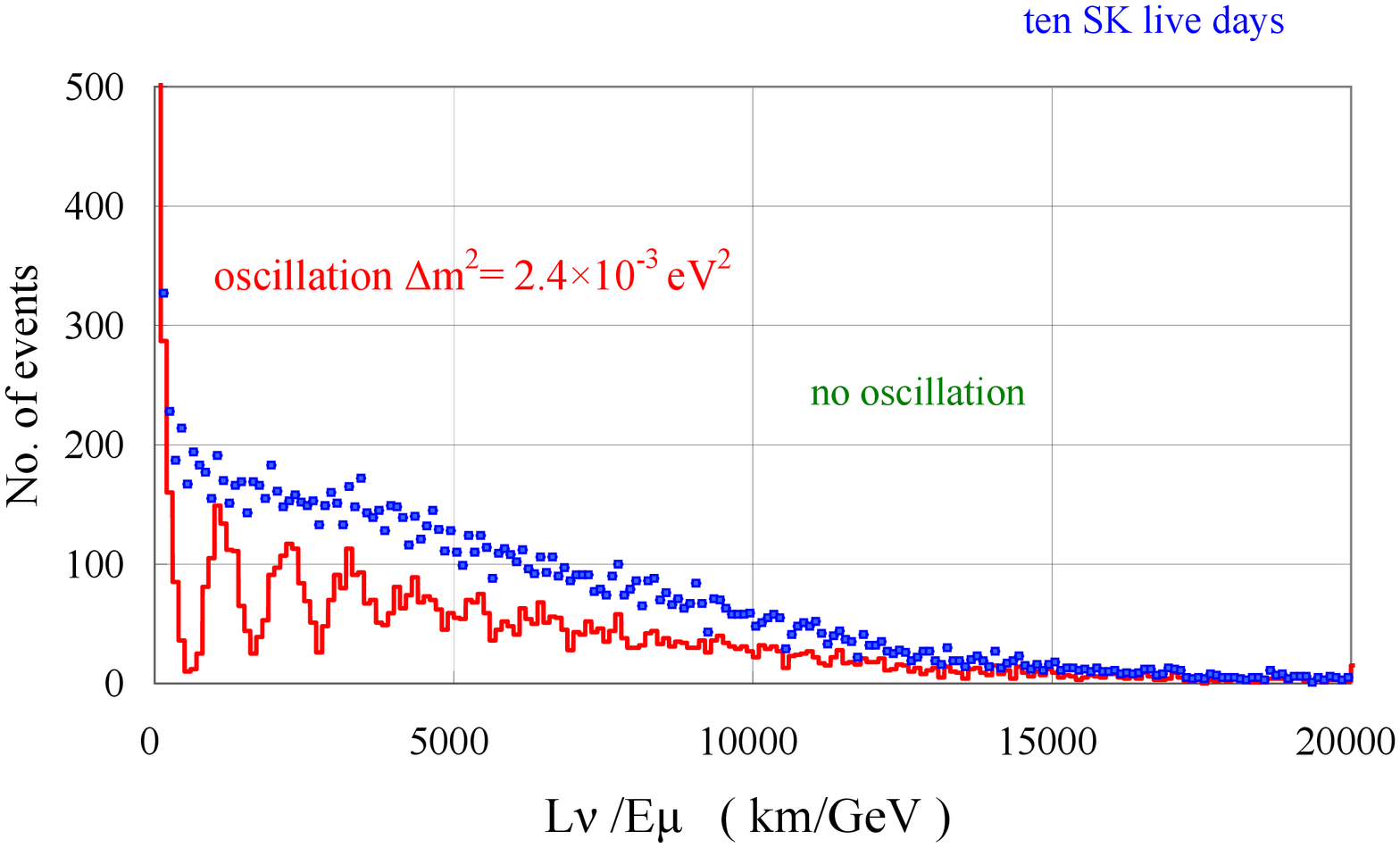}
}}
\vspace{-1.5cm}
\caption{$L_{\nu}/E_{\mu}$ distribution with 
and without oscillation for 14892 days (10 SK live days).}
\label{figL056}
\vspace{-0.5cm}
\rotatebox{90}{%
\resizebox{0.4\textwidth}{!}{%
  \includegraphics{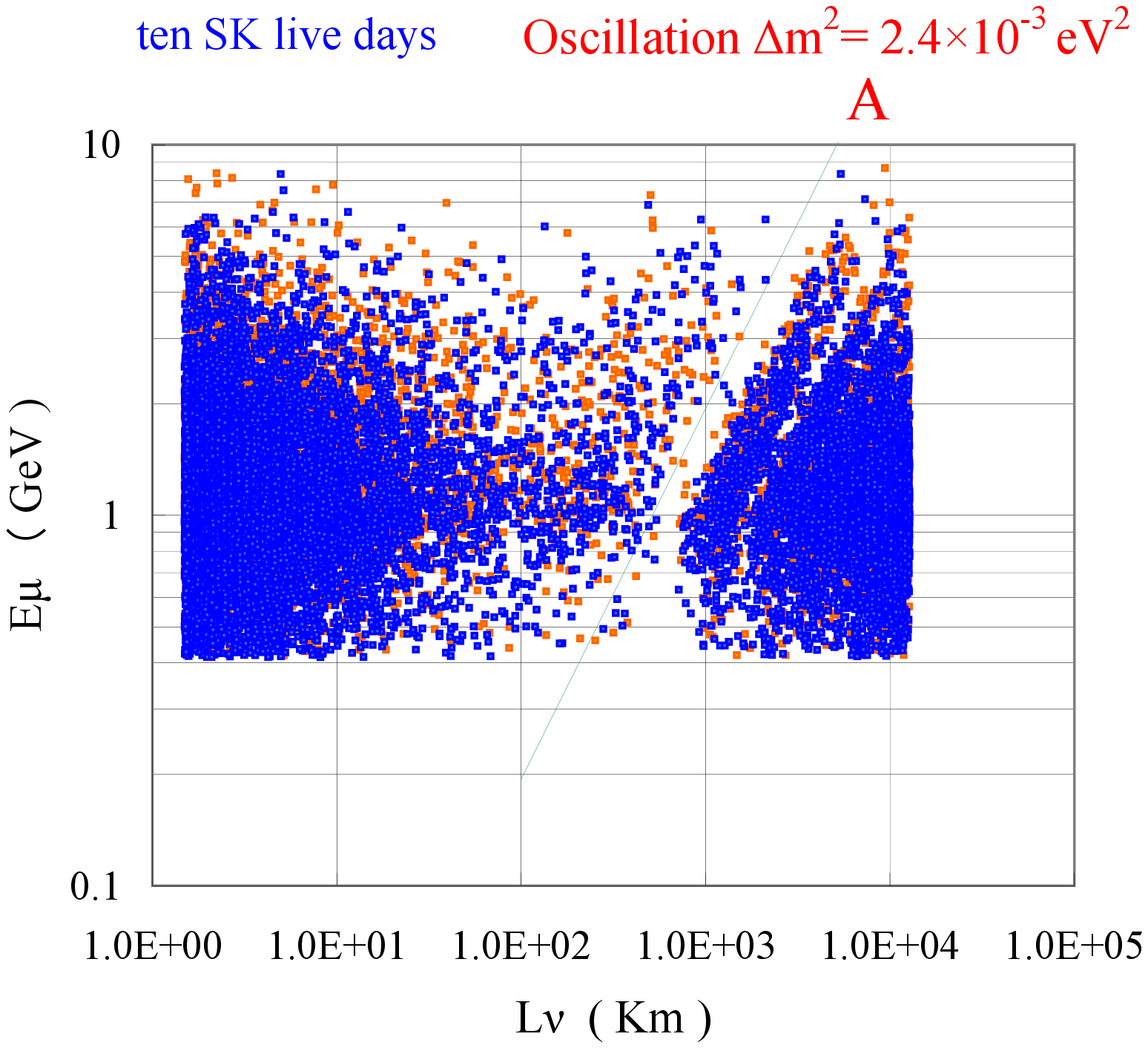}
}}
\vspace{-1.5cm}
\caption{Correlation diagram between $L_{\nu}$ and $E_{\mu}$
 with oscillation for 14892 days (10 SK live days).}
\label{figJ043}

\end{center}
\end{figure}

\subsubsection{For null oscillation}
%\hspace*{0.2cm}

 In Figure~\ref{figJ041}, we give $L_{\nu}/E_{\mu}$ distribution without 
oscillation for 37230 live days (25 SK live days)
of Super-Kamiokande Experiment to consider the statistical fluctuation 
effect as precisely as possible. 

It is seen from the comparison of 
Figure~\ref{figJ041} with Figure~\ref{figJ016} for $L_{\nu}/E_{\mu}$ 
distribution
that there is no appreciable difference between them and it denotes that 
the transform from $E_{\mu}$ to $E_{\nu}$
(see Figures~19 in \cite{Konishi2})
 does not cause any appreciable change in $L/E$ distribution.
 In other words, the appreciable change are caused by the transform of $L_{\mu}$ to $L_{\nu}$ 
(see Figures~11 to 13 ,and/or Figures~16 to 18 in \cite{Konishi2}).

\subsubsection{For oscillation (SK oscillation parameters)}
%\hspace*{1cm}

Here, we compare Figure~\ref{figJ042},
$L_{\nu}/E_{\mu}$ distribution for 37230 live days
(25 SK live days), with Figure~\ref{figJ021}, corresponding 
$L_{\nu}/E_{\nu}$ distribution.
 Combined Figure~\ref{figJ041} with Figure~\ref{figJ042},
 we give $L_{\nu}/E_{\mu}$ distributions with and without oscillation 
are given in a linear scale in Figure~\ref{figL056}.
 Being different from Figure~\ref{figJ024} for 
$L_{\nu}/E_{\nu}$ distribution, 
$L_{\nu}/E_{\mu}$ distribution with oscillation in Figure~\ref{figL056}
  begins to make the maximum oscillation pattern collapse after the 
first maximum oscillation.
 This fact corresponds to the situation that in 
$L_{\nu}/E_{\mu}$ distributions, the transform of 
$E_{\mu}$ from $E_{\nu}$ makes it difficult to form
the "envelope-like" relation between $L_{\nu}/E_{\mu}$ distributions 
with and without oscillation after the first maximum oscillation.
 It is clear from the comparison of these figures that 
$L_{\nu}/E_{\mu}$ distribution coincides almost with 
$L_{\nu}/E_{\nu}$ distribution around the first maximum oscillation,
but the former become to deviate from the latter after the 
second maximum oscillation.
 It reflects the correlation between $E_{\mu}$ and $E_{\nu}$
(see Figures~19 in \cite{Konishi2}).
 The situation that the first maximum oscillation can be "observed" is 
understandable from the existence of the vacant region indicated 
by arrow A in Figure~\ref{figJ043}, too.
 However, it is needless to say that both $L_{\nu}/E_{\nu}$ and 
$L_{\nu}/E_{\mu}$ distributions cannot be observed because of 
the neutrality of $L_{\nu}$.  

\subsection{$L_{\mu}/E_{\mu}$ distribution}

\begin{figure}
\begin{center}
\vspace{-2cm}
%\end{center}
%\end{figure}
%\begin{figure}
%\begin{center}
\rotatebox{90}{%
\resizebox{0.4\textwidth}{!}{%
  \includegraphics{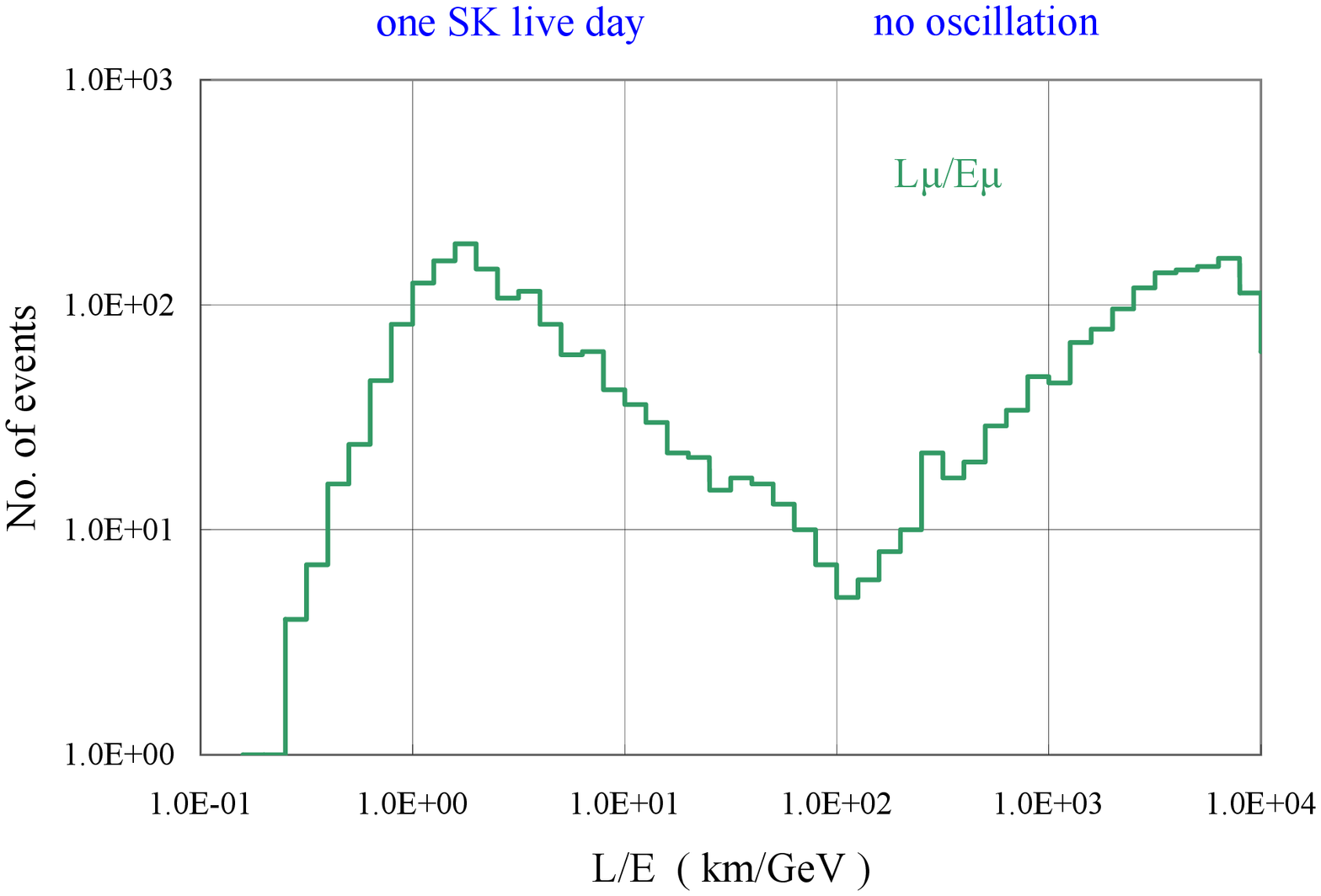}
}}
\vspace{-1.7cm}
\caption{$L_{\mu}/E_{\mu}$ distribution without oscillation
for 1489.2 live days (one SK live day).}
\label{figJ027}       % Give a unique label
\vspace{-0.7cm}
\rotatebox{90}{%
\resizebox{0.38\textwidth}{!}{%
  \includegraphics{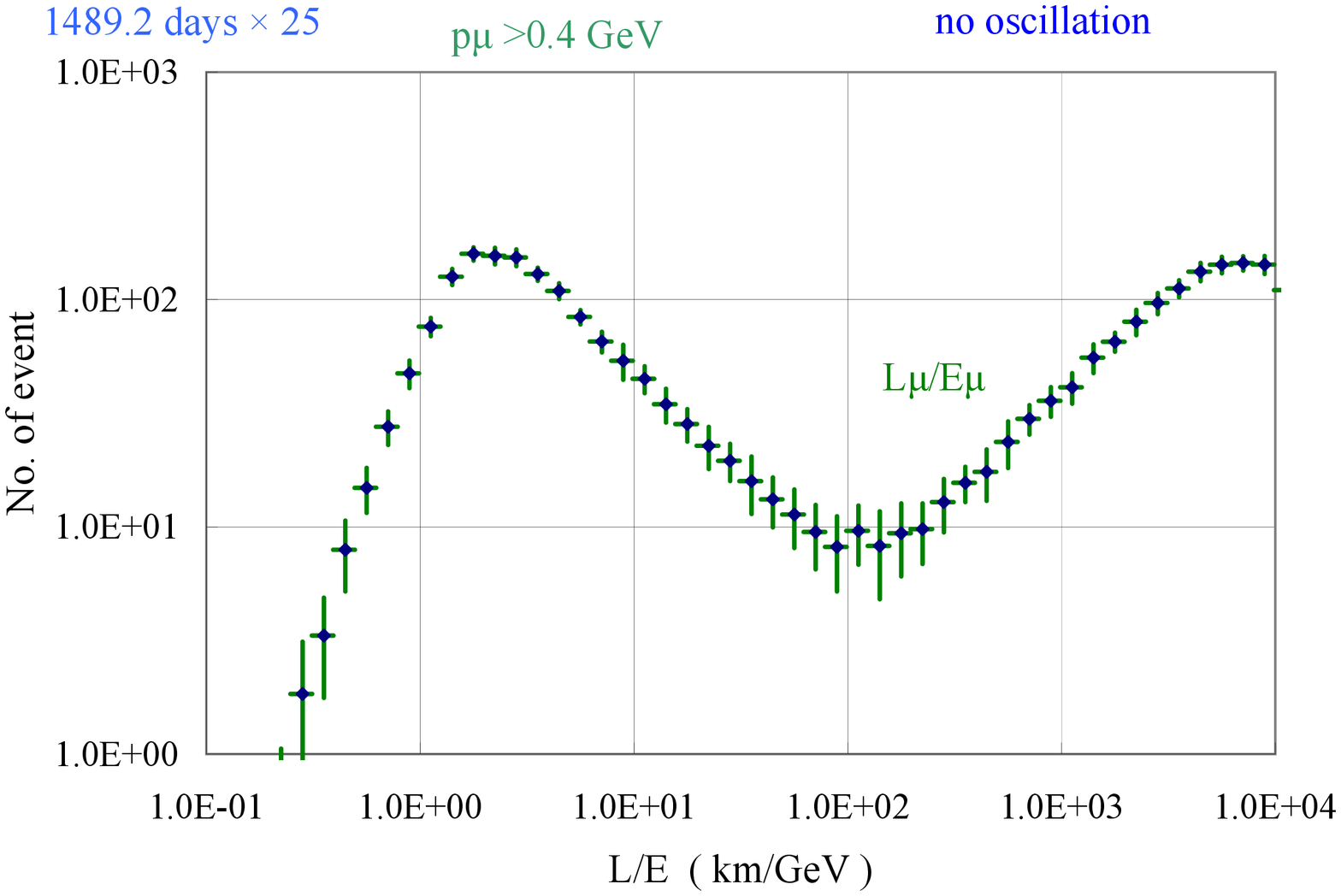}
%LmuEmufcNO_ten.eps}
}}
\vspace{-1.7cm}
\caption{$L_{\mu}/E_{\mu}$ distribution without oscillation
for 37230 live days (25 SK live days).}
\label{figJ028}
%\end{center}
%\end{figure}

%\begin{figure}
\vspace{-0.7cm}
\rotatebox{90}{%
\resizebox{0.4\textwidth}{!}{%
  \includegraphics{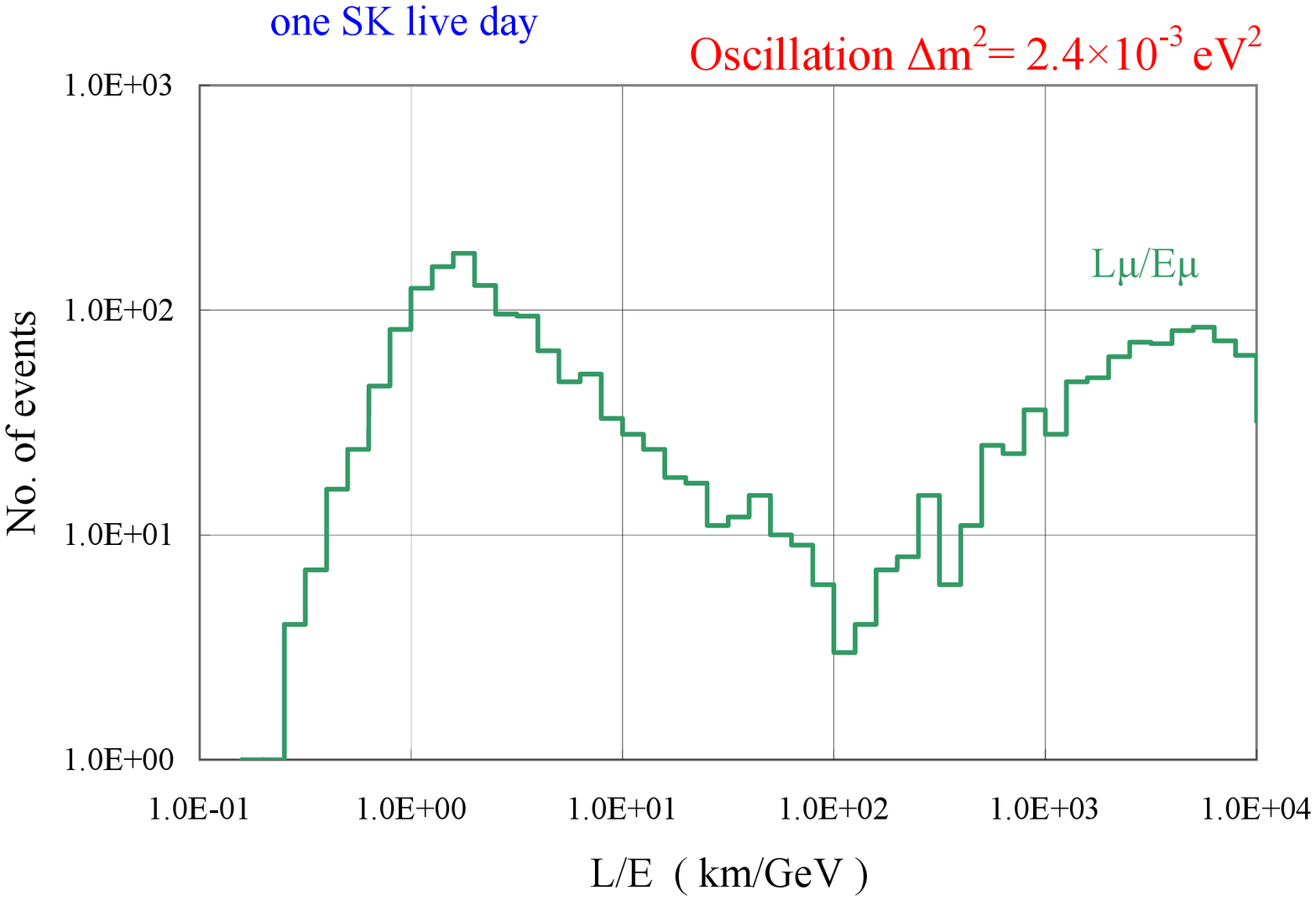}
}}
\vspace{-1.7cm}
\caption{$L_{\mu}/E_{\mu}$ distribution with oscillation
for 1489.2 live days (one SK live day).}
\label{figJ029}       % Give a unique label
\vspace{-0.7cm}
\rotatebox{90}{%
\resizebox{0.4\textwidth}{!}{%
  \includegraphics{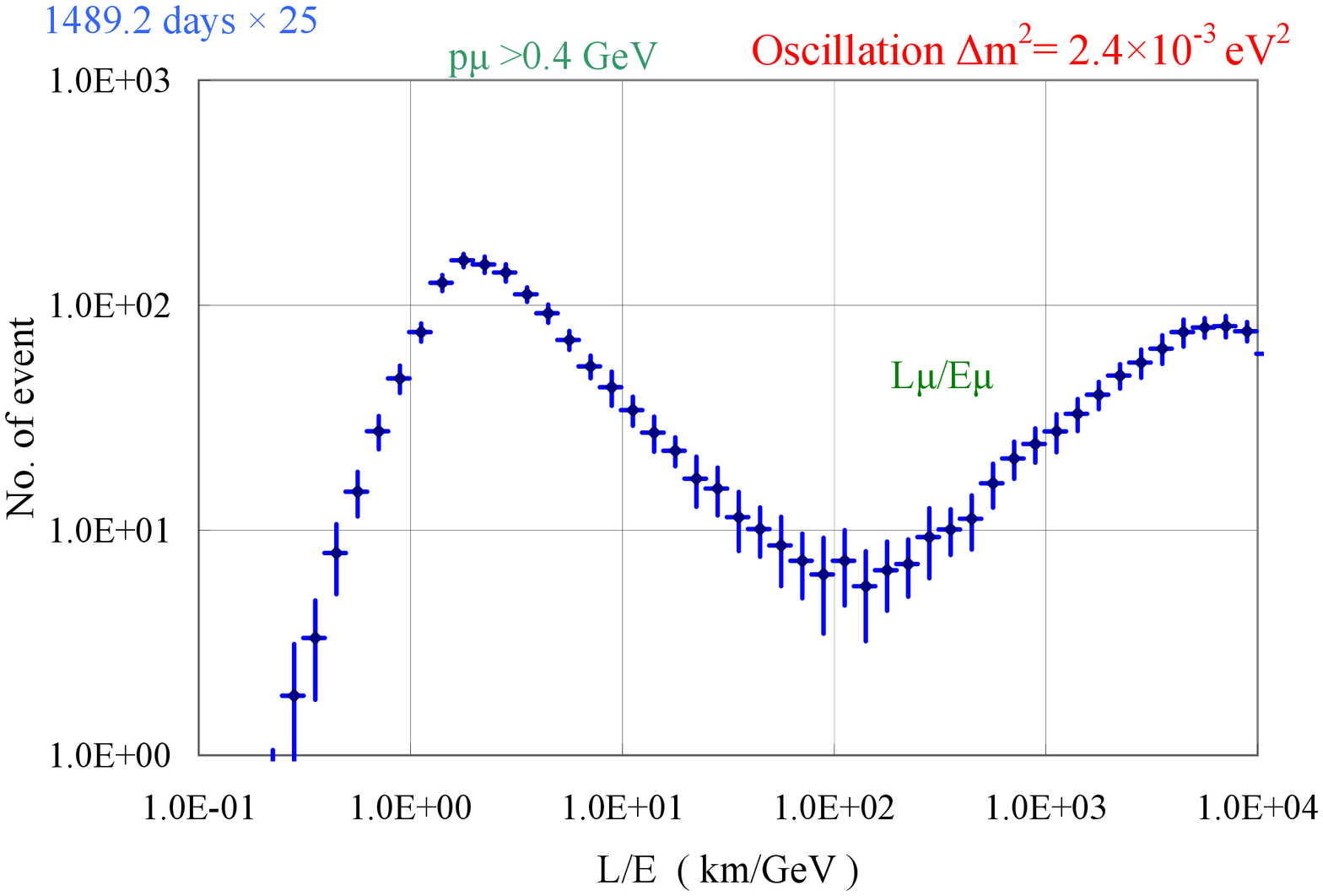}
}}
\vspace{-1.7cm}
\caption{$L_{\mu}/E_{\mu}$ distribution with oscillation
for 37230 live days (25 SK live days).}
\label{figJ030}
%\vspace{-0.2cm}
\end{center}
\end{figure}

The physical quantities measured by
Super-Kamiokande Collaboration  
 are $L_{\mu}$ and $E_{\mu}$,
 but neither $L_{\nu}$ and $E_{\nu}$.
In this sense, we carefully examine the validity of the survival 
probability for a given flavor whose variables are $L_{\mu}$ and $E_{\mu}$
, but neither $L_{\nu}$ and $E_{\nu}$. 
In other words, we examine whether we can find the maximum oscillations 
on $L_{\mu}/E_{\mu}$ distribution or not, 
because the existence of the maximum oscillation in 
$L_{\mu}/E_{\mu}$ distribution is 
exactly the same as the existence of the survival probability for a 
given flavor whose variable are $L_{\mu}$ and $E_{\mu}$ and vice-versa.

%
%As well known, Super-Kamiokande Collaboration put the crucial 
%assumption in their $L/E$ analysis. 
%The first one is $L_{\nu}\approx L_{\mu}$ 
%({\it the SK assumption on the direction})
% and the other is the estimation of 
%$E_{\nu}$ by $E_{\mu}$ (see Eq.(8.15) in \cite{Ishitsuka}).
% These two procedures inevitably cause uncertainties around the 
%interpretation of the neutrino oscillation. In this sense, one cannot 
%expect more fruitful ones than what they obtain the analysis of 
%$L_{\mu}/E_{\mu}$. In this sense, we examine $L_{\mu}/E_{\mu}$ 
%distribution with oscillation carefully.  

\subsubsection{For null oscillation}

 In Figure~\ref{figJ027}, we give one sample for one SK live day
 (1489.2 live days) from the totally 37230 live days (25 SK live days) 
events, each of which has 1489.2 live days.
 Figure~\ref{figJ028} shows the average distribution accompanied by the 
statistical uncertainty bar (not experimental error bar).
 It is clear from these figures that the existence of the dip or bottom, 
namely
the sinusoidal character, means the contribution merely from 
horizontal-like contribution, having no relation with any neutrino
 oscillation character, as they must be.   

\subsubsection{For oscillation (SK oscillation parameters)}

In Figures~\ref{figJ029} and \ref{figJ030}, 
we give the $L_{\mu}/E_{\mu}$ distributions with 
 oscillation for 1489.2 live days (one SK live day) and 37230 live days 
(25 SK live days), respectively. 
In Figure~\ref{figJ029}, we may observe the uneven histogram,
 something like
 curious bottoms coming from neutrino oscillation. However, 
in Figure~\ref{figJ030} where 
the statistics is 25 times as much as that of 
Figure~\ref{figJ029}, the histogram
 becomes smoother and such bottoms disappear, which 
turns out finally for the bottoms to be pseudo. 

\begin{figure}
\begin{center}
\vspace{-2cm}
\hspace*{-1cm}
\rotatebox{90}{%
\resizebox{0.4\textwidth}{!}{%
  \includegraphics{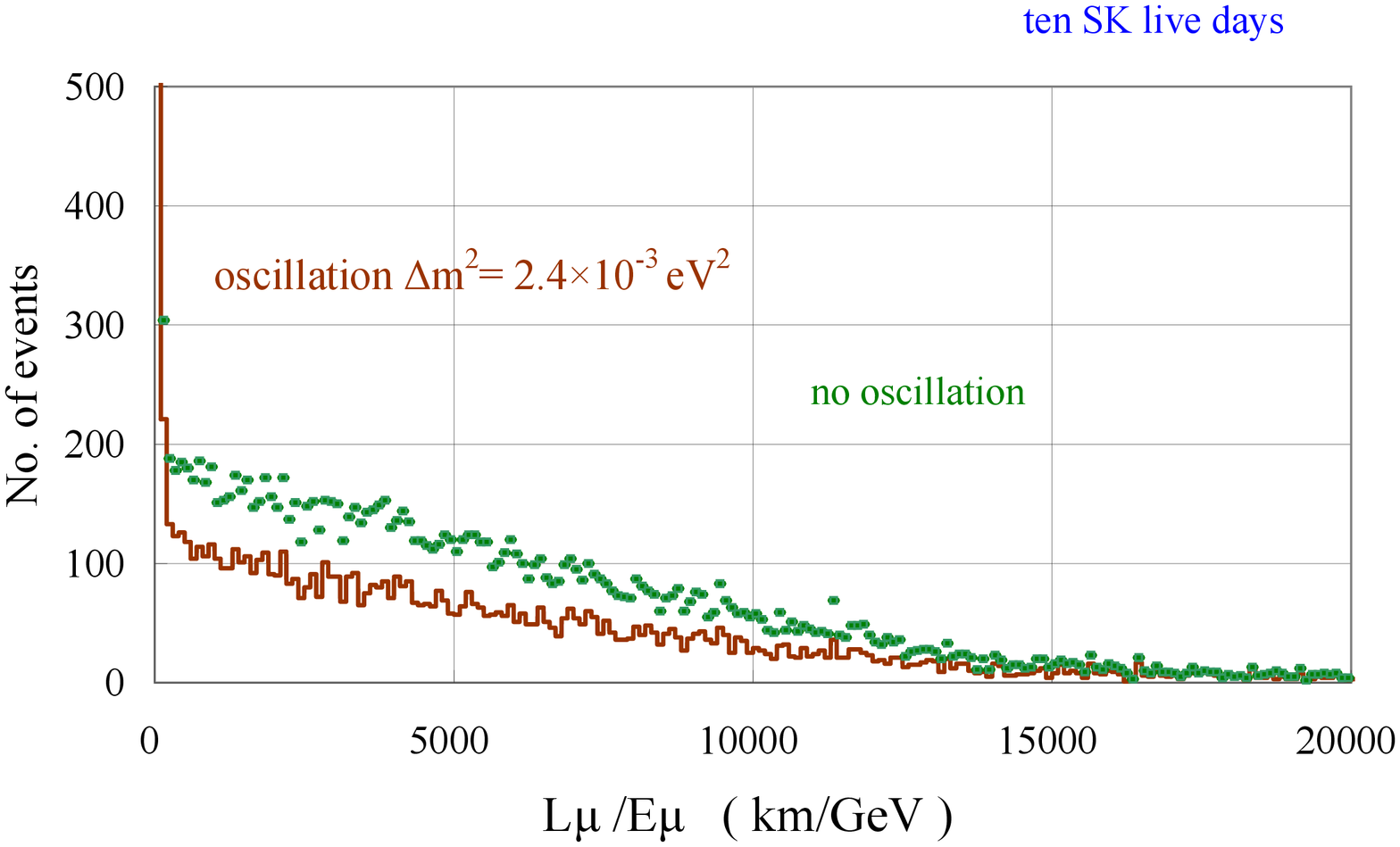}
}}
\vspace{-1.5cm}
\caption{$L_{\mu}/E_{\mu}$ distribution
with and without oscillation for 14892 live days (10 SK live days).}
\label{figJ033}

\vspace{-0.5cm}
\hspace*{-1cm}
\rotatebox{90}{%
\resizebox{0.45\textwidth}{!}{%
  \includegraphics{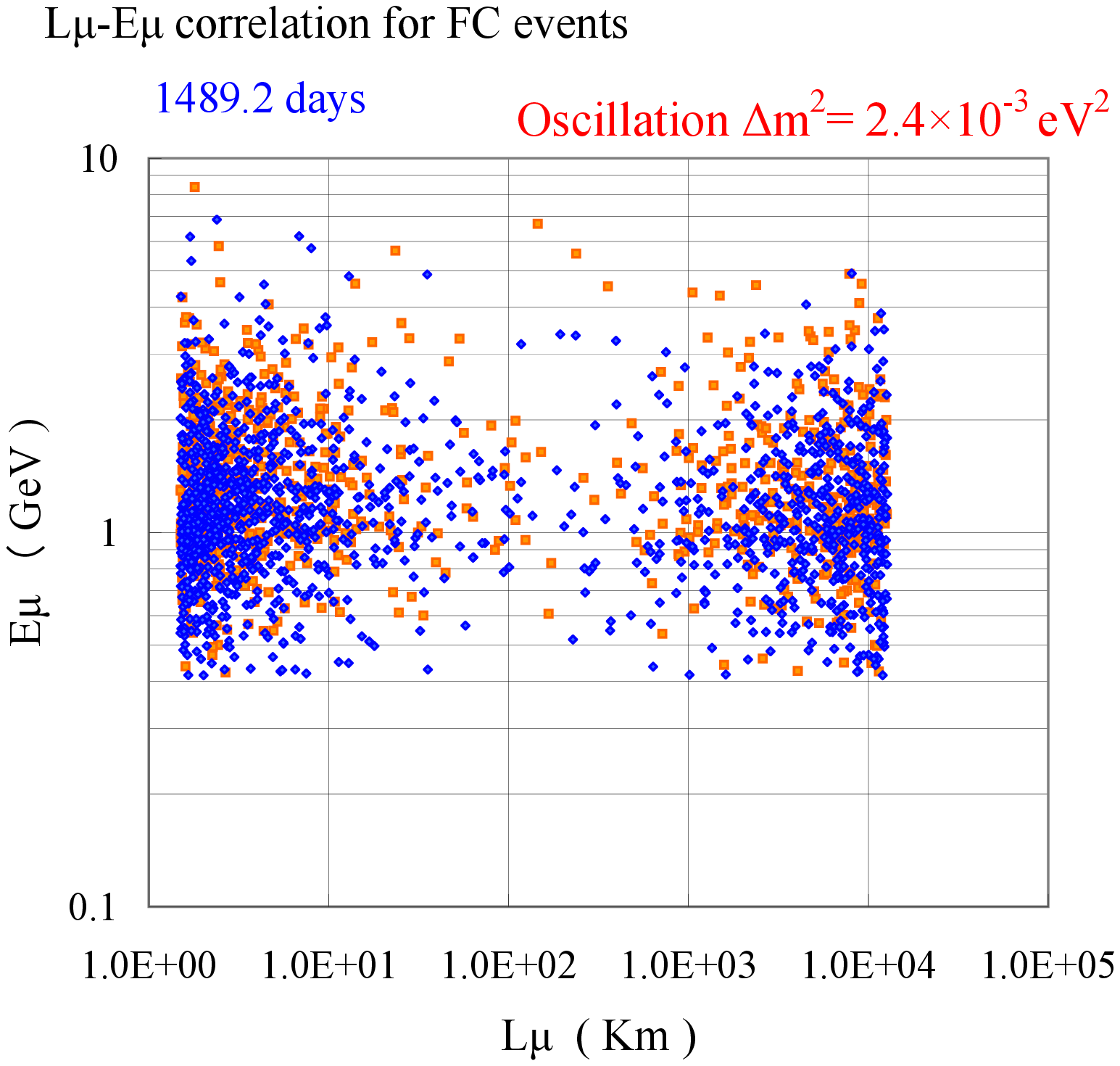}
}}
\vspace{-1.5cm}
\caption{Correlation diagram between $L_{\mu}$ and $E_{\mu}$
with oscillation for 1489.2 live days (one SK live day).}
\label{figJ031}
\vspace{-0.5cm}
\hspace*{-1cm}
\rotatebox{90}{%
\resizebox{0.45\textwidth}{!}{%
  \includegraphics{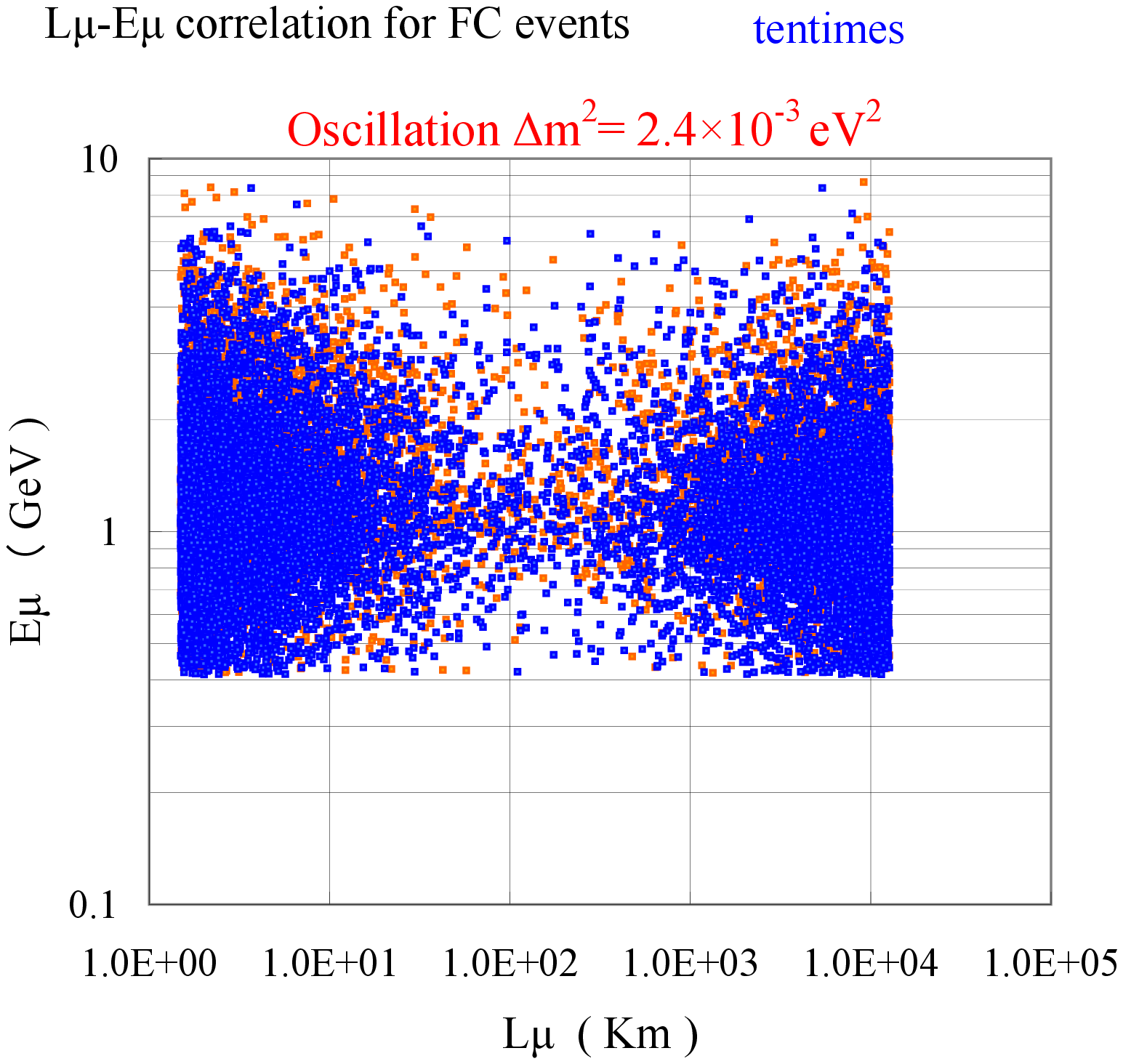}
}}
\vspace{-1.5cm}
\caption{Correlation diagram between $L_{\mu}$ and $E_{\mu}$
with oscillation for 14892 live days (10 SK live days).}
\label{figJ032}
\end{center}
\end{figure}

In order to examine the existence of the maximum oscillations in $L/E$ 
distribution, it is better to express $L/E$ distribution in a linear 
scale, but not in a logarithmic scale, 
as shown in Figure~\ref{figJ024} for $L_{\nu}/E_{\nu}$ distribution.
 In Figure~\ref{figJ033}, we give $L_{\mu}/E_{\mu}$ distribution with 
and without oscillation for 10 SK live days (14892 live days).
 It is clear from the comparison of Figure~\ref{figJ033} with 
Figure~\ref{figJ024} for $L_{\nu}/E_{\nu}$ 
distribution that we cannot find any maximum oscillation in 
$L_{\mu}/E_{\mu}$ distribution with oscillation.
 Also, we find the $L_{\nu}/E_{\nu}$ distribution without oscillation 
forms an envelop of the corresponding one with oscillation 
in Figure~\ref{figJ024}, as it must be, 
while we cannot find such a relation on $L_{\mu}/E_{\mu}$ distribution
in Figure~\ref{figJ033}.
 This denotes that $L_{\mu}/E_{\mu}$ cannot be the variables of the 
survival probability for a given flavor.  
 In order to confirm the lack of the maximum oscillations in 
$L_{\mu}/E_{\mu}$ distribution, we give a correlation diagram between $L_{\mu}$ and $E_{\mu}$ for one SK live days in Figure~\ref{figJ031} 
and that for 10 SK live days in Figure~\ref{figJ032}, respectively.
 If the maximum oscillations really exist in 
$L_{\mu}/E_{\mu}$ distribution, then we can expect to find the vacant 
regions for $L_{\mu}$ and $E_{\mu}$ diagrams
 in Figures~\ref{figJ031} and \ref{figJ032},
 as shown clearly in $L_{\nu}$ and $E_{\nu}$diagrams of 
Figures~\ref{figJ022} and \ref{figJ023}. 
However, we cannot find anything like vacant regions in 
Figures~\ref{figJ031} and \ref{figJ032} at all.

\begin{figure}
\begin{center}
\vspace{-2cm}
\hspace*{-1cm}
\rotatebox{90}{%
\resizebox{0.4\textwidth}{!}{%
  \includegraphics{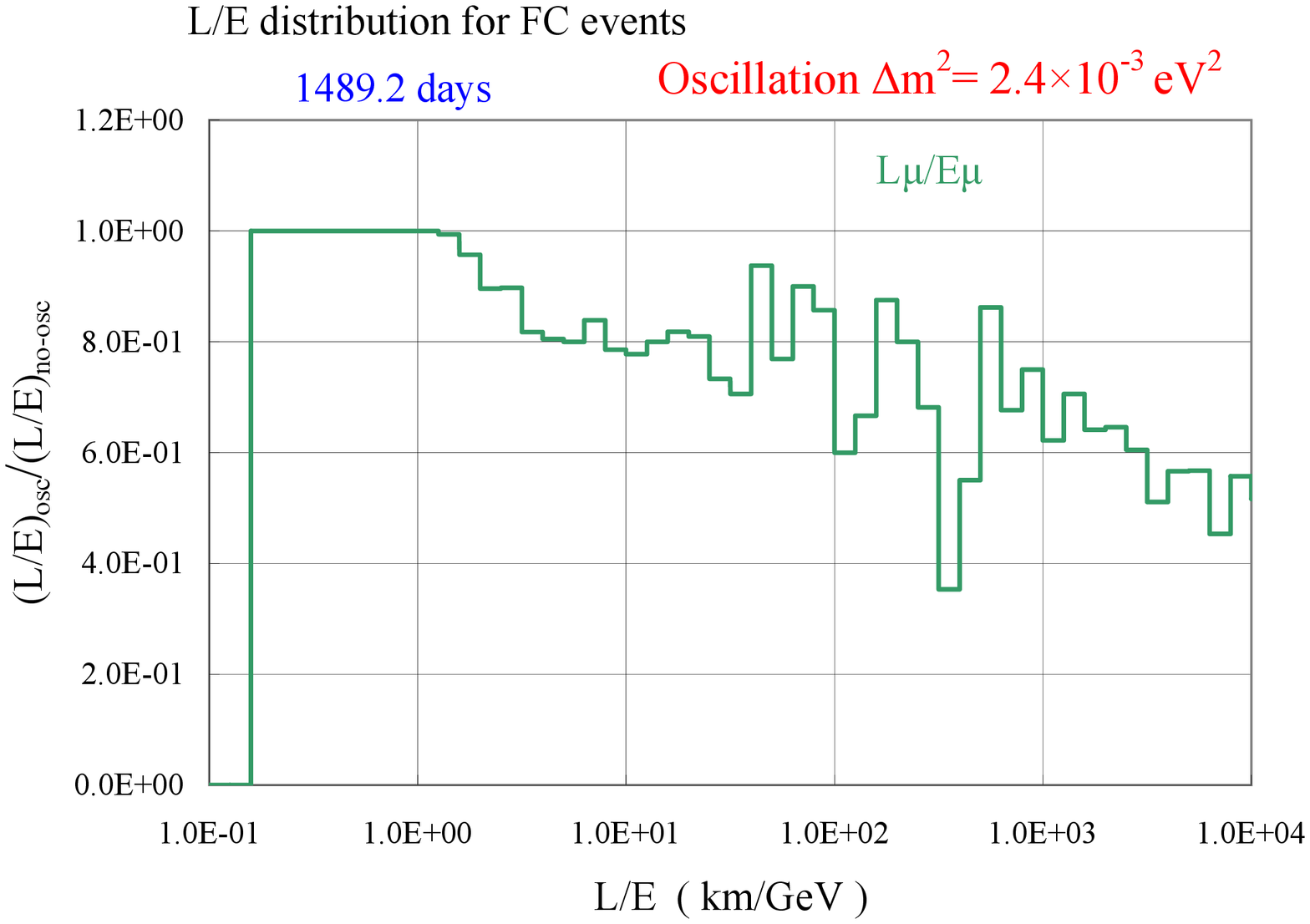}
}}
\vspace{-1.5cm}
\caption{The ratio of 
$(L_{\mu}/E_{\mu})_{osc}/(L_{\mu}/E_{\mu})_{null}$
 for 1489.2 live days (one SK live day).}
\label{figJ034}

\vspace{-0.7cm}
\hspace*{-1cm}
\rotatebox{90}{%
\resizebox{0.4\textwidth}{!}{%
  \includegraphics{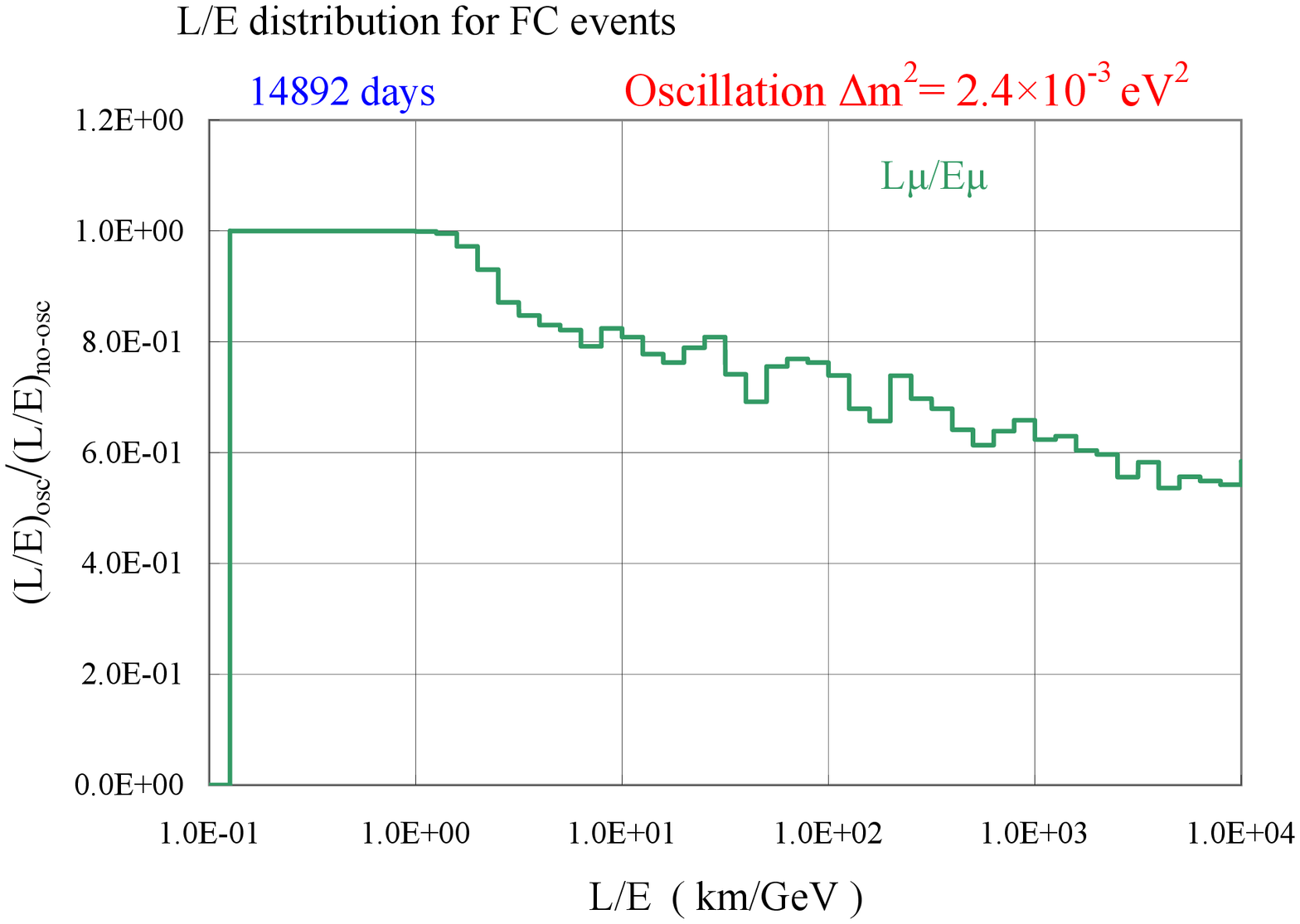}
}}
\vspace{-1.5cm}
\caption{The ratio of 
$(L_{\mu}/E_{\mu})_{osc}/(L_{\mu}/E_{\mu})_{null}$
 for 14892 live days (10 SK live days).}
\label{figJ035}
%\vspace{-0.5cm}
\end{center}
\end{figure}

In Figure~\ref{figJ034}, we show one example of
$(L_{\mu}/E_{\mu})_{osc}/(L_{\mu}/E_{\mu})_{null}$
for one SK live day among all possible sets of ratios.
  We may find pseudo dips in the figure.
 In Figure~\ref{figJ035}, 
we give those for ten SK live days,
whose statistics is larger than that of Figure~\ref{figJ034}
by 10 times, such pseudo dips disappear here.
Thus the histogram becomes a rather decreasing function of 
$L_{\mu}/E_{\mu}$ in Figure~\ref{figJ035}. 
When we further make statistics higher, the survival probability for
$L_{\mu}/E_{\mu}$ distribution should be a monotonously decreasing 
function of $L_{\mu}/E_{\mu}$, whithout showing any 
characteristics of the maximum oscillation,  
which is in remarkable contrast to Figures~\ref{figJ025} or \ref{figJ026}
for $L_{\nu}/E_{\nu}$ distribution.

 Summarized Figures~\ref{figJ029} to \ref{figJ035}, we say that 
$L_{\mu}/E_{\mu}$ distribution cannot give the 
maximum oscillations in any sense. 
This denotes that $L_{\mu}/E_{\mu}$ distributions are not 
constructed based on the survival probability for a given flavor 
which is the fundamental principle for neutrino oscillation.

\subsection{$L_{\mu}/E_{\nu}$ distribution}
Instead of analyzing $L_{\mu}/E_{\mu}$ distribution,
 Super-Kamiokande Collaboration have analyzed 
$L_{\mu}/E_{\nu}$ distribution where 
$E_{\nu}$ is approximated as the polynomial of 
$E_{\nu}$ (See,Eq.(7) in the preceding paper\cite{Konishi2}).
Consequently, we examine the $L_{\mu}/E_{\nu}$ distribution. 

\subsubsection{For null oscillation}

\begin{figure}
\begin{center}
\vspace{-2cm}
\rotatebox{90}{%
\resizebox{0.4\textwidth}{!}{%
  \includegraphics{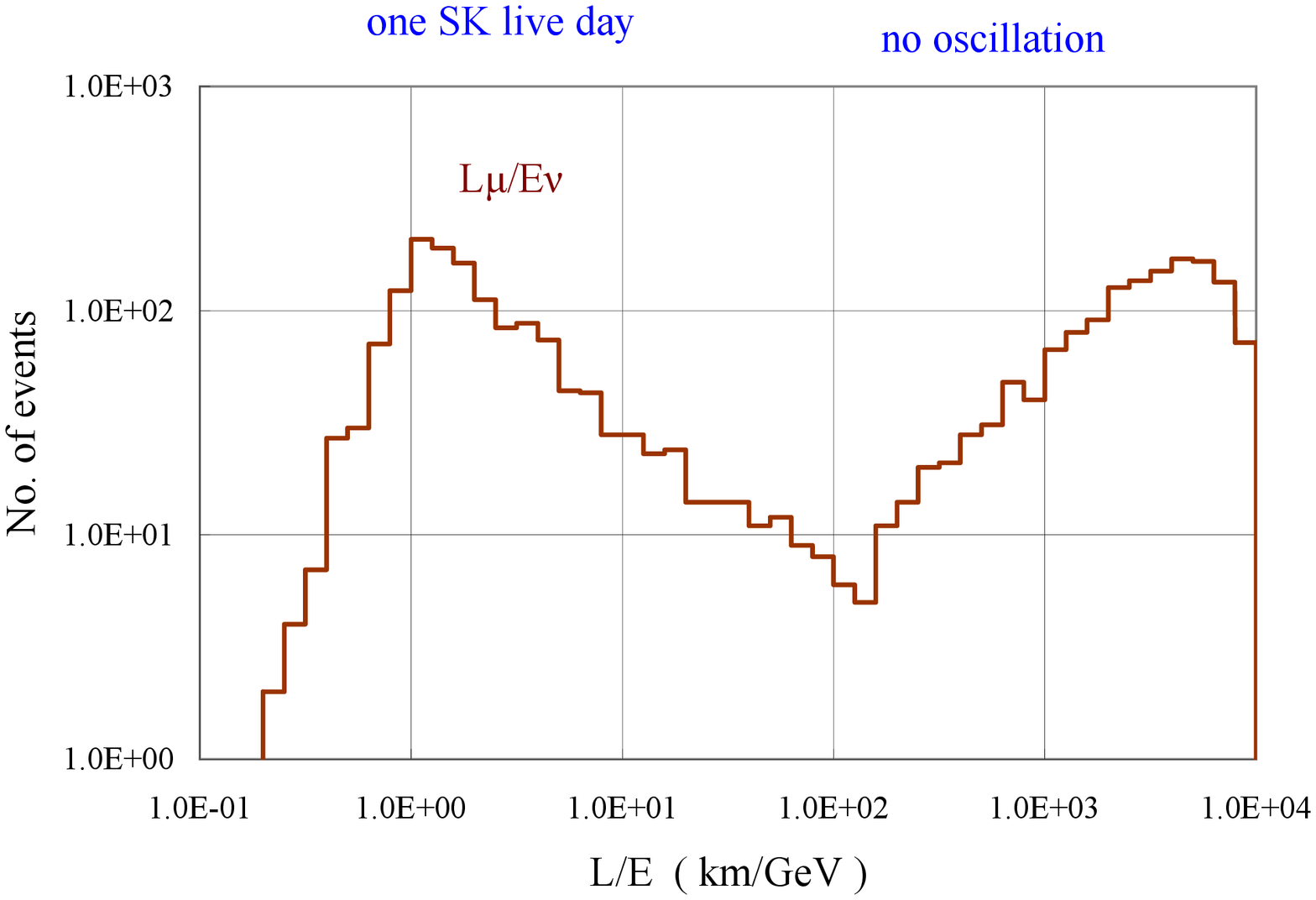}
}}
\vspace{-1.5cm}
\caption{$L_{\mu}/E_{\nu}$ distribution without oscillation
for 1489.2 live days (one SK live day).}
\label{figJ036}
\vspace{-0.7cm}
\rotatebox{90}{%
\resizebox{0.4\textwidth}{!}{%
  \includegraphics{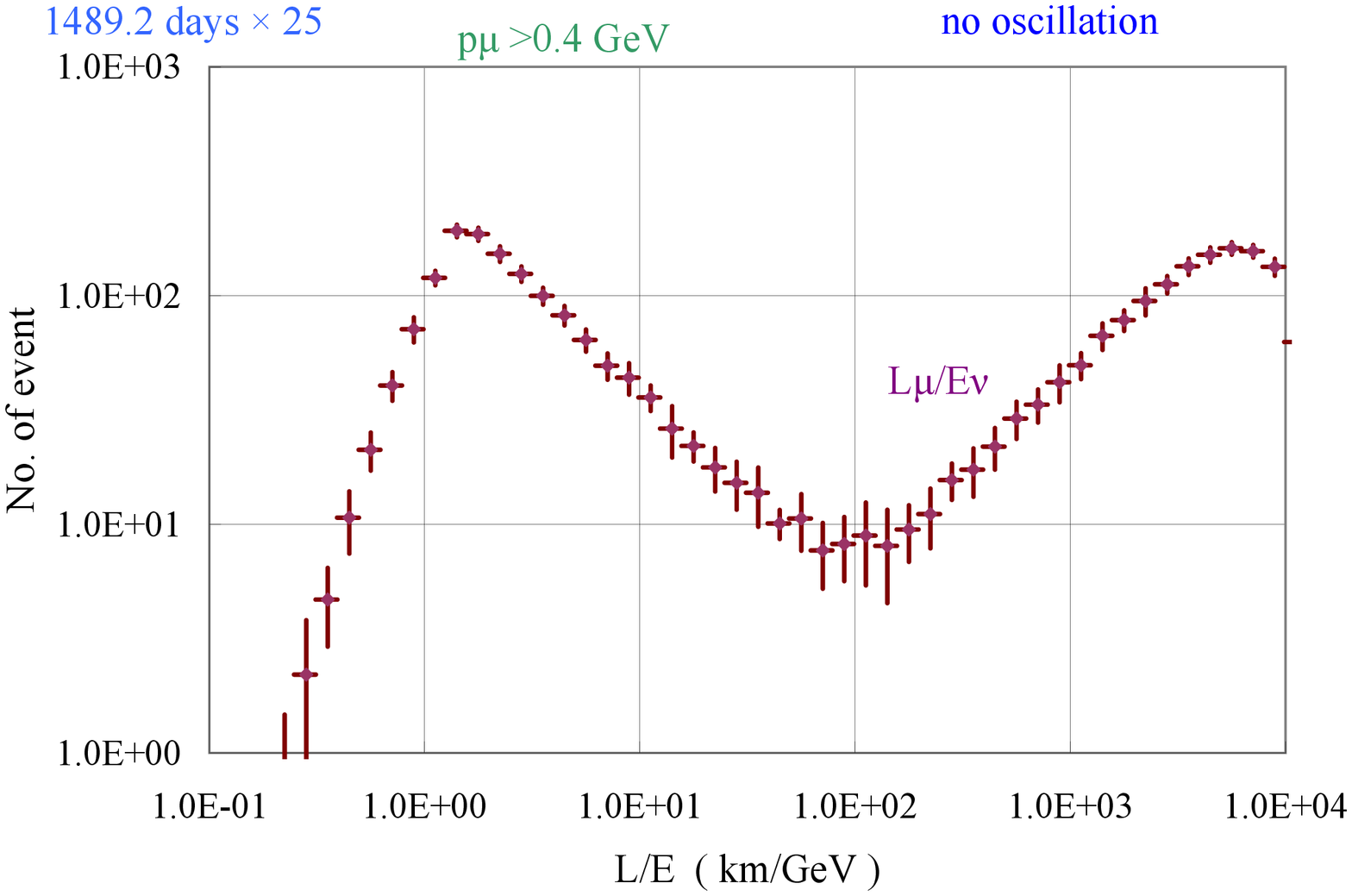}
}}
\vspace{-1.5cm}
\caption{$L_{\mu}/E_{\nu}$ distribution without oscillation
for 37230 live days (25 SK live days).}
\label{figJ037}
\end{center}
\end{figure}

In Figures~\ref{figJ036} and \ref{figJ037},
 we give $L_{\mu}/E_{\nu}$ distributions without 
oscillation for 1489.2 live days (one SK live day) and 37230 live days
(25 SK live days), respectively.
Comparing Figure~\ref{figJ036} with Figure~\ref{figJ037},
 the larger statistics makes the
 distribution smoother.
 Also, there is a sinusoidal-like bottom expressed in a logarithmic scale
which has no relation with neutrino oscillation.

\subsubsection{For oscillation (SK oscillation parameters)}

\begin{figure}
\begin{center}
\vspace{-2cm}
\rotatebox{90}{%
\resizebox{0.45\textwidth}{!}{%
  \includegraphics{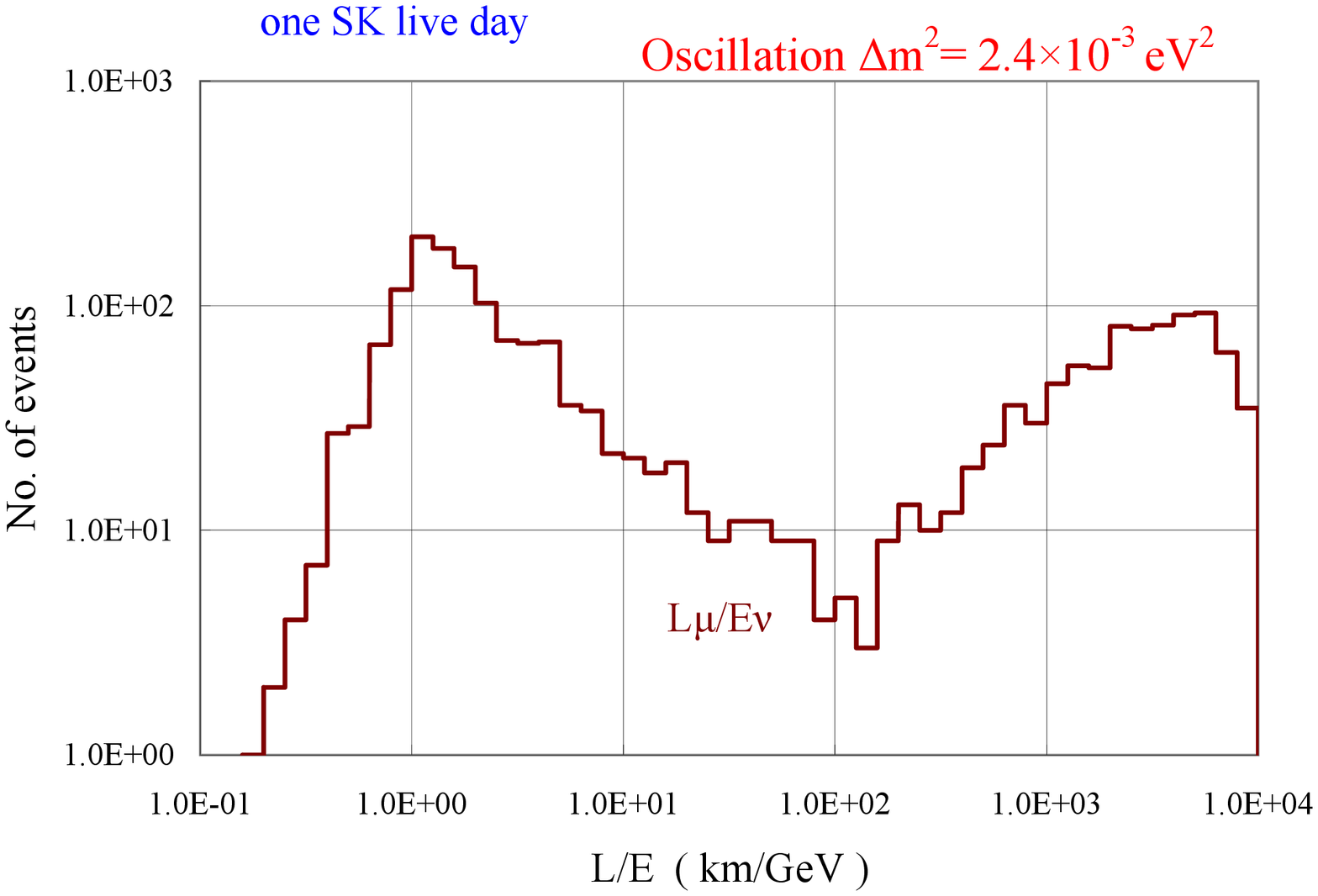}
}}
\vspace{-1.5cm}
\caption{$L_{\mu}/E_{\nu}$ distribution with oscillation
for 1489.2 live days (one SK live day).}
\label{figJ038}
\vspace{-0.5cm}
\rotatebox{90}{%
\resizebox{0.45\textwidth}{!}{%
  \includegraphics{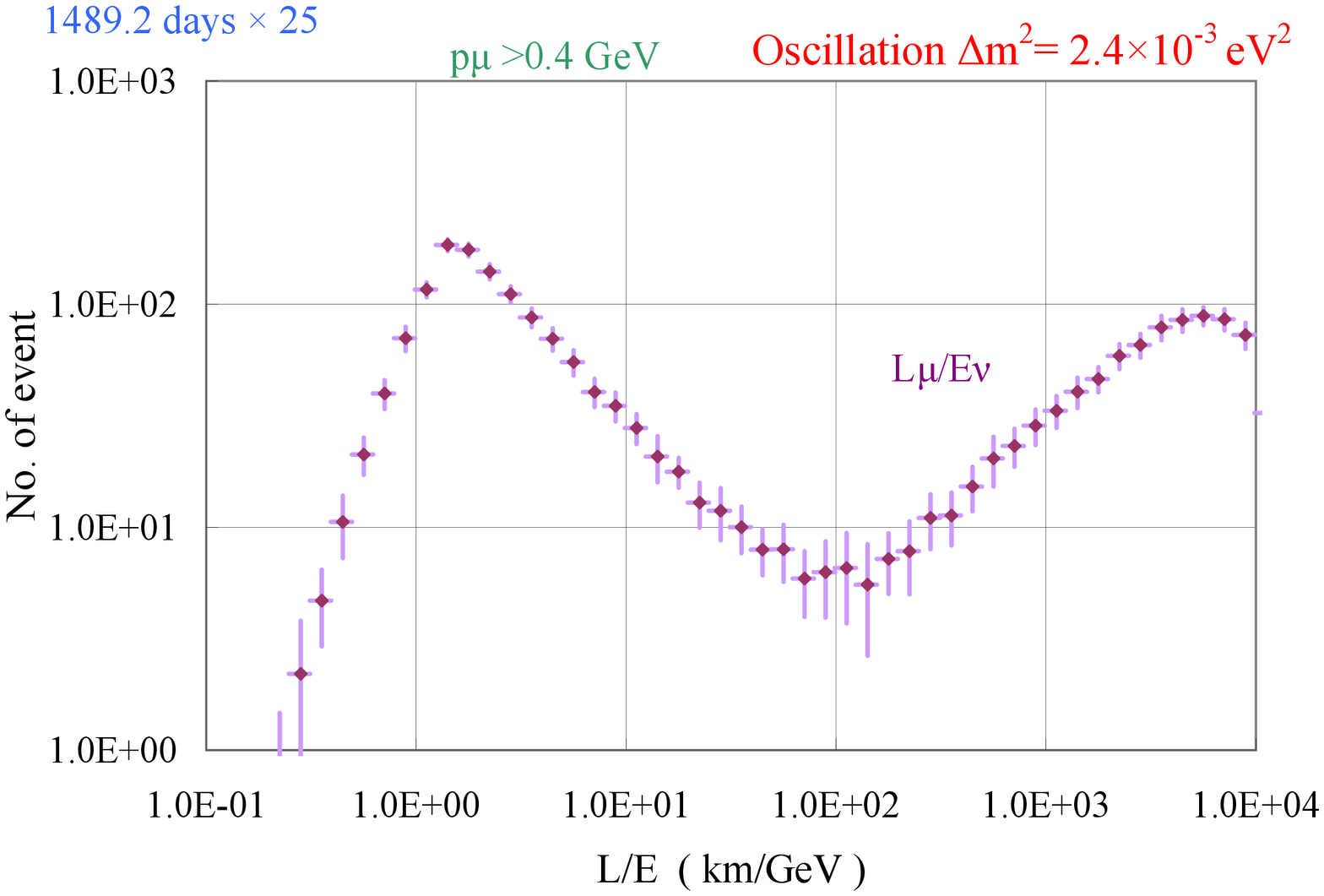}
}}
\vspace{-1.5cm}
\caption{$L_{\mu}/E_{\nu}$ distribution with oscillation
for 37230 live days (25 SK live days).}
\label{figJ039}
%\end{center}
%\end{figure}
%\begin{figure}
%\begin{center}
\vspace{-0.5cm}
\rotatebox{90}{%
\resizebox{0.45\textwidth}{!}{%
  \includegraphics{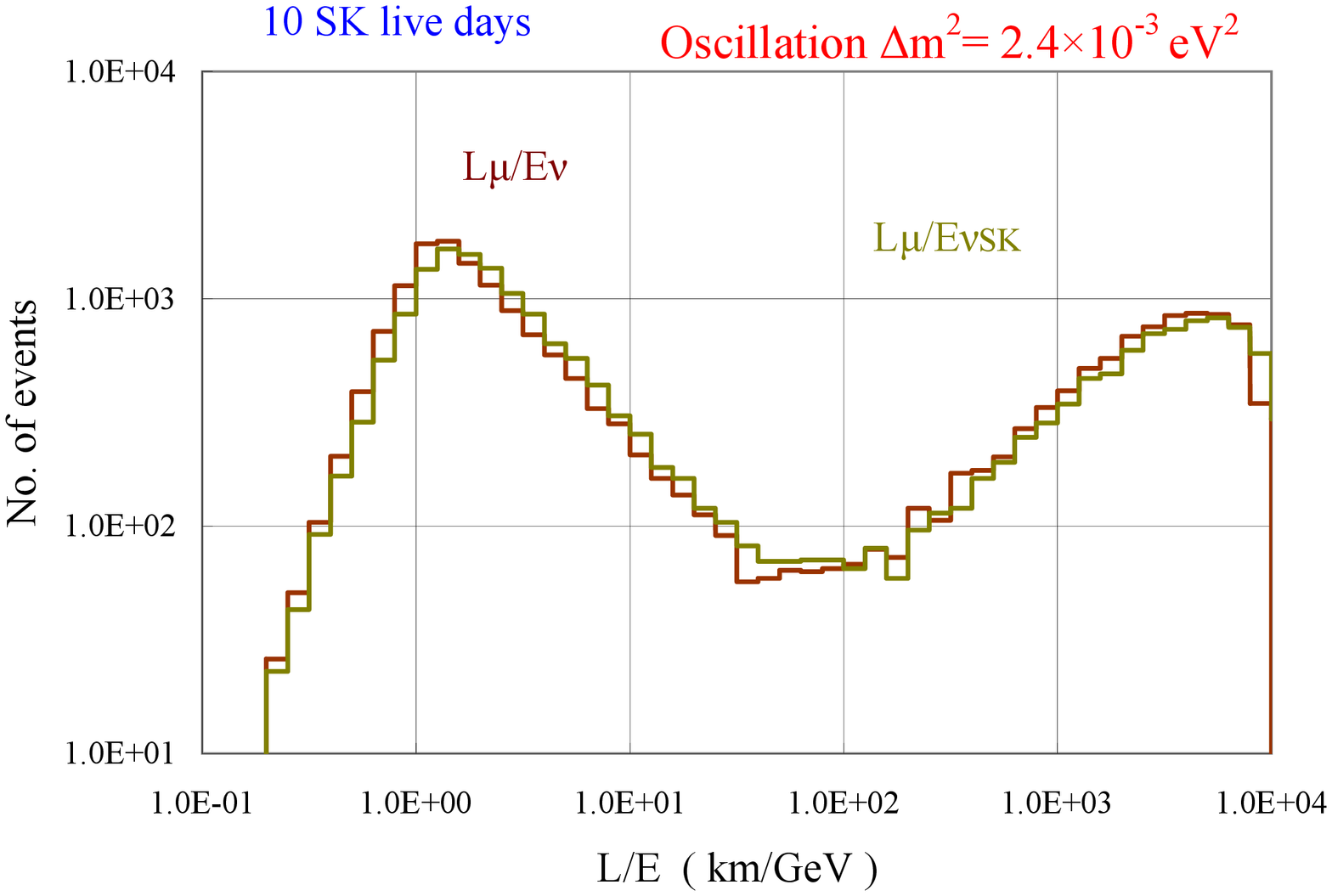}
}}
\vspace{-1.5cm}
\caption{$L_{\mu}/E_{\nu,SK}$ distribution in
comparison with $L_{\mu}/E_{\nu}$ distribution with oscillation 
for 14892 day (10 SK live days).}
\label{figJ040}
\end{center}
\end{figure}

In Figures~\ref{figJ038} and \ref{figJ039}, 
we give the $L_{\mu}/E_{\nu}$ distribution with 
oscillation for 1489.2 live days (one SK live day) and 37230 live days
(25 SK live days), respectively. 
In Figure~\ref{figJ038},
 we may find something like a bottom 
%which corresponds to the first maximum oscillation
 near $\sim$200 (km/GeV).
However, such the dip disappears, by making the statistics larger as 
shown in Figure~\ref{figJ039}. 
At any rate, we cannot find any indication on the maximum oscillation 
from these figures.


\subsubsection{$L_{\mu}/E_{\nu,SK}$ distribution for the oscillation}

 Instead of $E_{\nu}$ which is correctly sampled from the corresponding 
probability functions, 
let us utilize $E_{\nu,SK}$ which is obtained from the "approximate" 
formula (Eq.(6)) in the preceding paper\cite{Konishi2}).

 We express $E_{\nu}$ described in Eq.(6)
utilized by Super-Kamiokande Collaboration as $E_{\nu,SK}$ 
to discriminate from our $E_{\nu}$ obtained in the stochastic manner 
correctly.

In Figure~\ref{figJ040}, 
we give $L_{\mu}/E_{\nu,SK}$ distribution for 14892 live days 
(10 SK live days),
comparing with $L_{\mu}/E_{\nu}$ distribution.
It is understood from the comparison that there is no significant 
difference between $L_{\mu}/E_{\nu,SK}$ distribution and 
$L_{\mu}/E_{\nu}$ one in a logarithmic scale.
This fact tells us that the "aproximate" formula for $E_{\nu}$ used by 
Super-Kamiokande Collaboration 
%which is not suitable for the treatment of the stochastic quantities, 
does not produce so significant difference in the logarithmic scale,
which may be accidental.
Although the approximated foumula is not suitable for the treatment of
stochastic quantities (see discussion in 3.3 \cite{Konishi2}),
the result is understandable from 
Figure~14 in the preceding paper\cite{Konishi2},
because there is no significant difference between the real 
distribution (correlation) and the "aproximate" formula apparently.
%Figure~\ref{figH014}.
 Also, we can conclude that we do not 
find any hole corresponding to the maximum oscillation in 
%$L_{\mu}/E_{\nu}$ or 
$L_{\mu}/E_{\nu,SK}$ distributions.
The reason why the Figure~\ref{figJ039} can not show such dip structure 
 as shown in Figures~\ref{figJ018} and \ref{figJ019},  
 comes from the situation that the role of $L_{\nu}$
is much more crucial than that of $E_{\nu}$ in the $L/E$ analysis.
Namely, $L_{\nu}$ cannot be replaced by $L_{\mu}$ at all.
% Also, see the discussion in the following subsection 4.4.    

In Figure~\ref{figJ040}, we give the comparison of 
$L_{\mu}/E_{\nu}$ distribution with $L_{\mu}/E_{\nu,SK}$ one.

The apparent small difference between $L_{\mu}/E_{\nu,SK}$ distribution
and $L_{\mu}/E_{\nu}$ one in Figure~\ref{figJ040}
may come from that $L_{\mu}$ plays an effective role in comparison with 
$E_{\nu}$ or $E_{\nu,SK}$, in spite of the situation that there are 
non-negligible differences between $E_{\nu}$ or $E_{\nu,SK}$ 
(see Figure~19 of the preceding paper\cite{Konishi2}).

\begin{figure}
\begin{center}
\vspace{-2cm}
\rotatebox{90}{%
\resizebox{0.45\textwidth}{!}{%
  \includegraphics{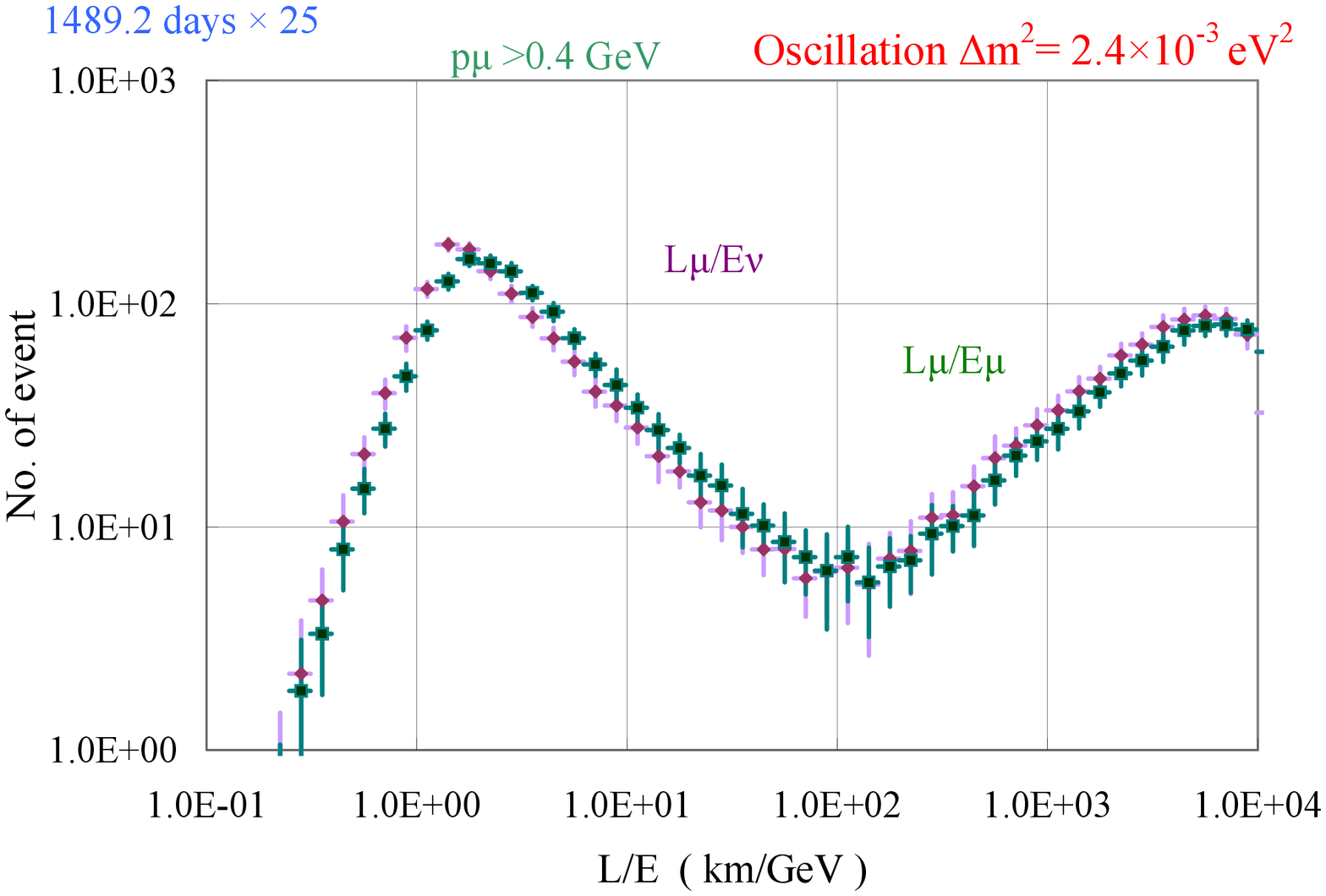}
}}
\vspace{-1.5cm}
\caption{Comparison between 
$L_{\mu}/E_{\nu}$ distribution and $L_{\mu}/E_{\mu}$ distribution 
 with oscillation for 37230 days (25 SK live days).}
\label{figK045}
\vspace{-0.5cm}
\rotatebox{90}{%
\resizebox{0.45\textwidth}{!}{%
  \includegraphics{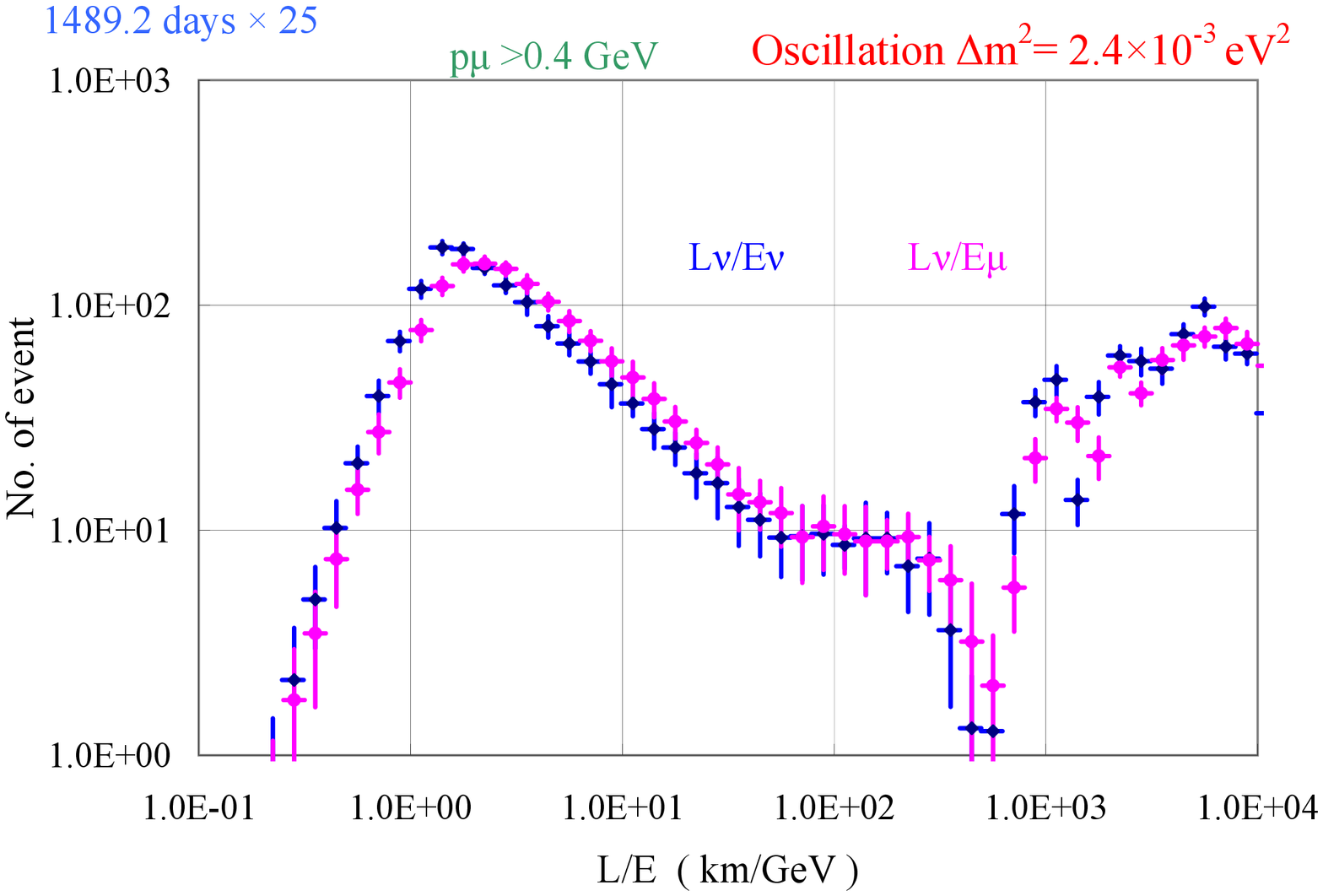}
}}
\vspace{-1.5cm}
\caption{Comparison between 
$L_{\nu}/E_{\nu}$ distribution and $L_{\nu}/E_{\mu}$ distribution 
 with oscillation for 37230 days (25 SK live days).}
\label{figK044}
%\end{center}
%\end{figure}
%\begin{figure}
%\begin{center}
\vspace{-0.5cm}
\rotatebox{90}{%
\resizebox{0.45\textwidth}{!}{%
  \includegraphics{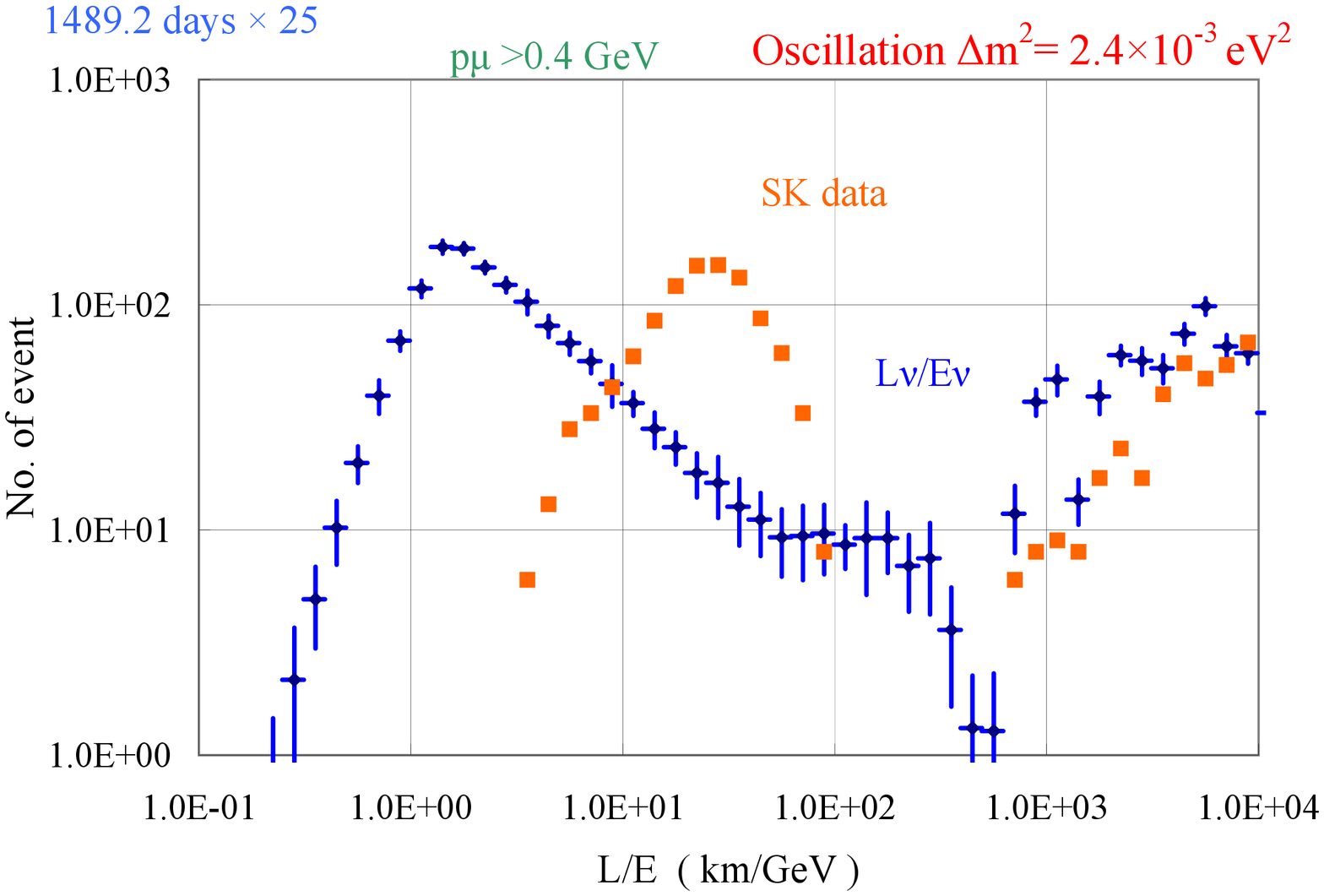}
}}
\vspace{-1.5cm}
\caption{The comparison of $L/E$ distribution for single-ring muon 
events due to QEL among {\it Fully Contained Events} with the 
corresponding one by the Super-Kamiokande
Experiment.}
\label{figK046}
\end{center}
\end{figure}

In Figure~\ref{figK045}, we compare $L_{\mu}/E_{\nu}$ distribution with $L_{\mu}/E_{\nu}$ one. 
The pretty overlapping between them in a logarithmic scale denotes that 
$L_{\mu}$ play an important role while the energies concerned only play 
the secondary role. The similar situation is expected in $L_{\nu}$.
In Figure~\ref{figK044}, we compare $L_{\nu}/E_{\mu}$ distribution with 
$L_{\nu}/E_{\nu}$ one. 
It is clear from the figure that we find the first maximum oscillations 
in both distributions on nearly same locations. 
Another clearer situation is found in the comparison of both 
distributions expressed in a linear scale shown as 
Figures~\ref{figJ024} and \ref{figL056}. 
It become clear from the comparison of Figures~\ref{figJ040},
\ref{figK045} and \ref{figK044} that the flight length, 
either $L_{\nu}$ or $L_{\mu}$, plays a decisively important role 
in any $L/E$ distribution, compared with the energies concerned,
 either $E_{\nu}$ or $E_{\mu}$, as it should be.

%%%%%%%%%%%%%%%%%%%%%%%%%%%%%%%%%%%%%%%%%%%%%%%
\section{Comparison of $L/E$ Distribution in the 
Super-Kamiokande Experiment with our Results}

In our classification, $L/E$ distribution by Super-Kamiokande 
Collaboration 
should be compared with our $L_{\mu}/E_{\mu}$ distribution 
because they measure $L_{\mu}$and $E_{\mu}$ directly, 
or it should be compared
with our $L_{\mu}/E_{\nu}$ distribution because they get 
$E_{\nu}$ through the transform from $E_{\mu}$. 
However, they assert that they measure $L_{\nu}$ on 
{\it the SK assumption on the direction}. 
Consequently, we compare here their $L/E$ distribution with our 
$L_{\nu}/E_{\nu}$ one at first.
In Figure~\ref{figK046},
we compare our $L_{\nu}/E_{\nu}$ distribution with their single ring muon 
events among {\it Fully Contained Events} should be compared\footnote
{We read out {\it Fully Contained Events} among total events
from Super-Kamiokande Collaboration \cite {Ashie2}\cite {Ashie1}.
Since we are only interested in single ring muon events,
events with the highest quality, excluding
{\it Partially Contained Events} for our analysis. 
}
as corresponding ones.

There are two important matters to be examined in the $L/E$ distribution 
related to the shapes between ours and theirs which can be discussed 
without entering the details for technical and experimental condition or 
criteria around their experiments.
The first one is related to the location and its shape for the first 
maximum oscillation.
And the second one is related to the location which give the maximum 
frequency of the events concerned.
 In the first one, we can observe the first maximum oscillation at 
$L_{\nu}/E_{\nu} = 515$ km/GeV (see Figures~\ref{figJ018} to 
\ref{figJ026})
 exactly in our computer numerical experiment under the oscillation 
parameters obtained by Super-Kamiokande Collaboration. 
Furthermore, we can observe clearly the second, the third or more higher 
order maximum oscillations at the anticipated locations (see Eq.(3) and 
Figure \ref{figJ024}) in our numerical experiment.
 Also, the shapes of those maximum oscillations are rather sharp 
which comes from the specified oscillation parameters obtained 
by Super-Kamiokande Collaboration. 
On the other hand, they obtain a broader region for the absence of the
 neutrino events as the result of the first maximum oscillation
such as $100<L/E<800$ (km/GeV).
  Such a broader region may contradict the concept of the survival 
probability for a given flavor under the specification of their 
oscillation parameters, taking account of the results from the 
analysis of the single ring muon events among 
{\it Fully Contained Events}, the highest quality events among all events 
to be analyzed by them. 

 Now, we examine the remarkable difference between ours and theirs as for 
the locations of the maximum frequencies for the events.
  As shown in Figure~\ref{figK046}, 
we give it as $1.0<L_{\nu}/E_{\nu}<1.26$ (km/GeV),
 while they give $20<L_{\nu}/E_{\nu}<25$ (km/GeV) which is 
larger than ours by one order of the magnitude.

\begin{figure}
\begin{center}
\vspace{-1cm}
\hspace*{-0.5cm}
\rotatebox{90}{%
\resizebox{0.4\textwidth}{!}{%
  \includegraphics{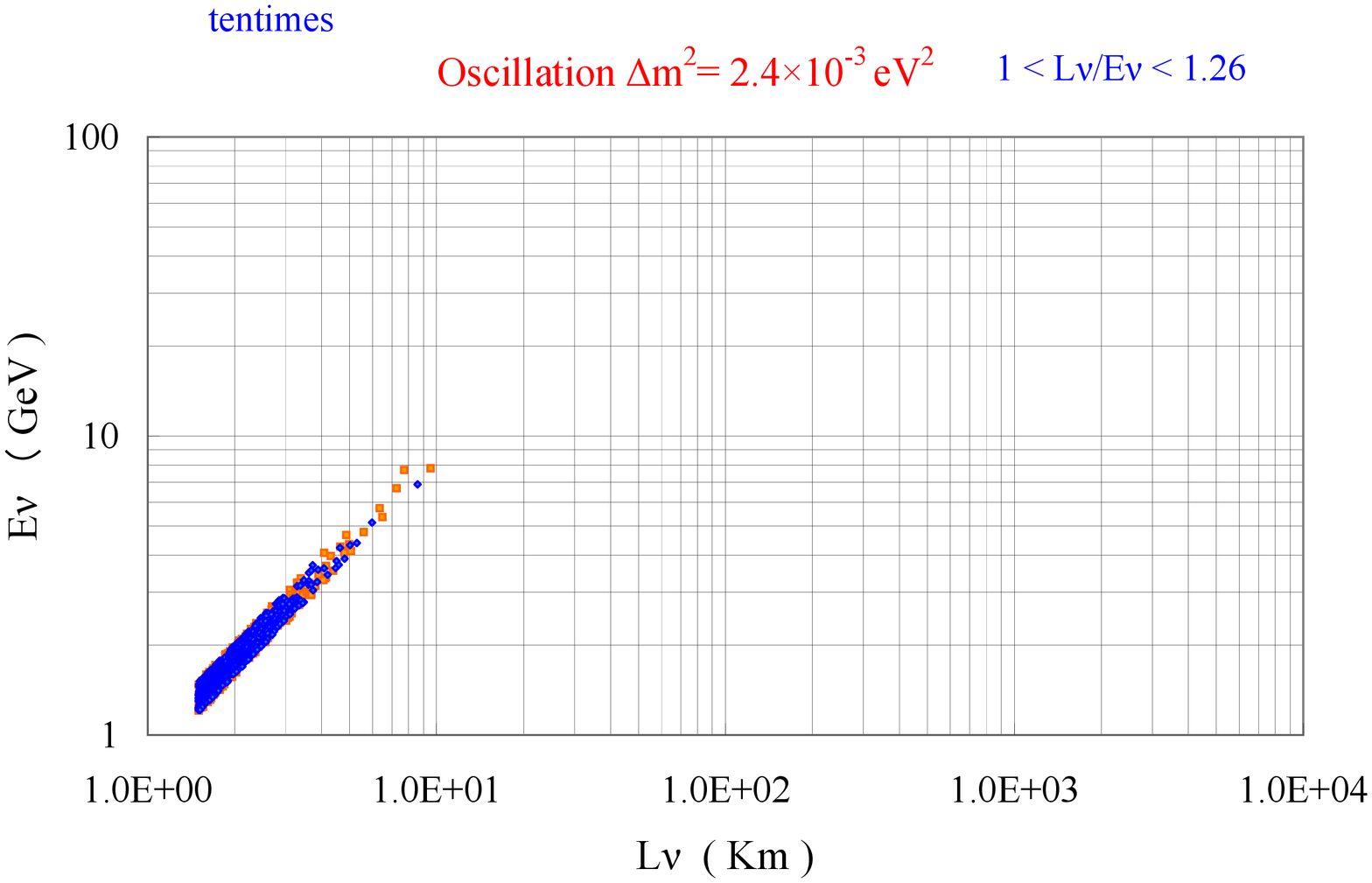}
}}
\vspace{-1.5cm}
\caption{Correlation diagram between $L_{\nu}$ and $E_{\nu}$ for 
$1.0<L_{\nu}/E_{\nu}<1.26 $ (km/GeV) which corresponds to the maximum frequency of the neutrino events for $L_{\nu}/E_{\nu}$
distribution in our computer numerical experiment for 14892 live days
 (10 SK live days).}
\label{figK047}
\vspace{-0.5cm}
\hspace*{-0.5cm}
\rotatebox{90}{%
\resizebox{0.4\textwidth}{!}{%
  \includegraphics{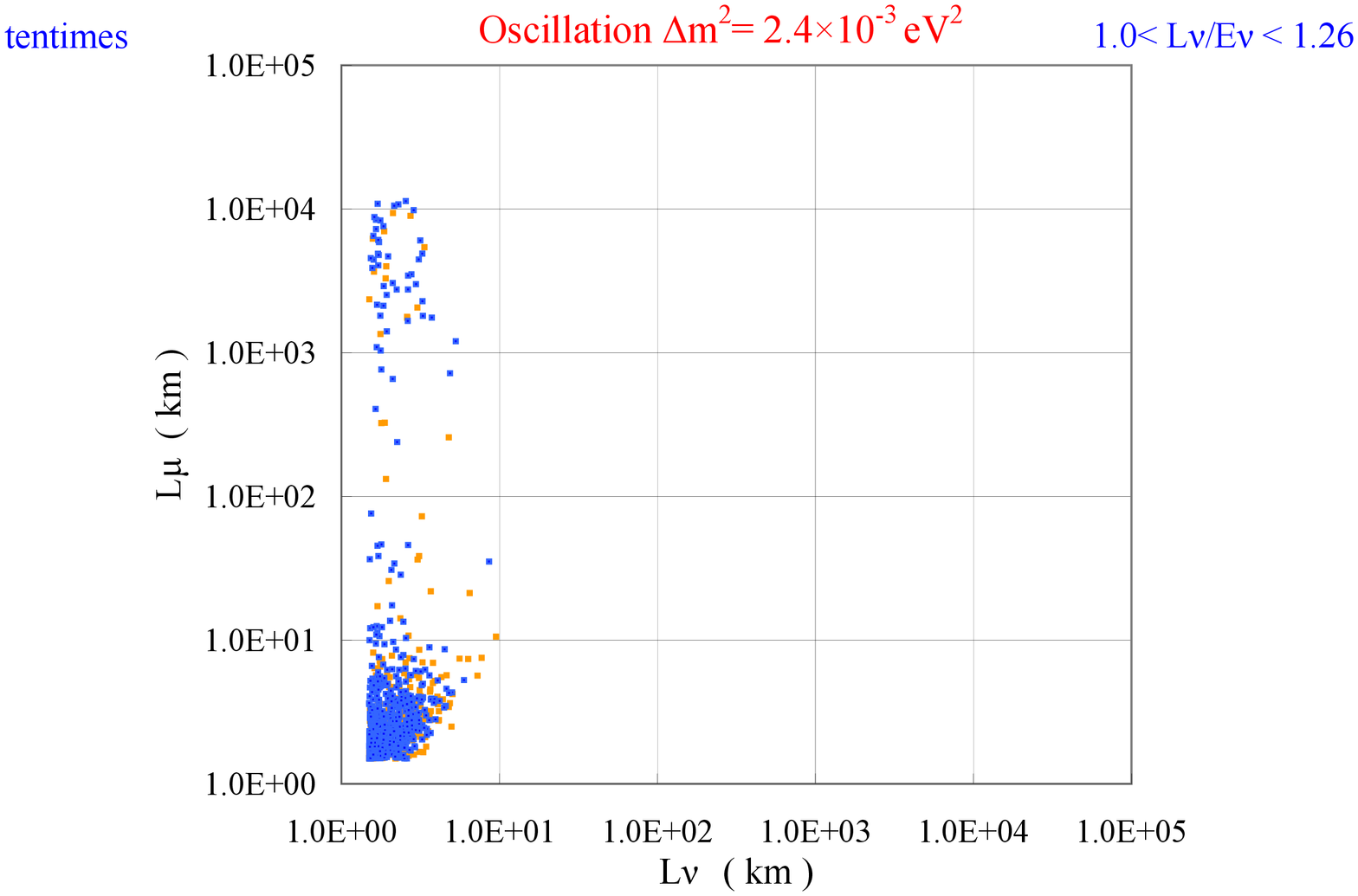}
}}
\vspace{-1.5cm}
\caption{Correlation diagram between $L_{\nu}$ and $L_{\mu}$ for 
$1.0<L_{\nu}/E_{\nu}<1.26 $ (km/GeV) 
under the neutrino oscillation parmeters obtained by 
Super-Kamiokande Collaboration
for 14892 live days (10 SK live days).}
\label{figK048}
\vspace{-0.5cm}
\hspace*{-0.5cm}
\rotatebox{90}{%
\resizebox{0.4\textwidth}{!}{%
  \includegraphics{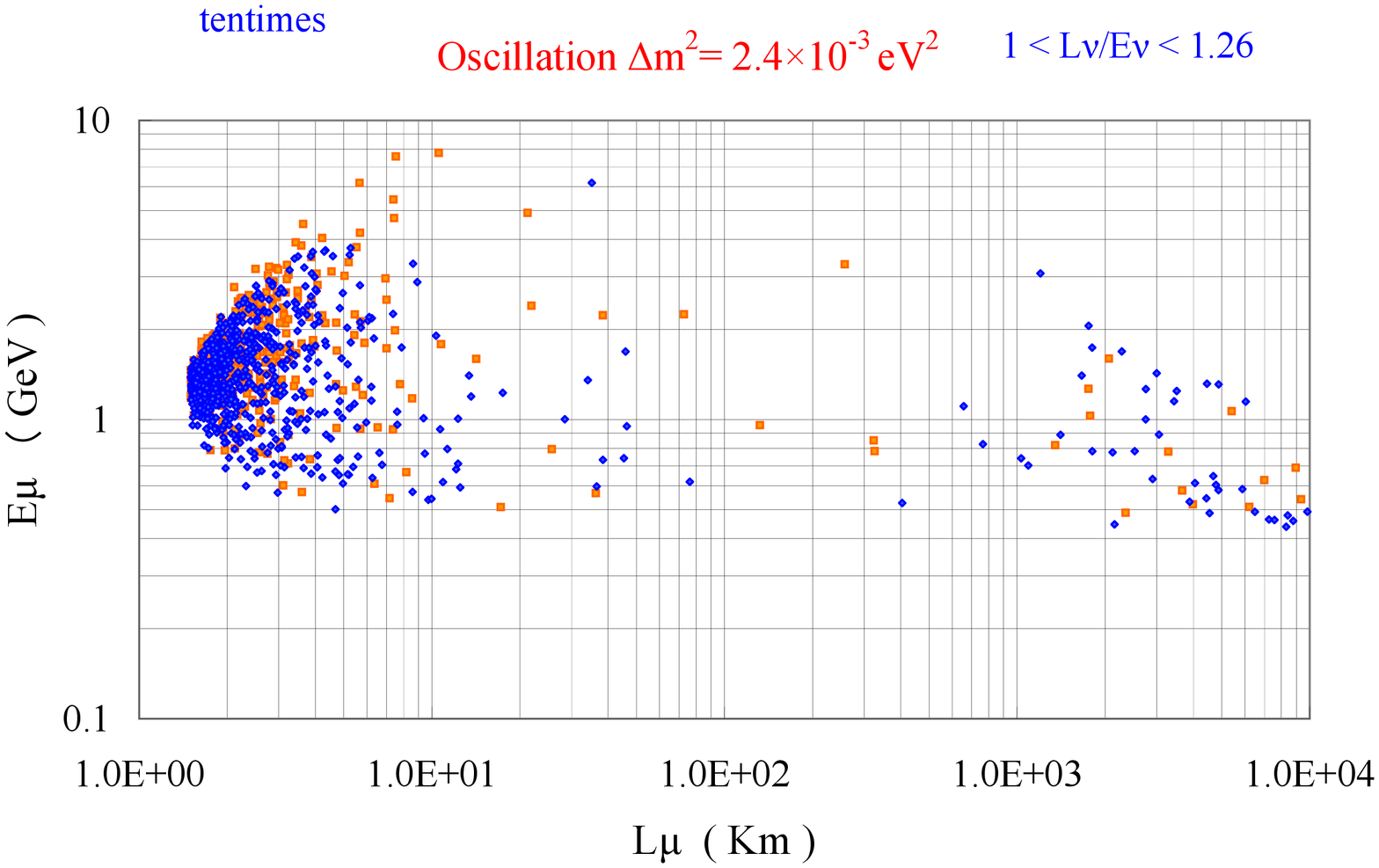}
}}
\vspace{-1.5cm}
\caption{Correlation diagram between $L_{\mu}$ and $E_{\mu}$ for 
$1.0<L_{\nu}/E_{\nu}<1.26 $ (km/GeV) which corresponds to the maximum frequency of the neutrino events for $L_{\nu}/E_{\nu}$
distribution in our computer numerical experiment for 14892 live days
(10 SK live days).}
\label{figK049}
\end{center}
\end{figure}

\begin{figure}
\begin{center}
\vspace{-1cm}
\hspace*{-0.5cm}
\rotatebox{90}{%
\resizebox{0.4\textwidth}{!}{%
  \includegraphics{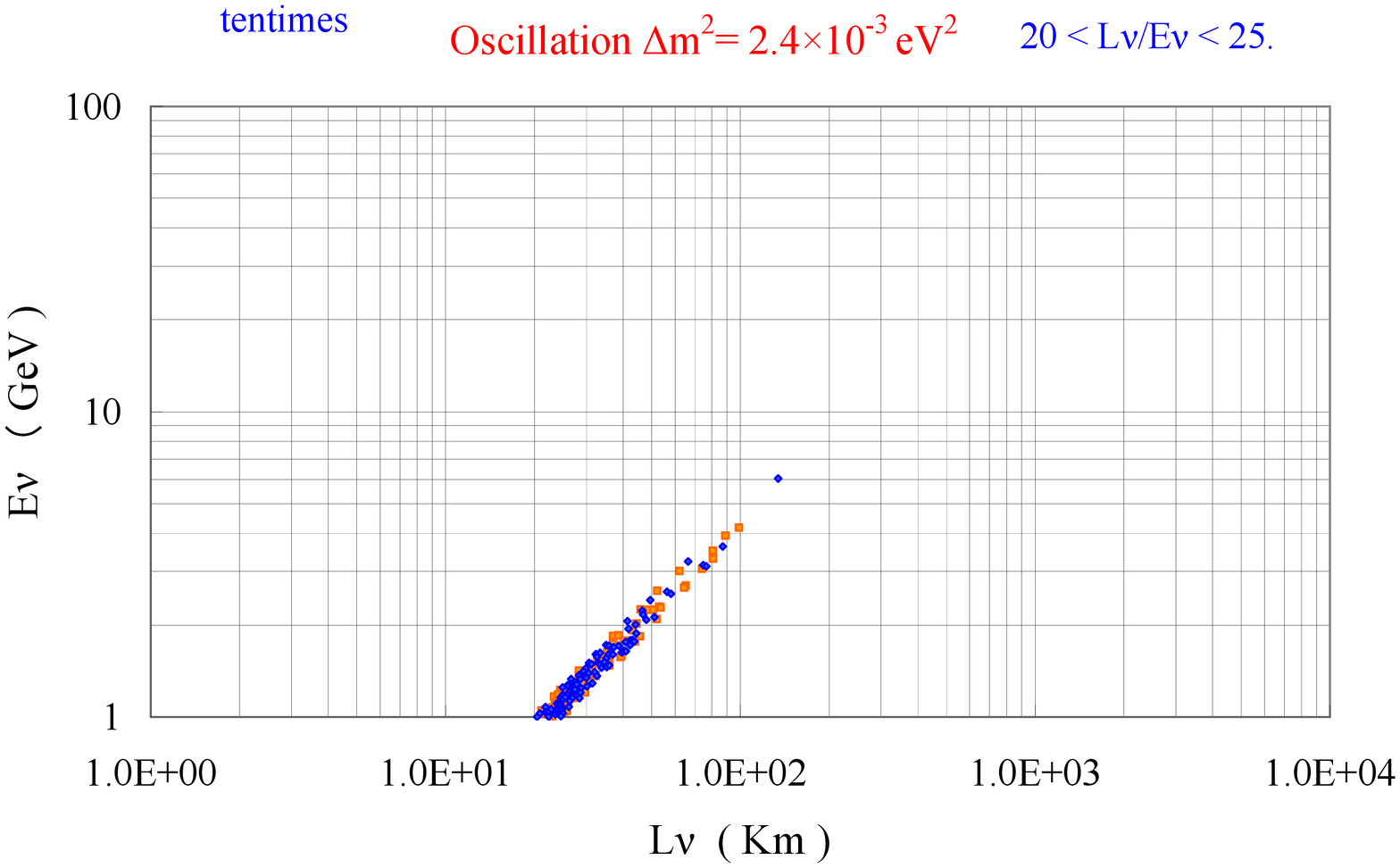}
}}
\vspace{-1.5cm}
\caption{Correlation diagram between $L_{\nu}$ and $E_{\nu}$ for 
$20<L_{\nu}/E_{\nu}<25 $ (km/GeV) which corresponds to the maximum frequency of the neutrino events for $L_{\mu}/E_{\nu}$
distribution in SK experiment for 14892 live days
(10 SK live days).}
\label{figK050}
\vspace{-0.5cm}
\hspace*{-0.5cm}
\rotatebox{90}{%
\resizebox{0.4\textwidth}{!}{%
  \includegraphics{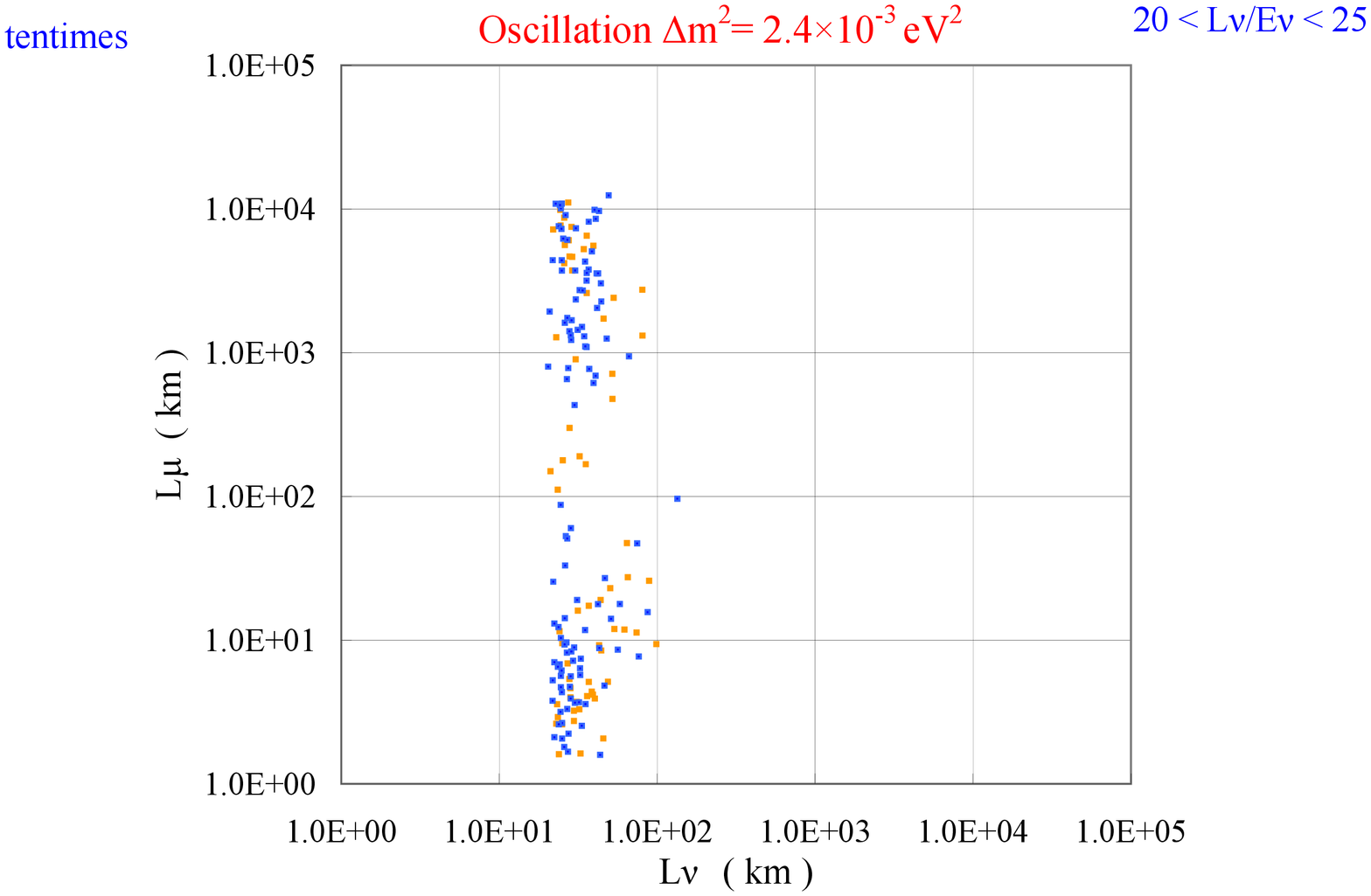}
}}
\vspace{-1.5cm}
\caption{Correlation diagram between $L_{\nu}$ and $L_{\mu}$ for 
$20<L_{\nu}/E_{\nu}<25 $ (km/GeV) 
under the neutrino oscillation parmeters obtained by 
Super-Kamiokande Collaboration
for 14892 live days (10 SK live days).}
\label{figK051}
\vspace{-0.5cm}
\hspace*{-0.5cm}
\rotatebox{90}{%
\resizebox{0.4\textwidth}{!}{%
  \includegraphics{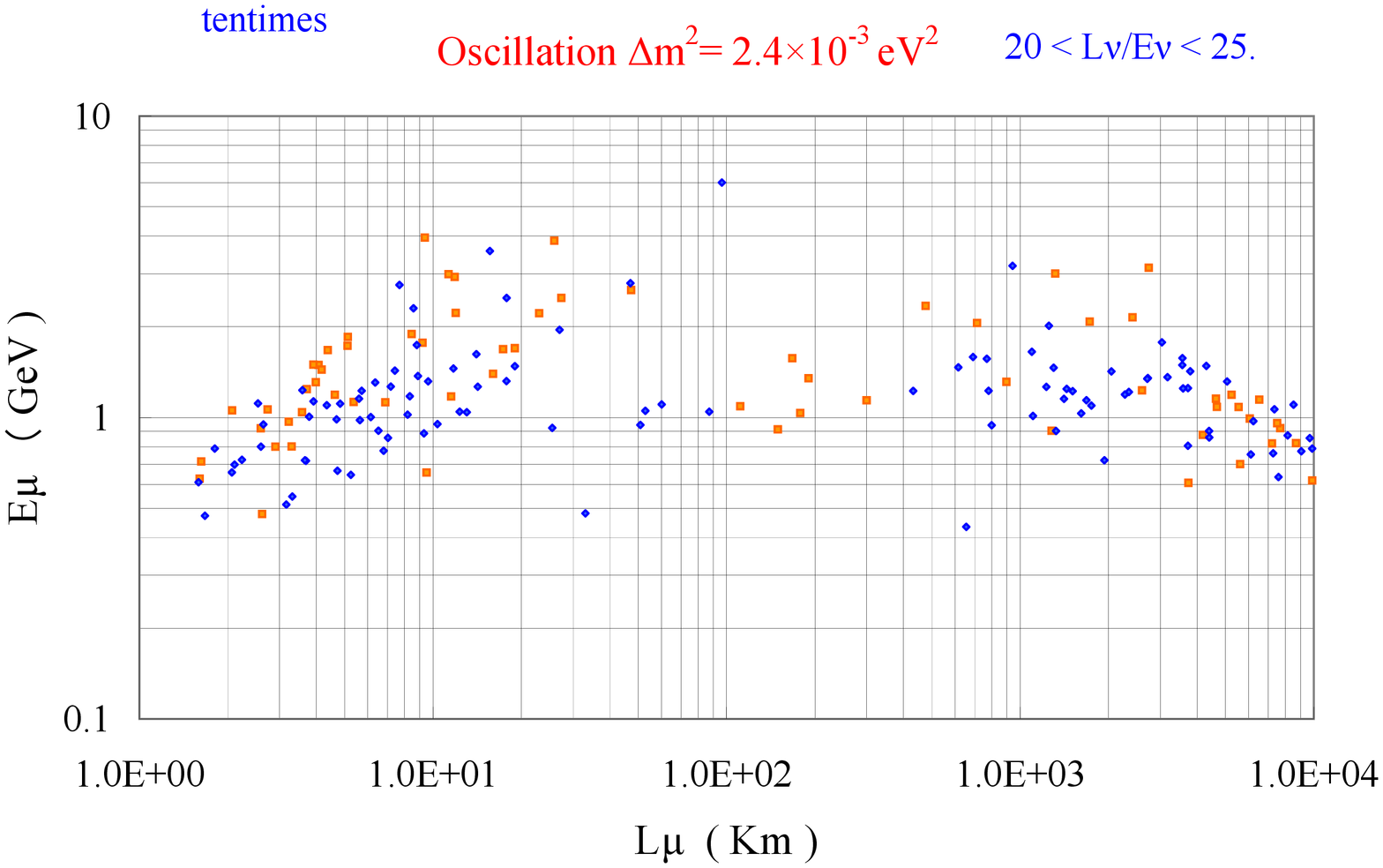}
}}
\vspace{-1.5cm}
\caption{Correlation diagram between $L_{\mu}$ and $E_{\mu}$ for 
$20<L_{\nu}/E_{\nu}<25 $ (km/GeV) which corresponds to the maximum frequency of the neutrino events for $L_{\nu}/E_{\nu}$
distribution in SK experiment for 14892 live days
(10 SK live days).}
\label{figK052}
\end{center}
\end{figure}

\begin{figure}
\begin{center}

\vspace{-1cm}
\hspace*{-0.5cm}
\rotatebox{90}{%
\resizebox{0.4\textwidth}{!}{%
  \includegraphics{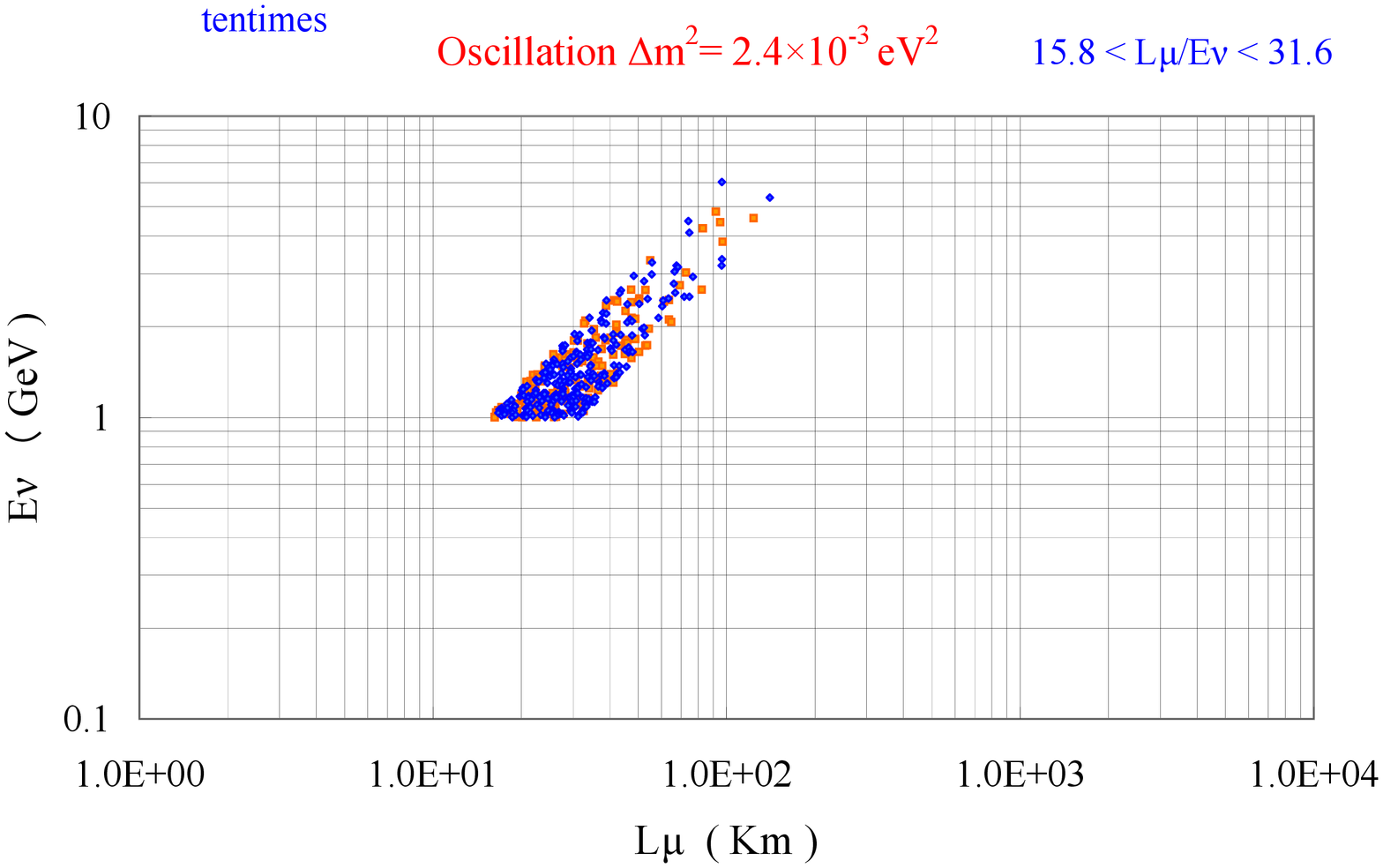}
}}
\vspace{-1.5cm}
\caption{Correlation diagram between $L_{\mu}$ and $E_{\nu}$ for 
$15.8<L_{\mu}/E_{\nu}<31.6 $ (km/GeV) which correspond to the maximum 
frequency of the neutrino events for $L_{\mu}/E_{\nu}$
distribution in SK experiment for 14892 live days
(10 SK live days).}
\label{figK053}
\vspace{-0.5cm}
\hspace*{-0.5cm}
\rotatebox{90}{%
\resizebox{0.4\textwidth}{!}{%
  \includegraphics{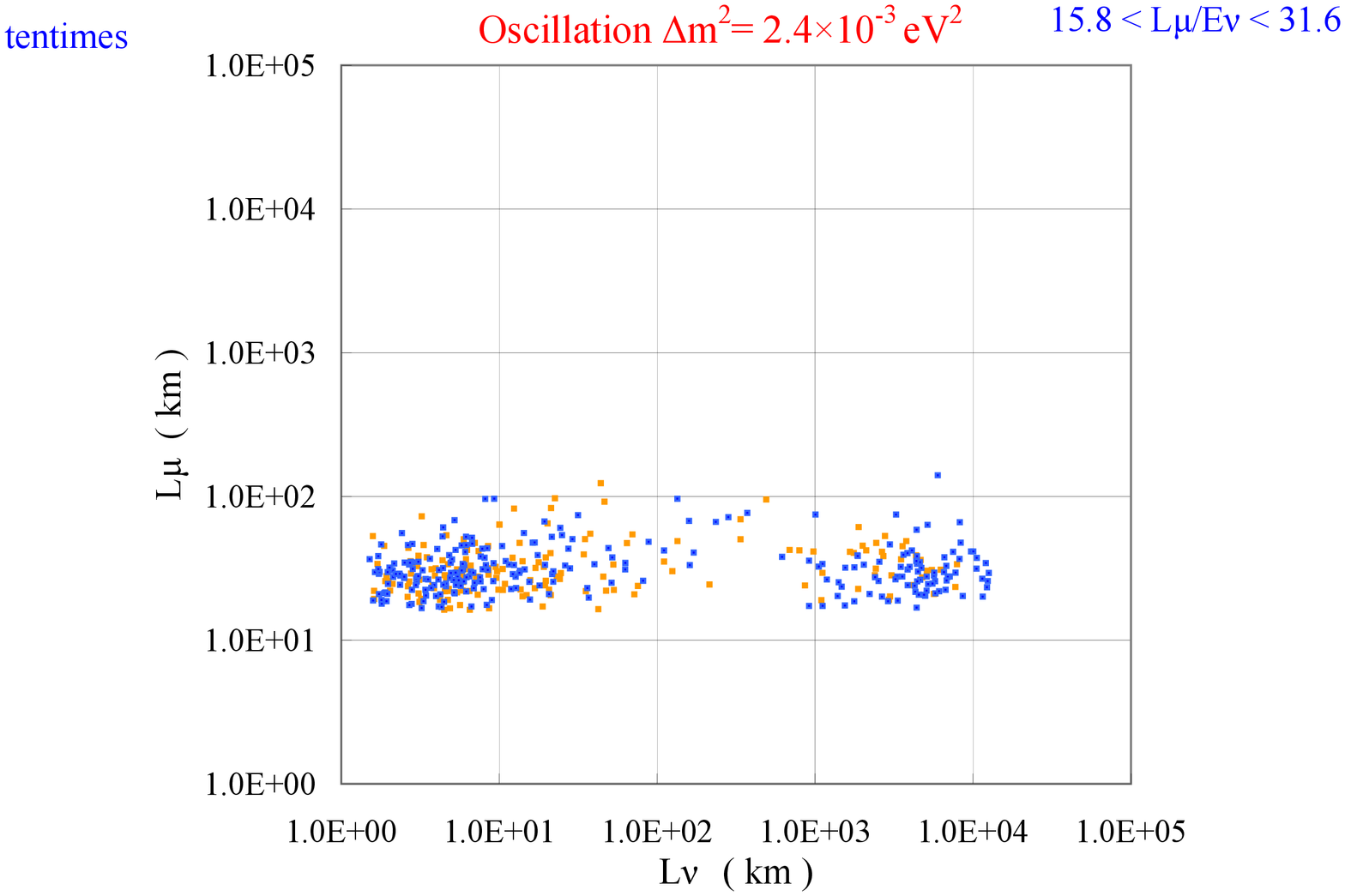}
}}
\vspace{-1.5cm}
\caption{Correlation diagram between $L_{\nu}$ and $L_{\mu}$ for 
$15.8<L_{\mu}/E_{\nu}<31.6 $ (km/GeV) 
under the neutrino oscillation parmeters obtained by 
Super-Kamiokande Collaboration
for 14892 live days (10 SK live days).}
\label{figK054}
\vspace{-0.5cm}
\hspace*{-0.5cm}
\rotatebox{90}{%
\resizebox{0.4\textwidth}{!}{%
  \includegraphics{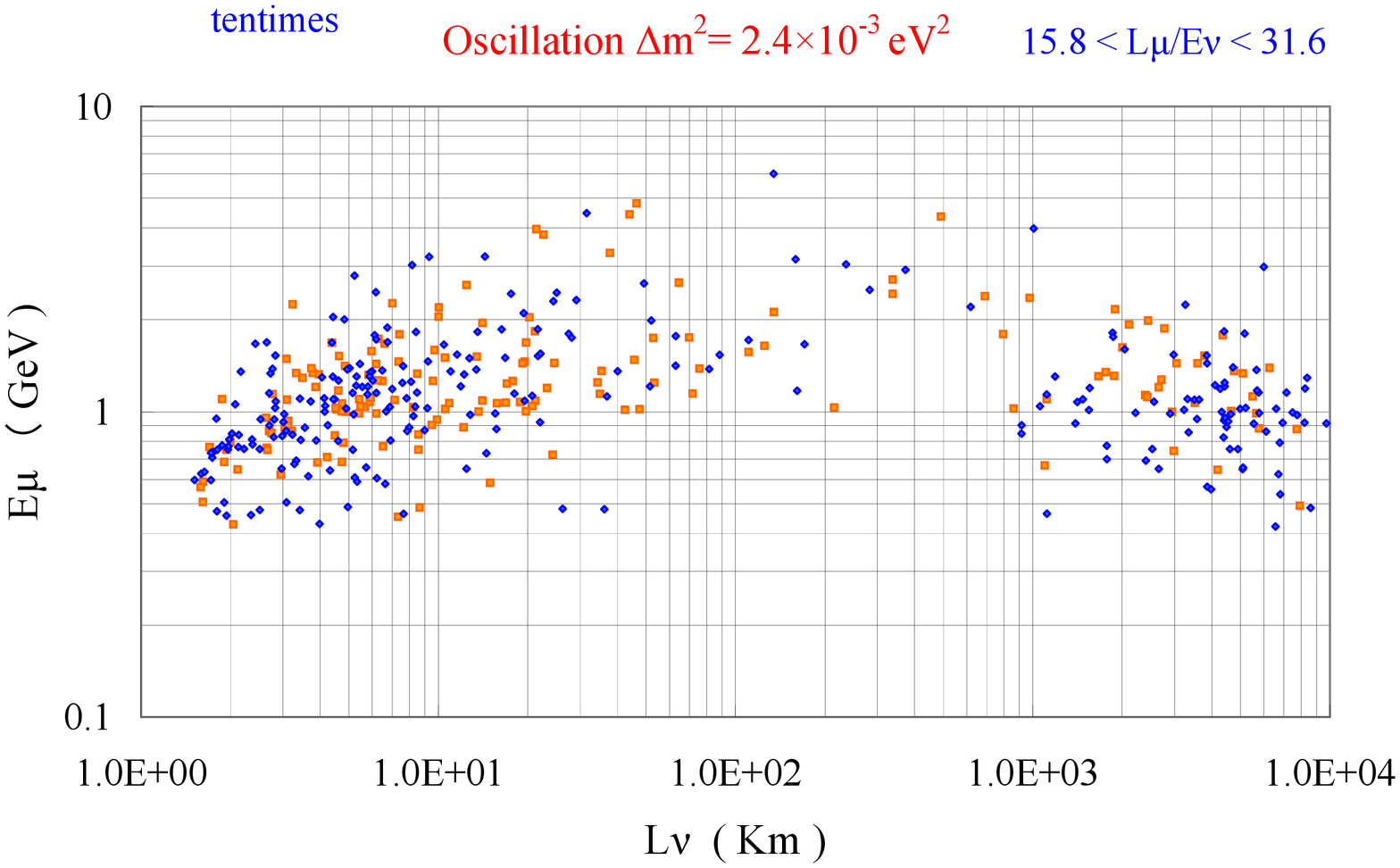}
}}
\vspace{-1.5cm}
\caption{Correlation diagram between $L_{\nu}$ and $E_{\mu}$ for 
$15.8<L_{\mu}/E_{\nu}<31.6 $ (km/GeV) which correspond to the maximum 
frequency of the neutrino events for $L_{\mu}/E_{\nu}$
distribution in SK experiment for 14892 live days
(10 SK live days).}
\label{figK055}
\end{center}
\end{figure}

 Here, at first, in our computer numerical experiment, we discuss the 
correlation between $L$ and $E$ at the location for the maximum frequency for 
the events and the similar correlation at the corresponding location 
where Super-Kamiokande Collaboration give their maximum frequency.
 Next, we clarify what happen at the location for the maximum frequency 
for the events in their experiment.

 In the following discussion, we designate neutrino events  
with $135^\circ<\theta_{\nu}<180^\circ$
as {\it vertical like events}, 
 events with $90^\circ<\theta_{\nu}<135^\circ$
as {\it horizontal like events}  
and events with $88^\circ<\theta_{\nu}<92^\circ$
as {\it exclusively horizontal events}, respectively.

In Figures~\ref{figK047} to \ref{figK049}, we give the correlations for 
the maximum frequency for the events 
in our $L_{\nu}/E_{\nu}$ distribution.  
Also, in Figures~\ref{figK050} to \ref{figK052},
 we  give the similar 
correlations for the location in our $L_{\nu}/E_{\nu}$ distribution 
which correspond to the maximum frequency of the events obtained by 
Super-Kamiokande Collaboration.
In Figure~\ref{figK047}, 
we give the correlation between $L_{\nu}$ and $E_{\nu}$
for the interval $1.0<L_{\nu}/E_{\nu}<1.26 $ (km/GeV)
at the maximum frequency for the events in our computer numerical 
experiment. 
It is clear from the figure that all incident neutrino events have
 the values of $L_{\nu}$ less than 10 km, 
corresponding to the {\it vertical-like},
{\it horizontal like} and the 
{\it exclusively horizontal} downward neutrinos,
 taking account of the transform between 
$L_{\nu}$($L_{\mu}$) and $cos\theta_{\nu}$($cos\theta_{\mu}$), 
as they must be. 
All these neutrinos cover all incident directions as 
the downward and they correspond to the maximum frequency for the events. 
 It is clear from Figure~\ref{figK050} that the incident 
neutrino events are concentrated into 
the interval of from 20 to 100 km for the value of $L_{\nu}$ 
which correspond to the {\it horizontal like events} and the 
{\it exclusively horizontal events}, not including 
{\it vertical like events},
 taking the account of the relation between $L_{\nu}$($L_{\mu}$) 
and $cos\theta_{\nu}$($cos\theta_{\mu}$). 
It is quite natural that events number there is smaller than that 
for the maximum frequency which includes the vertical like events.

In Figure~\ref{figK048}, we give the correlation between
$L_{\nu}$ and $L_{\mu}$ for the same intervals as  
in Figure~\ref{figK047}.
  It is clear from the figure that the majority of the events is 
concentrated into the squared region with 
$L_{\nu}<10$~km and $L_{\mu}<10$~km.
This denotes that the downward incident neutrinos 
produce muons in the forward direction 
irrespective of scattering angles.
 At the same time, it should be noticed 
from the figure that the non-negligible parts of the downward incident 
neutrino events produce the upward muons 
($\sim$1000 to 10000 km) due to the backscattering as well as the 
azimuthal angle effect due to QEL for both 
{\it horizontal like} and 
{\it exclusively horizontal events}
which is clearly shown in Figure~\ref{figK049}, too.
 These upward muons may be surely 
identified as the products of the upward neutrinos in the analysis 
performed by the Super-Kamiokande Collaboration.

 Furthermore, from the comparisons 
of Figure~\ref{figK048} with Figure~\ref{figK051} and 
of Figure~\ref{figK049} with Figure~\ref{figK052}, 
it is easily understood that 
{\it exclusively horizontal neutrinos} which occupy 
the majority in Figures~\ref{figK051} to \ref{figK052} 
are more influenced 
by the effects of both the backscattering and the azimuthal angle 
in QEL, compared with the cases 
in Figures~\ref{figK048} and \ref{figK049}. 
This denote {\it exclusively horizontal like neutrino} (downward)
 produce upward muons (backward direction) 
with higher probability compared with both  
{\it vertical like events} and {\it horizontal like events}. 
Thus, from the comparison of 
Figures~\ref{figK047} to \ref{figK049} with 
Figures~\ref{figK050} to \ref{figK052}, 
it is concluded that there is no contradiction for the 
interpretation of all figures between the maximum frequency 
for the event in our $L_{\nu}/E_{\nu}$ distribution 
and the corresponding distribution at the location where 
Super-Kamiokande Collaboration give their maximum frequebcy.

 Now, we examine the reliability of the maximum frequency of the events 
obtained by Super-Kamiokande collaboration as shown in 
Figure~\ref{figK046}.
 They assert that they measure 
the directions of the incident neutrinos by measuring those of muons under {\it the SK assumption on the direction}.  
However, what they measure really are the directions of 
the muons, but not those of the corresponding neutrinos due to their 
neutrality. 
Consequently, here, we examine the 
$L_{\mu}/E_{\nu}$ distribution in detail, 
not $L_{\nu}/E_{\nu}$ distribution for checking the experimental data 
obtained by Super-Kamiokande Collaboration. 
Strictly speaking, they measure $L_{\mu}$ and $E_{\mu}$, 
not $L_{\mu}$ and $E_{\nu}$. 
However, as they transform the original $E_{\mu}$
 to $E_{\nu}$ (Eq.(7) in Part1\cite{Konishi2}), we interpret 
they "measure" $L_{\mu}$ and $E_{\nu}$.
 Thus, we examine our $L_{\mu}/E_{\nu}$ distribution for the interval
% 15.8  Lmu/Enu 31.6 (km/Gev) 
$15.8<L_{\nu}/E_{\nu}<31.6 $ (km/GeV)
where Super-Kamiokande Collaboration give their maximum frequency 
of the events. 

In Figures~\ref{figK053} to \ref{figK055},
we give our correlations 
%among $L_{\nu}$, $L_{\mu}$,$E_{\nu}$ and $E_{\mu}$
 for the events at the location
$15.8<L_{\mu}/E_{\nu}<31.6 $ (km/GeV)  
 where Super-Kamiokande Collaboration give their maximum 
frequency for the events.
 In Figure~\ref{figK053}, we give the correlation between 
$L_{\mu}$ and $E_{\nu}$. 
In Figure~\ref{figK054}, we give the correlation between 
$L_{\nu}$ and $L_{\mu}$. 
In Figure~\ref{figK055}, 
we give the correlation between $L_{\nu}$ and $E_{\mu}$.  

 It is understood from these figures that here, neutrino events 
produce exclusively the downward muons which consist of 
{\it horizontal like events} and {\it exclusively horizontal events}, 
but not {\it vertical like events}, 
taking account of the transform from  
$L_{\nu}$($L_{\mu}$) to $cos\theta_{\nu}$ ($cos\theta_{\mu}$).   
Also, it is easily seen that such the muons are produce by the 
parent neutrinos whose zenith angle distribute over downward to 
upward widely.
 This fact shows without doubt that one cannot decide the direction 
of the incident neutrino even the maximum frequency for the events 
from the measurement of the produced muons.   

Here, we comment to the recent work on $L/E$ analysis by 
Dufours\cite{Dufours}, a member of 
Super-Kamiokande Collaboration\footnote
{One may take a notice that it is not appropriate to cite Ph.D due to 
their nature of "unpublished". However, many Ph.D thesis around 
Super-Kamiokande have been publised and there are no any contradictions 
between their Ph.D theses and SK papers and the detailed 
descriptions are exclusively found in these Ph.D theses.   
}.
 In her paper, she has carried out the Monte Carlo simulation with 
neutrino oscillation around $L/E$ analysis and has obtained a 
beautiful agreement between the experimental data and her Monte Carlo 
results. 
This seems to be the first Monte Carlo simulation with oscillation in 
Super-Kamiokande Collaboration, since before this work, Super-Kamiokande 
Collaboration have been comparing their experimental results with their 
Monte Carlo simulation without oscillation and have estimated neutrino 
oscillation parameters from the difference between them. 
In order to keep the 
consistency with the usual Monte Carlo simulations without oscillation 
performed by Super-Kamiokande Collaboration,
 she must have been carried out her Monte 
Carlo simulation with oscillation under 
{\it the SK assumption on the direction},
because {\it the SK assumption on the direction} is the cornerstone
throughout their analysis on neutrino oscillation. 
It is too clear that the results obtained by us contradict her 
result, even if considering the difference that we examine 
{\it the Fully Contained Events} only,
while she has examined {\it the Partially Contained Events}
 in addition to {\it the Fully Contained Events}.
Furthermore, it may be unnatural that she has obtained extremely 
beautiful agreement with the experimental data.       

\begin{figure}
\begin{center}
\vspace{-1cm}
\hspace*{-0.5cm}
\rotatebox{90}{%
\resizebox{0.4\textwidth}{!}{%
  \includegraphics{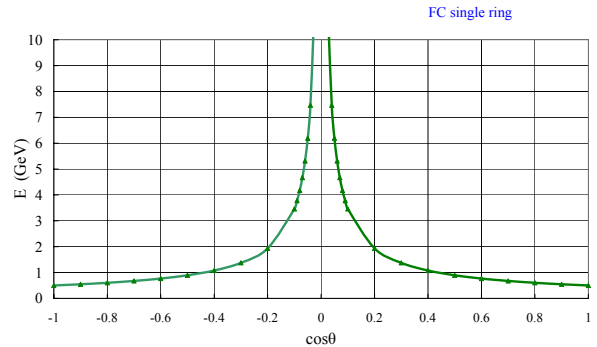}
}}
\vspace{-1.5cm}
\caption{
The excluded region for $L/E$ analysis of FC single-ring events by Super-Kamiokande.}
\label{figM58}
\end{center}
\end{figure}

\begin{figure}
\begin{center}
\vspace{-1.0cm}
\hspace*{-0.5cm}
\rotatebox{90}{%
\resizebox{0.35\textwidth}{!}{%
  \includegraphics{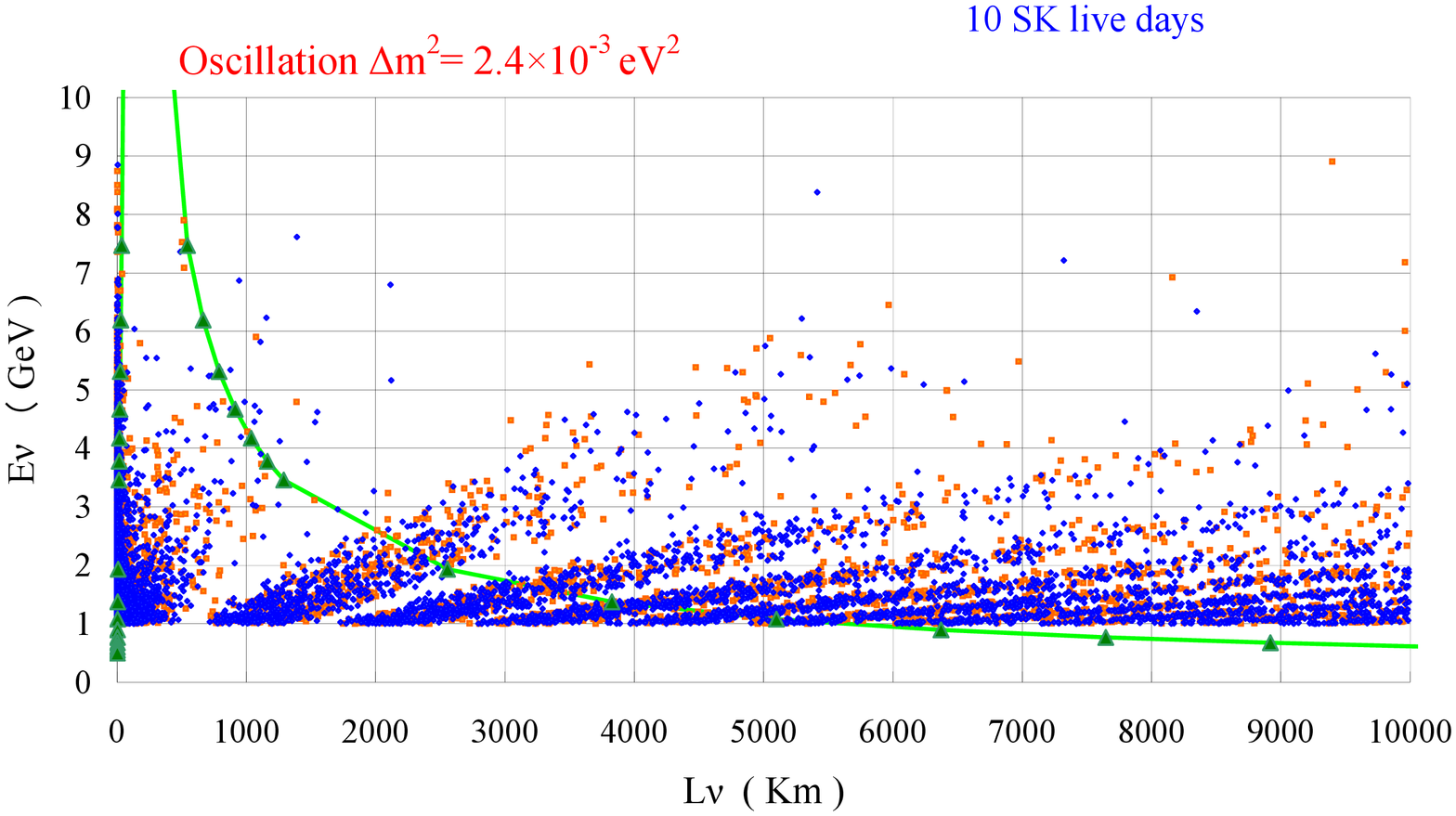}
}}
\vspace{-1.7cm}
\caption{
 The relation between the excluded region and the correlation between
 $L_{\nu}$ and $E_{\nu}$ for 10 SK live days.}
\label{figM59}
\vspace{-0.8cm}
\hspace*{-1.0cm}
\rotatebox{90}{%
\resizebox{0.4\textwidth}{!}{%
  \includegraphics{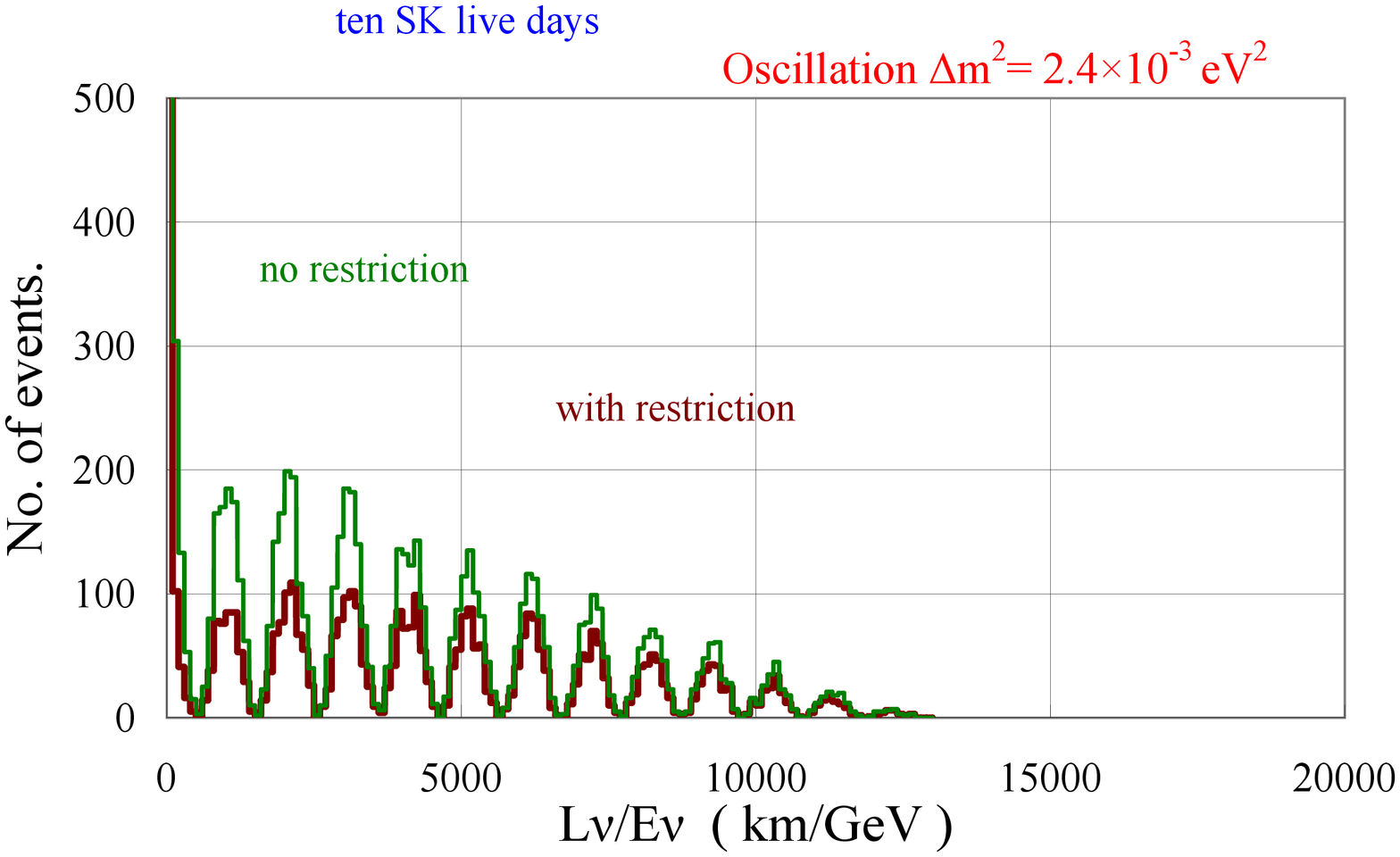}
}}
\vspace{-1.5cm}
\caption{
 $L_{\nu}/E_{\nu}$ distributions with and without the restriction 
imposed by Super-Kamiokande for 10 SK live days.}
\label{figM60}
%\end{center}
%\end{figure}

%\begin{figure}
%\begin{center}
%\vspace{-1cm}
\hspace*{-0.5cm}
\rotatebox{90}{%
\resizebox{0.35\textwidth}{!}{%
  \includegraphics{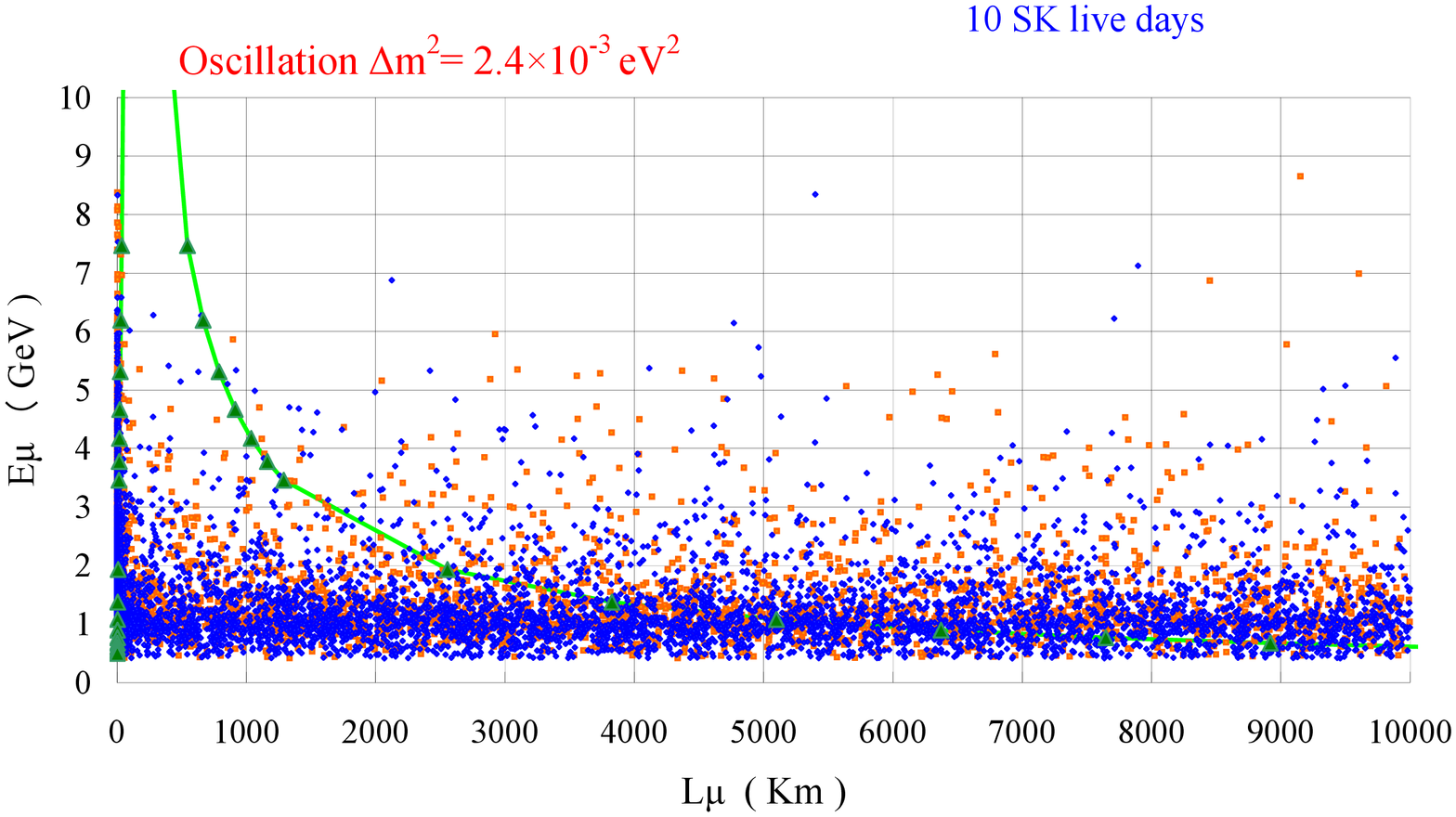}
}}
\vspace{-1.7cm}
\caption{
The relation between the excluded region and the correlation between
 $L_{\mu}$ and $E_{\mu}$ for 10 SK live days.}
\label{figM61}
\vspace{-0.8cm}
\hspace*{-1.0cm}
\rotatebox{90}{%
\resizebox{0.4\textwidth}{!}{%
  \includegraphics{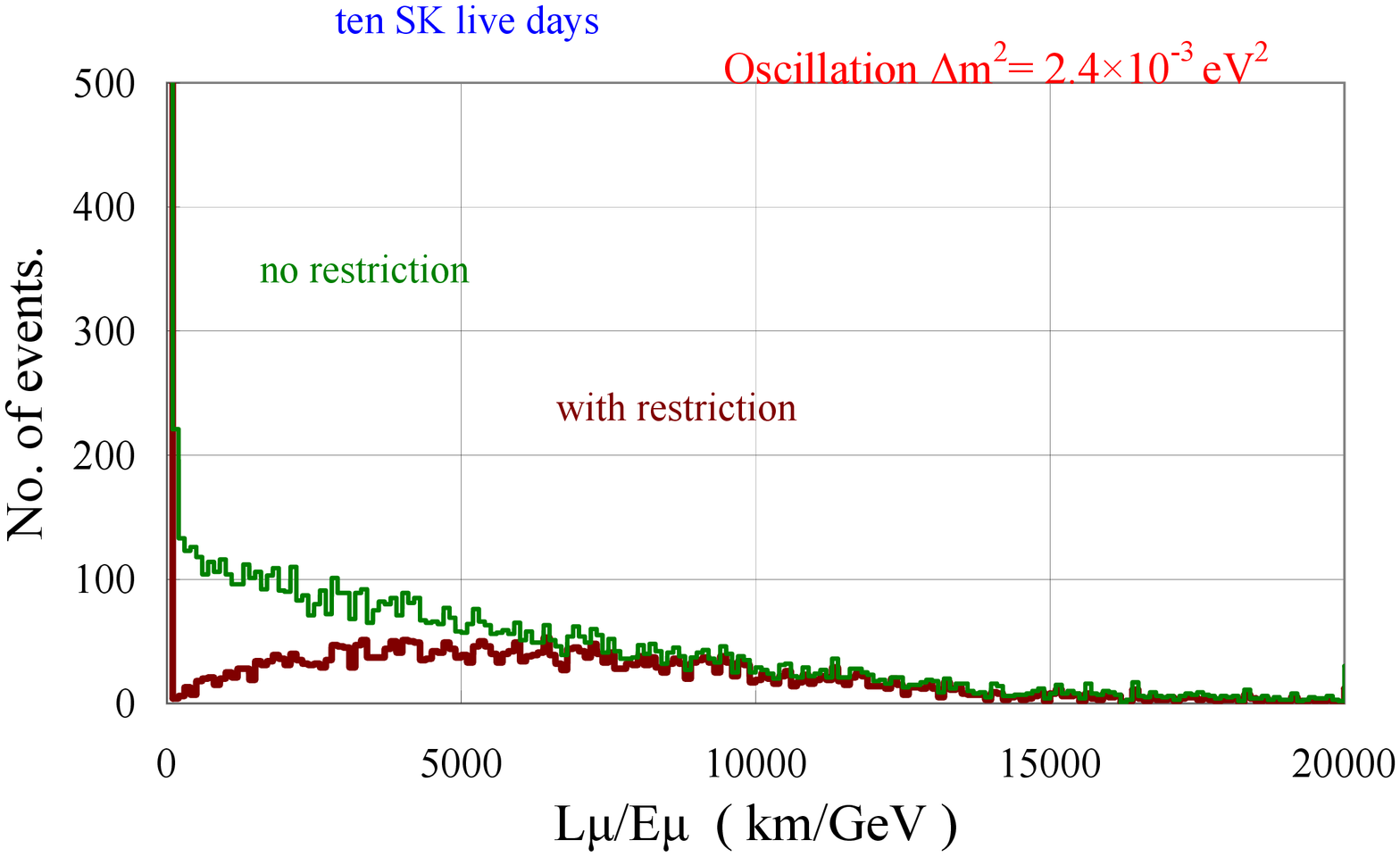}
}}
\vspace{-1.5cm}
\caption{
$L_{\mu}/E_{\mu}$ distributions with and without the restriction 
imposed by Super-Kamiokande for 10 SK live days.}
\label{figM62}
%\vspace{-1cm}
\end{center}
\end{figure}

Finally, we examine the data selection procedure made by Super-Kamiokande 
(hereafter called as SK data selection procedure) which is imposed upon 
their {\it Fully Contained Events} in the single ring muon events.
 They introduce such a procedure to exclude ambiguity mainly coming from 
horizontal like events. 
They exclude single ring muon events as {\it Fully Contained Events}
which exist within the region described in Figure 2(a) 
in their paper \cite{Ashie1}. 
We reproduce it in Figure~\ref{figM58}. 
The region enclosed by two lines is the excluded region in their analysis.
Here, it should be emphasized that we need not exclude the horizontal like 
neutrino events at all in our $L/E$ analysis against the SK analysis, 
because we have not any experimental errors even in our horizontal-like 
neutrino events due to the nature of our computer numerical 
experiment.         
However, it is not vain to examine whether the SK data selection 
procedure is appropriate or not for our analysis in our computer 
numerical experiment.  
  In Figure~\ref{figM59}, we give correlation diagram between
 $L_{\nu}$ and $E_{\nu}$ with oscillation together with the lines for 
exclusion given by Supper-Kamiokande Collaboration. 
The lines for exclusion in Figure~\ref{figM59} are transformed from 
the lines in Figure~\ref{figM58}. 
The region enclosed by two lines in Figure~\ref{figM59} are the 
excluded region. 
Namely, the events concerned within the excluded region should be 
subtracted from the original distribution.  
Thus, $L_{\nu}/E_{\nu}$ distribution of the resultant events after 
subtraction procedure (with restriction in the figure) 
is shown in Figure~\ref{figM60} together that without subtraction
(no restriction in the figure).
 It is clear from the figure that $L_{\nu}/E_{\nu}$ distribution
with the SK data selection procedure keeps still the 
characteristics of a series of the maximum oscillations in spite of the 
decrease of the events in smaller value of $L_{\nu}/E_{\nu}$. 
Here, we show $L_{\nu}/E_{\nu}$ distribution under the SK data 
selection procedure in a linear scale to make their physical image clearer.
It is needless to say that we need not introduce the SK data selection 
procedure into our computer numerical experiment due to the 
"complete experiment" by the definition, 
but even if we introduce it,
 Figure~\ref{figM60} shows that the essential characteristics of 
our $L_{\nu}/E_{\nu}$ distribution is never changed.  

Similarly we examine the case of $L_{\mu}/E_{\mu}$ distribution. 
Figure~\ref{figM61} for $L_{\mu}/E_{\mu}$ distribution corresponds 
to Figure~\ref{figM59} for $L_{\nu}/E_{\nu}$ distribution. 
Also in Figure~\ref{figM62},  
$L_{\mu}/E_{\mu}$ distribution corresponds to 
Figure~\ref{figM60} for $L_{\nu}/E_{\nu}$ distribution.

 It is clear from Figure~\ref{figM62} that the $L_{\mu}/E_{\mu}$ 
distribution with the SK data selection procedure 
do not show anything like the maximum oscillation in the same way as the 
original $L_{\mu}/E_{\mu}$ distribution.  

Summarized from Figures~\ref{figM58} to \ref{figM62}, 
we could not extract the neutrino oscillation 
parameters from $L_{\mu}/E_{\mu}$ distribution, 
even if we apply the SK data selection 
procedure to the original $L_{\mu}/E_{\mu}$ distribution, 
while we can keep the essential character of the maximum oscillations 
in $L_{\nu}/E_{\nu}$ analysis, even if we apply to the SK 
data selection procedure to the original $L_{\nu}/E_{\nu}$
distribution. 
Finally, we should say again that we need not introduce the SK data 
selection procedure into our computer numerical experiment due to the 
nature of no error experiment.

%%%%%%%%%%%%%%%%%%%%%%%%%%%%%%%%%%%%%%%%%%%%%%%%%%%%%%%

\section{Conclusion}
 The determination of the neutrino oscillation parameters is entirely 
based on the survival probability for a given flavor.
Consequently, it is inevitable to decide $L_{\nu}$ and $E_{\nu}$
as precisely as possible, when one wants specify neutrino oscillation 
parameters. 
However, in cosmic ray experiments, one may not measure both 
$L_{\nu}$ and $E_{\nu}$ due to their neutralities,
 in addition to because of the incapability for the 
determination of the directions of the incident neutrinos due to the 
essential nature of cosmic ray beams. 
Therefore, cosmic ray physicists are forced to assume the direction of
 incident neutrinos a priori, when they do not carry out computer 
numerical experiment as their second experiment. 
In the case of Super-Kamiokande Collaboration, they assume that the 
direction of incident neutrino is approximately the same as that of 
emitted lepton ({\it the SK assumption on the direction}). 
In the preceding paper\cite{Konishi2},
 we have verified that {\it the SK assumption on the direction} 
does not hold even if approximately.
 The essential conclusion obtained by the present paper is 
that in principle one may not specify the neutrino oscillation parameters 
from the cosmic ray experiments due to the unknown directions of the 
incident neutrinos.
 Our verification that Super-Kamiokande Collaboration cannot specify the 
neutrino oscillation parameters through their $L/E$ analysis at least, 
and consequently, our conclusion
can be applied to any type of experiment of cosmic ray physics where the 
direction of the neutrino, in principle, cannot be determined. 
Our conclusion tells us that only accelerator physics can specify the 
neutrino oscillation parameters reliably, if the neutrino oscillation 
really exists. 

 Deduction of our conclusion is as follows:
\begin{itemize}
\item[(1)]
There are much uncertainty factors in cosmic ray physics, compared with 
the accelerator physics due to the original nature of cosmic ray. 
Consequently, in spite of such a difficulty, if one wants to carry out 
the experiment with high precision on neutrino oscillation, we should 
focus the simplest and clearest "target" by which 
one get high quality information on neutrino oscillation.  
With such a motivation, 
we choose the single ring muon events due to QEL which they 
occur inside the detector and terminate inside the detector
 ({\it Fully Contained Events}). 
Here, the kind of neutrino concerned is clear
 (electron neutrino or muon neutrino).
 The energy of emitted lepton and its direction can be estimated reliably 
( from the standpoint of Super-Kamiokande at least).
 The circumstance around our computer numerical experiment is modeled 
after real Super-Kamiokande experiment in essential points. 
We have analyze the single ring muon events due to QEL as 
{\it Fully Contained Events} obtained by our computer 
numerical experiment.
 
\item[(2)]
We have carefully and in detail examined the validity of 
{\it the SK assumption on the direction} which is the 
"cornerstone stone" for their analysis around neutrino 
oscillation.
 As the result, we have clarified that 
{\it the SK assumption on the direction} 
does not hold even if approximately.  
Also, we examine the validity of the unique relation 
between $E_{\nu}$ and $E_{\mu}$ expressed in the polynomial obtained by 
Super-Kamiokande Collaboration and we have clarified that the unique 
relation between them  does not hold.
These two improper treatments originate from the situation
that they do not consider 
characteristics of the stochastic processes concerned seriously.
However, the unreliability on the directions of the incident neutrinos 
influences final result in a fatal manner, 
while unreliability 
on the energy estimation does not provide the significant 
errors compared with the former. 
The concrete summaries are given in (3).

\item[(3)]
Due to the nature of the computer numerical experiment, assuming neutrino 
oscillation parameters obtained by Super-Kamiokande Collaboration, 
we carry out all possible combination of $L/E$ analysis, namely, 
$L_{\nu}/E_{\nu}$ analysis, $L_{\nu}/E_{\mu}$ analysis,
$L_{\mu}/E_{\nu}$ analysis and $L_{\mu}/E_{\mu}$ analysis, based 
on the survival probability for a given flavor whose variables are
$L_{\nu}$and $E_{\nu}$.
 Among four $L/E$ analyses, only $L_{\nu}/E_{\nu}$ analysis has 
reproduced the existence of the maximum oscillations, not only the first 
maximum oscillation but also the second, the third, the fourth and so on. 
The confirmation of a series of the maximum oscillations, such as ,the 
first, the second, the third and so on in $L_{\nu}/E_{\nu}$ analysis 
shows that our computer numerical experiments have been carried out 
in a correct manner. 
 The $L_{\nu}/E_{\mu}$ analysis has reproduced the first maximum 
oscillation roughly, but cannot reproduce the maximum oscillation after 
the second. 
Both $L_{\mu}/E_{\nu}$ and $L_{\mu}/E_{\mu}$ analyses cannot reproduce 
any characteristics of the maximum oscillation at all.  
Notice that Super-Kamiokande Collaboration have carried out 
either $L_{\mu}/E_{\nu}$ analysis or $L_{\mu}/E_{\mu}$ analysis,
neither $L_{\nu}/E_{\nu}$ analysis nor $L_{\nu}/E_{\mu}$ analysis. 
Combined with the item (1), these facts tell us that the decisive variable 
in the survival probability for neutrino oscillation is 
$L_{\nu}$ but neither $L_{\mu}$, nor $E_{\nu}$, nor $E_{\mu}$. 
Thus, our verification that {\it the SK assumption on the direction} 
does not hold even approximately requests urgently
  Super-Kamiokande Collaboration to re-analyse their results around 
the zenith angle distributions for neutrino events 
which have been regarded as the establishment of the existence of the 
neutrino oscillation, because their analysis on neutrino oscillation 
(atmospheric neutrino) is entirely based on the survival probability 
and {\it the SK assumption on the direction}.  

Finally, we would like to emphasize the importance of the cosmic ray 
study in order to avoid any misunderstandings.
This characteristics of the cosmic ray study never make it lose its 
raison-d'etre.  
The main role of cosmic ray physics is to grasp qualitatively the 
essential of something like new.  Up to now, 
cosmic ray study has been contributing to find new indications in 
fundamental physics and from now on,
it will be continue to fulfill its role.

\end{itemize}

%%%%%%%%%%%%%%%%%%%%%%%%%%%%%%%%%%%%%%%%%%%%%%%%%%%%%%%%%%%%%%%%%%%%%%%%%%%%%%%%%%%%
%\noindent {\bf APPENDICES}\\
%{\bf Exact Monte Carlo Simulations in Various Cases}\\


\bibliographystyle{model1a-num-names}

\begin{thebibliography}{00}
  \bibitem{Konishi2} Konishi,E {\it et al.}, "Part~1" the paper submitted with the present one.
 \bibitem{Ashie2} Ashie,Y. {\it et al.}, Phys. Rev. D {\bf 71} (2005) 112005.
  \bibitem{Honda} Honda, M., {\it et al.}, \  Phys.\ Rev. D {\bf 52} (1996) 4985.\\
 Honda, M., {\it et al.}, \  Phys.\ Rev. D {\bf 70} (2004)043008-1. 
  \bibitem{Ashie1} Ashie,Y {\it et al.}, Phys.Rev.Lett.{\bf93}
(2004)101801-1.
  \bibitem{Konishi1} Konishi,E {\it et al.}, arXiv hep-ex/0808.0664v2.
  \bibitem{Dufours} Dufours, F M, Ph.D thesis(2009), Boston University. 
  \bibitem{Ishitsuka} Ishitsuka, M, Ph.D thesis (2004), University of Tokyo. 

\end{thebibliography}
 
%  \bibitem{Konishi} Konishi,E {\it et al.}, arXiv hep-ex/0808.0664v2
%  \bibitem{part1} Konishi,E.,Minorikawa,Y.,Galkin,V.I.,Ishiwata,M.,
%Nakamura,I.,Kato,M. and Misaki,A arXiv:hep-ex/0808.0664v2





\end{document}